\RequirePackage{fix-cm}
\documentclass[smallcondensed]{svjour3}     
\RequirePackage{amsmath}
\smartqed  
\usepackage{graphicx}
\usepackage{mathptmx}      
\usepackage{graphicx}
\usepackage{epstopdf}
\usepackage{etoolbox}
\usepackage{lmodern}
\usepackage[nottoc,notlot,notlof]{tocbibind}
\usepackage{amssymb}
\usepackage{yhmath}
\usepackage{placeins}
\usepackage{commath}
\usepackage{multirow}
\usepackage{array}
\usepackage{natbib}
\usepackage{caption}
\usepackage[Symbolsmallscale]{upgreek}
\usepackage{enumitem}

\usepackage{color}
\def \figwidth {3.6cm}
\newcolumntype{C}[1]{>{\centering\let\newline\\\arraybackslash\hspace{0pt}}m{#1}}
\setcitestyle{aysep={},yysep={;}}

\makeatletter
\renewcommand\paragraph{\@startsection{paragraph}{4}{\z@}%
            {-2.5ex\@plus -1ex \@minus -.25ex}%
            {1.25ex \@plus .25ex}%
            {\normalfont\normalsize\itshape}}
\makeatother
 \journalname{Machine Learning}
\begin{document}
\title{Adaptive Multiple Importance Sampling for Gaussian Processes
}


\author{Xiaoyu Xiong         \and
        V\'aclav \v{S}m\'idl  \and 
        Maurizio Filippone
}


\institute{Xiaoyu Xiong \at
              School of Computing Science, University of Glasgow, UK \\
              \email{x.xiong.1@research.gla.ac.uk}           
           \and
           V\'aclav \v{S}m\'idl \at
              Institute of Information Theory and Automation, Prague, Czech Republic \\
              \email{smidl@utia.cas.cz}
            \and
           Maurizio Filippone \at
             Department of Data Science, EURECOM, Biot, France \\
              \email{maurizio.filippone@eurecom.fr}
}

\date{}
\maketitle

\begin{abstract}
In applications of Gaussian processes where quantification of uncertainty is a strict requirement, it is necessary to accurately characterize the posterior distribution over Gaussian process covariance parameters.
Normally, this is done by means of standard Markov chain Monte Carlo (MCMC) algorithms.
Motivated by the issues related to the complexity of calculating the marginal likelihood that can make MCMC algorithms inefficient, this paper develops an alternative inference framework based on Adaptive Multiple Importance Sampling (AMIS).
This paper studies the application of AMIS in the case of a Gaussian likelihood, and proposes the Pseudo-Marginal AMIS for non-Gaussian likelihoods, where the marginal likelihood is unbiasedly estimated.
The results suggest that the proposed framework outperforms MCMC-based inference of covariance parameters in a wide range of scenarios and remains competitive for moderately large dimensional parameter spaces.
\end{abstract}


\section{Introduction}

Gaussian Processes (GPs) have been proved to be a successful class of statistical inference methods for data analysis in several applied domains, such as pattern recognition \citep{Rasmussen06,Bishop06,FilipponeIEEETPAMI14}, neuroimaging \citep{FilipponeAOAS12}, signal processing \citep{KimIEEETAFFC14}, Bayesian optimization \citep{Jones98}, and emulation and calibration of computer codes \citep{Kennedy01}. 
The features that make GPs appealing are the nonparametric formulation that yields the possibility to flexibly model data, and the parameterization that makes it possible to gain some understanding on the system under study.
These properties hinge on the parameterization of the GP covariance function and on the way GP covariance parameters are optimized or inferred. 

It is established that optimizing covariance parameters can severely affect the ability of the model to quantify uncertainty in predictions \citep{Neal99,FilipponeIEEETPAMI14,FilipponeAOAS12,Taylor12}.
Therefore, in applications where this is undesirable, it is necessary to accurately characterize the posterior distribution over covariance parameters and propagate this source of uncertainty forward to predictions.
This task is particularly challenging when dealing with GPs.
Inference of GP covariance parameters is generally analytically intractable, but a further complication arises from the difficulties associated with the repeated computation of the so called marginal likelihood, which is necessary when employing any standard inference method.
In particular, in the case of a Gaussian likelihood, the marginal likelihood is computable but extremely costly due to the cubic scaling with the number of input vectors.
When the likelihood function is not Gaussian, e.g., in classification, in ordinal regression, or in Cox-processes, the marginal likelihood is not even computable analytically.

In response to the challenges above, a large body of the literature develops approximate inference methods \citep{Williams98,Opper00,Kuss05,Rasmussen06,Nickisch08,Hensman15}, which, although successful in many cases, give no guarantees on the effect on the quantification of uncertainty in practice.
In the direction of quantify uncertainty without introducing any bias, in the literature there have been attempts to employ Markov chain Monte Carlo (MCMC) techniques; we can broadly divide such attempts in works that propose reparameterization techniques \citep{Neal99,Murray10,Vanhatalo07,FilipponeML13}, or methods that carry out inference based on unbiased computations of the marginal likelihood \citep{FilipponeIEEETPAMI14,FilipponeICPR14,Murray15}. 
Although these approaches proved successful in a variety of scenarios, employing MCMC algorithms may lead to inefficiencies; for instance, optimal acceptance rates for popular MCMC algorithms such as the Metropolis-Hastings (MH) algorithm (around $25\%$ \citep{Roberts97b}) and the Hybrid Monte Carlo (HMC) algorithm (about $65\%$ \citep{Beskos13,Neal11}) indicate that several expensive computations are wasted.
Furthermore, it is established that introducing adaptivity into MCMC proposal mechanisms to improve efficiency may lead to convergence issues \citep{Andrieu01}.

In this paper we develop a general framework to learn GPs aimed at overcoming the aforementioned limitations of MCMC methods for GPs, where expectations under the posterior distribution over covariance parameters are carried out by means of the Adaptive Multiple Importance Sampling (AMIS) algorithm \citep{Cornuet12}.
The application of this framework to the Gaussian likelihood case, although novel, is relatively straightforward given that the likelihood is computable.
In the case of non-Gaussian likelihoods, the impossibility to compute the likelihood exactly, motivates us to propose a novel version of AMIS where the likelihood is only (unbiasedly) estimated.
Inspired by the Pseudo-Marginal MCMC approaches \citep{Andrieu09}, we therefore propose the Pseudo-Marginal AMIS (PM-AMIS) algorithm, and provide a theoretical analysis showing under which conditions PM-AMIS yields expectations under covariance parameters without introducing any bias.
The proposed PM-AMIS is an instance of the Importance Sampling squared (IS$^2$) algorithms \citep{Pitt12,Tran14} that are gaining popularity as practical Bayesian inference methods.

Summarizing, the main contribution of this work are:
(i) the application of AMIS to learn GPs with any likelihoods;
(ii) a theoretical analysis of PM-AMIS;
(iii) an extensive comparison of convergence speed with respect to computational complexity of AMIS versus MCMC methods.

The results demonstrate the value of our proposal.
In particular, the results indicate that AMIS is competitive with MCMC algorithms in terms of convergence speed over computational cost when calculating expectations under the posterior distribution over covariance parameters.
Furthermore, the results suggest that AMIS is a valid alternative to MCMC algorithms even in the case of moderately large dimensional parameter spaces, which is common when employing richly parameterized covariances (e.g., Automatic Relevance Determination (ARD) covariances \citep{MacKay94}).
Overall, the results suggest a promising direction to speedup inference over GP covariance parameters given that importance sampling-based inference methods, unlike MCMC algorithms, are inherently parallel.

The paper is organised as follows. In section 2 we provide a brief review of GP regression and Bayesian inference. Section 3 presents the proposed Adaptive Multiple Importance Sampling for Gaussian Processes. Section 4 reports the experiments and the results. In section 5, we conclude the paper.

\section{Bayesian Gaussian Processes}

\subsection{Gaussian Processes}
Let $\mathbf{X}$ be a set of $n$ input vectors $\mathbf{x}_i \in \mathbb{R}^d (1\leq i \leq n)$ and let $\mathbf{y}$ be the vector consisting of the corresponding observations $y_i$. 
In most GP models, the assumption is that observations are conditionally independent given a set of $n$ latent variables.
Such latent variables are modeled as realizations of a function $f(\mathbf{x})$ at the input vectors $\mathbf{x}_1, \dots, \mathbf{x}_n$, i.e., $ \mathbf{f} = \{f(\mathbf{x}_1), \dots, f(\mathbf{x}_n)\}$. 
Latent variables are used to express the likelihood function, that under the assumption of independence becomes $p(\mathbf{y}\mid \mathbf{f}) =\prod_{i=1}^{n}p(y_i\mid f_i)$, with $p(y_i\mid f_i)$ depending on the particular type of data being modeled (e.g., Gaussian for regression, Bernoulli for probit classification with probability $P(y_i = 1) = \Upphi(f(\mathbf{x}_i))$ where $\Upphi$ is defined as the cumulative normal distribution).

What characterizes GP models is the way the latent variables are specified.
In particular, in GP models the assumption is that the function $f(\mathbf{x})$ is distributed as a GP, which implies that the latent function values $\mathbf{f}$ are jointly distributed as a Gaussian $ p(\mathbf{f} \mid \boldsymbol{\theta})
 \sim {\cal N} (\mathbf{0},\mathbf{K})$, where $\mathbf{K}$ is the covariance matrix.
The entries of the covariance matrix $\mathbf{K}$ are specified by a covariance (kernel) function with hyperparameters $\boldsymbol{\theta}$ between pair of input vectors. 
In this work, two kinds of covariance function are considered. The first is the RBF (Radial Basis Function) kernel defined as:
\begin{equation}\label{eq:RBF_kernel}
  k(\mathbf{x}_i, \mathbf{x}_j) = \sigma \exp\left\{-\frac{1}{\tau^2}\parallel\mathbf{x}_i - \mathbf{x}_j \parallel^2\right\}  
 \end{equation}
The parameter $\tau$ defines the length-scale of the interaction between the input vectors, $\sigma$ represents the marginal variance for each latent variable. 
The second is the ARD kernel, which takes the form:
\begin{equation}\label{eq:ARD_kernel}
  k(\mathbf{x}_i, \mathbf{x}_j) = \sigma \exp\left\{- \sum_{r=1}^{d}\frac{1}{{\tau_r}^2}(\mathbf{x}_{i(r)} - \mathbf{x}_{j(r)})^2\right\}  
 \end{equation} 
The advantage of the ARD kernel is that it accounts for the influence of each feature on the prediction of $y$, with larger values of parameters ($\tau_1, ... , \tau_d$) indicating a higher influence of the corresponding features \citep{KimIEEETAFFC14}.
For simplicity of notation, in the remainder of the paper we will denote by $\boldsymbol{\theta}$ the vector of all kernel parameters.

When making predictions, using a point estimate of $\boldsymbol{\theta}$ has been reported to potentially underestimate the uncertainty in predictions or yield inaccurate assessment of the relative influence of different features \citep{FilipponeAOAS12, FilipponeIEEETPAMI14, Bishop06}. 
Therefore, a Bayesian approach is usually adopted to overcome these limitations, which entails characterizing the posterior distribution over covariance parameters. 
In order to do so, it is necessary to employ methods, such as MCMC, that require computing the marginal likelihood every time $\boldsymbol{\theta}$ is updated.
We now discuss the challenges associated with the computation of the marginal likelihood for the special case of a Gaussian likelihood and the more general case of a non-Gaussian likelihood.

\subsubsection{Gaussian likelihood}
In the GP regression setting, the observations $ \mathbf{y}$ are modeled to be Gaussian distributed with mean of $\mathbf{f}$ (latent variables) and covariance $\lambda\mathbf{I}$, where $\mathbf{I}$ denotes the identity matrix and $\lambda$ is the variance of noise on the observations. 
In this setting, the likelihood $p(\mathbf{y} \mid \mathbf{f})$ and the GP priors $ p(\mathbf{f} \mid \boldsymbol{\theta})$ form a conjugate pair, so latent variables can be integrated out of the model leading to $p(\mathbf{y}\mid \mathbf{X},\mathbf{\boldsymbol\theta}) \sim {\cal N} (0,\mathbf{C})$, where $\mathbf{C} = \mathbf{K} + \lambda\mathbf{I}$.
This gives a direct access to the log-marginal likelihood
$$
\log[p(\mathbf{y} | \boldsymbol{\theta})] = -\frac{1}{2} \log|\mathbf{C}| - \frac{1}{2} \mathbf{y}^{\top} \mathbf{C}^{-1} \mathbf{y} + \mathrm{const.}
$$
Although computable, the log-marginal likelihood requires computing the log determinant of $\mathbf{C}$ and solve a linear system involving $\mathbf{C}$.
These calculations are usually carried out by factorizing the matrix $\mathbf{C}$ using the Cholesky decomposition, giving $\mathbf{C} = \mathbf{L}\mathbf{L}^{\top}$, with $\mathbf{L}$ lower triangular. 
The Cholesky algorithm requires $O(n^3)$ operations, but after that computing the terms of the marginal likelihood requires at most $O(n^2)$ operations \citep{Rasmussen06}.

\subsubsection{Non-Gaussian likelihood}
In the case of non-Gaussian likelihoods, the likelihood $p(\mathbf{y} \mid \mathbf{f})$ and the GP prior $p(\mathbf{f} \mid \boldsymbol{\theta})$ are no longer conjugate.
As a consequence, it is not possible to solve the integral needed to integrate out latent variables
$$
p(\mathbf{y} | \boldsymbol{\theta}) = \int p(\mathbf{y} | \mathbf{f}) p(\mathbf{f} | \boldsymbol{\theta}) d\mathbf{f}
$$
and this requires some form of approximation.
A notable example is GP probit classification, which is what we explore in detail in this paper, where the observations $\mathbf{y}$ are assumed to be Bernoulli distributed with success probability given by:
\begin{equation}\label{eq:success_probab_of_probit}
  p(y_i  \mid f_i) = \Upphi(y_if_i)
 \end{equation}
For GPs with non-Gaussian likelihood, there have been several proposals on how to carry out approximation to integrate out the latent variables, or to avoid approximations altogether.
The focus of this paper is on methods that do not introduce any bias in the calculation of expectation under the posterior over covariance parameters, so we will discuss these approaches in detail in the next sections.

\subsection{Bayesian inference of covariance parameters}

For simplicity of notation, we will denote the posterior distribution over covariance parameters by 
\begin{equation}\label{eq:posterior}
\pi(\mathbf{\boldsymbol\theta}) := p(\mathbf{\boldsymbol\theta} | \mathbf{y}, \mathbf{X}) = \frac{p(\mathbf{y}\mid \mathbf{\boldsymbol\theta})p(\mathbf{\boldsymbol\theta})}{\int p(\mathbf{y}\mid \mathbf{\boldsymbol\theta})p(\mathbf{\boldsymbol\theta}) d\mathbf{\boldsymbol\theta}}
\end{equation}
where $p(\mathbf{\boldsymbol\theta})$ encodes any prior knowledge on the parameters $\boldsymbol\theta$. 
Within the Bayesian framework, we are usually interested in calculating expectations of functions of $\boldsymbol{\theta}$ with respect to the posterior distribution, i.e., $E_{\pi(\boldsymbol\theta)}[h(\boldsymbol\theta)]$.
For instance, setting $h(\boldsymbol\theta) = p(y_\star\mid \boldsymbol\theta,\mathbf{x_\star}, \mathbf{y}, \mathbf{X})$, we obtain the predictive distribution for the label $y_\star$ associated with a new input vector $\mathbf{x}_\star$. 

The denominator needed to normalize the posterior distribution $\pi(\mathbf{\boldsymbol\theta})$ is intractable, so it is not possible to characterize the posterior distribution analytically.
Despite this complication, it is possible to resort to a Monte Carlo approximation to compute expectations under the posterior distribution of $\boldsymbol{\theta}$:
\begin{equation} \label{eq:monte_approx_MCMC}
E_{\pi(\boldsymbol\theta)}[h(\boldsymbol\theta)] \simeq \frac{1}{N}\sum_{i=1}^{N} h(\boldsymbol\theta^{(i)})
\end{equation}
where $\boldsymbol\theta^{(i)}$ denotes the $i$th of $N$ samples from $\pi(\mathbf{\boldsymbol\theta})$. 
However, as it is usually not feasible to draw samples from $\pi(\boldsymbol\theta)$ directly, usually MCMC methods are employed to generate samples from the posterior $\pi(\boldsymbol\theta)$.

An alternative way to compute expectations, is by means of importance sampling, which takes the following form:
\begin{equation} \label{eq:EXPEC_IS}
E_{\pi(\boldsymbol\theta)}[h(\boldsymbol\theta)] = \int h(\boldsymbol\theta) \frac{\pi(\boldsymbol\theta)}{q(\boldsymbol\theta)} q(\boldsymbol\theta)d\boldsymbol\theta
\end{equation} where $q(\boldsymbol\theta)$ is the importance distribution. The corresponding Monte Carlo approximation is of the form:
\begin{equation} \label{eq:monte_approx_IS}
E_{\pi(\boldsymbol\theta)}[h(\boldsymbol\theta)] \simeq \frac{1}{N}\sum_{i=1}^{N}h(\boldsymbol\theta^{(i)}) \frac{\pi(\boldsymbol\theta^{(i)})}{q(\boldsymbol\theta^{(i)})}
\end{equation} 
where now the samples $\boldsymbol\theta^{(i)}$ are drawn from the importance sampling distribution $q(\boldsymbol\theta)$.
The key to make this Monte Carlo estimator accurate is to choose $q(\boldsymbol\theta)$ to be similar to the function that needs to be integrated.
It is easy to verify that when $q(\boldsymbol\theta)$ is proportional to the function that needs to be integrated, the variance of the importance sampling estimator is zero.
Therefore, the success of importance sampling relies on constructing a tractable importance distribution $q(\boldsymbol\theta)$ that well approximates $h(\boldsymbol\theta) \pi(\boldsymbol\theta)$.
In the remainder of this paper we will discuss and employ methods that adaptively construct $q(\boldsymbol\theta)$.

Both Monte Carlo approximations in equations \eqref{eq:monte_approx_MCMC} and \eqref{eq:monte_approx_IS} converge to the desired expectation, and in practice, they can estimate the desired integral to a given level of precision \citep{Gelman92,Flegal08}.
The experimental part of this work is devoted to the study of the convergence properties of the expectation $E_{\pi(\boldsymbol\theta)}[h(\boldsymbol\theta)]$ with respect to the computational effort needed to carry out the Monte Carlo approximations in Equations \eqref{eq:monte_approx_MCMC} and \eqref{eq:monte_approx_IS}.

\subsection{Pseudo-Marginal MCMC for inference of covariance parameters}

Difficulties in computing the expectation in equation \eqref{eq:monte_approx_MCMC} arise when the likelihood is not Gaussian, thus leading to the the impossibility to calculate the marginal likelihood exactly, which is necessary to employ standard MCMC algorithms to draw from the posterior $\pi(\mathbf{\boldsymbol\theta})$. 
In cases where the marginal likelihood can be unbiasedly estimated, it is possible to resort to so called Pseudo-Marginal MCMC approaches.
Taking the Metropolis-Hastings algorithm as an example, it is possible to replace the exact calculation of the Hastings ratio
$$
\frac{p(\mathbf{y}\mid \mathbf{\boldsymbol\theta^{\prime}})p(\mathbf{\boldsymbol\theta^{\prime}})}{p(\mathbf{y}\mid \mathbf{\boldsymbol\theta})p(\mathbf{\boldsymbol\theta})}
$$
by an approximation where the marginal likelihood is only unbiasedly estimated 
$$
\frac{\tilde{p}(\mathbf{y}\mid \mathbf{\boldsymbol\theta^{\prime}})p(\mathbf{\boldsymbol\theta^{\prime}})}{\tilde{p}(\mathbf{y}\mid \mathbf{\boldsymbol\theta})p(\mathbf{\boldsymbol\theta})}
$$
where $\tilde{p}(\mathbf{y}\mid \mathbf{\boldsymbol\theta})$ denotes such an approximation.
Interestingly, the introduction of this approximation does not affect the properties of the MCMC approach that still yields samples from the correct posterior $\pi(\mathbf{\boldsymbol\theta})$. 
The effect of introducing this approximation is that the efficiency of the corresponding MCMC approach is reduced; this is due to the possibility that the algorithm accepts a proposal with an largely overestimated value of the marginal likelihood, making it difficult for any new proposals to be accepted.

By inspecting the GP marginal likelihood 
 \begin{equation} \label{eq:margin_lik_gpc}
   p(\mathbf{y} \mid \boldsymbol\theta) = \int  p(\mathbf{y} \mid \mathbf{f})p(\mathbf{f} \mid \boldsymbol{\theta})d\mathbf{f}
 \end{equation}
we observe that we can attempt to unbiasedly estimate this integral using importance sampling:
 \begin{equation} \label{eq:pm_estimator_by_IS}
   \tilde{p}(\mathbf{y} \mid \boldsymbol\theta)\simeq \frac{1}{N_{imp}}\sum_{i=1}^{N_{imp}}\frac{p(\mathbf{y}\mid \mathbf{f}_i)p(\mathbf{f}_i \mid \boldsymbol{\theta})}{q(\mathbf{f}_i \mid \mathbf{y}, \boldsymbol{\theta})}
 \end{equation} 
Here $N_{imp}$ is the number of samples $\mathbf{f}_i$ drawn from the importance density $q(\mathbf{f} \mid \mathbf{y}, \boldsymbol{\theta})$. 
The motivation for attempting this approximation is that in the literature there have been many efforts to construct accurate approximations to the posterior distribution over the latent variables $p(\mathbf{f} | \mathbf{y}, \boldsymbol{\theta}) \propto p(\mathbf{y}\mid \mathbf{f}_i)p(\mathbf{f}_i \mid \boldsymbol{\theta})$.
This suggests that we can hope that the resulting estimator for the marginal likelihood is accurate enough to introduce an acceptable amount of noise in the calculation of the Hastings ratio to make the resulting MCMC approach reasonably efficient.
In this paper, we investigate approximations $q(\mathbf{f} \mid \mathbf{y}, \boldsymbol{\theta})$ to the posterior obtained by the Laplace Approximation and Expectation Propagation algorithms \citep{Rasmussen06}.

\section{Adaptive Multiple Importance Sampling for Gaussian Processes} \label{sec:AMIS}

Inefficiencies arising from the use of MCMC methods to sample from the posterior distribution over covariance parameters are due to fact that several proposals are rejected.
In order to mitigate this issue, some adaptation mechanisms of the proposals can be used based on previous MCMC samples, but the chain resulting from the adaptivity is no longer Markovian.
As a result, elaborate ergodicity results are needed to establish convergence to the true posterior distribution \citep{Haario99,Haario01,Andrieu01}.

In response to this, \citet{Cappe04} proposed a universal adaptive sampling scheme called population Monte Carlo (PMC), where the difference from Sequential Monte Carlo (SMC) \citep{Doucet01} is that the target distribution becomes static. 
This method is reported to have better adaptivity than MCMC due to the fact that the use of importance sampling makes ergodicity not an issue. 
At each iteration of PMC, sampling importance resampling \citep{Rubin88} is used to generate samples that are assumed to be marginally distributed from the target distribution and hence the approach is unbiased and can be stopped at any time. 
Moreover, the importance distribution can be adapted using part (generated at each iteration) or all of the importance sample sequence. \cite{Douc07a,Douc07b} also introduced updating mechanisms for the weights of the mixture in the so called D-kernel PMC, which leads to a reduction either in Kullback divergence between the mixture and the target distribution or in the asymptotic variance for a function of interest. 
An earlier adaptive importance sampling strategy is illustrated in \cite{Oh92}. 

\citet{Cornuet12} proposed a new perspective of adaptive importance sampling (AIS), called adaptive multiple importance sampling (AMIS), where the difference with the aforementioned PMC methods is that the importance weights of all simulations, produced previously as well as currently, are re-evaluated at each iteration. 
This method follows the ``deterministic multiple mixture'' sampling scheme of \cite{Owen00}. 
The corresponding importance weight takes the form
\begin{equation} \label{eq:amisweight}
 w_i^t = f({\boldsymbol{\theta}}_i^t)/\frac{1}{\sum_{t=0}^{T-1}N_t}\sum_{t=0}^{T-1}N_tq_t({\boldsymbol{\theta}}_i^t; \widehat{\boldsymbol\gamma_t})
 \end{equation} where $T$ is the total number of iterations, $f(.)$ denotes the target distribution $\pi(.)$ up to a constant, i.e., $\pi(.) \propto f(.) $, $q_t(.)$ denotes the importance density at iteration $t$ with sequentially updated parameters $\widehat{\boldsymbol\gamma_t}$, ${\boldsymbol{\theta}}_i^t$ are samples drawn from $q_t(.)$ with $0 \leq t \leq T - 1 $, $1 \leq i \leq N_t$.  

The fixed denominator in \eqref{eq:amisweight} gives the name ``deterministic multiple mixture''. 
The motivation is that this construction can achieve an upper bound on the asymptotic variance of the estimator without rejecting any simulations \citep{Owen00}. 
In AMIS, the parameters $\boldsymbol\gamma$ of a parametric importance function $q_t(\boldsymbol{\theta}; \boldsymbol\gamma)$ are sequentially updated using the entire sequence of weighted importance samples, based on efficiency criteria such as moment matching, minimum Kullback divergence with respect to the target or minimum variance of the weights (see, e.g., \cite{Ortiz00} for stochastic gradient-based optimization of these efficiency criteria).
This leads to a sequence of importance distributions, $q_1(\boldsymbol{\theta}; \widehat{\boldsymbol\gamma_1}),..., q_T(\boldsymbol{\theta}; \widehat{\boldsymbol\gamma_T})$. 
Algorithm 1 gives the pseudo code of the generic AMIS algorithm.

\begin{table}[t]
\rule{\textwidth}{1pt}
\textbf{Algorithm 1} Generic AMIS as analyzed by \cite{Cornuet12} \\
\rule{\textwidth}{0.5pt} 
\begin{itemize}
\item At iteration $t = 0$,
\begin{enumerate}
\setlength\itemsep{0.5em}
\item Generate $N_0$ independent samples ${\boldsymbol{\theta}}_i^0 (1\leq i \leq N_0)$ from the initial importance density $q_0(\boldsymbol{\theta}; \widehat{\boldsymbol\gamma_0})$
\item For $1\leq i \leq N_0$, compute $\delta_i^0 = N_0 q_0(\boldsymbol{\theta}_i^0$; $\widehat{\boldsymbol\gamma_0}), w_i^0 = f(\boldsymbol{\theta}_i^0) \Big / q_0(\boldsymbol{\theta}_i^0; \widehat{\boldsymbol\gamma_0})$
\item Estimate $\widehat{\boldsymbol\gamma_1}$ of $q_1(\boldsymbol{\theta}; \widehat{\boldsymbol\gamma_1})$ using the weighted samples $(\{\boldsymbol{\theta}_1^0, w_1^0\}$, ..., $\{\boldsymbol{\theta}_{N_0}^0, w_{N_0}^0\})$ and a well-chosen efficiency criterion for estimation.
\end{enumerate}

\item At iteration $t = 1$, ..., $T-1$,
\begin{enumerate}
\setlength\itemsep{0.5em}
\item Generate $N_t$ independent samples ${\boldsymbol{\theta}}_i^t (1\leq i \leq N_t)$ from $q_t(\boldsymbol{\theta}; \widehat{\boldsymbol\gamma_t})$

\item For $1\leq i \leq N_t$, compute the multiple mixture at $\boldsymbol{\theta}_i^t$
\begin{equation*}
\delta_i^t = N_0 q_0(\boldsymbol{\theta}_i^t; \widehat{\boldsymbol\gamma_0}) +  \sum_{l=1}^{t}N_lq_l(\boldsymbol{\theta}_i^t; \widehat{\boldsymbol\gamma_t})
\end{equation*}
and derive the importance weights of $\boldsymbol{\theta}_i^t$
\begin{equation*}
w_i^t = f(\boldsymbol{\theta}_i^t) \Big / \Big[\delta_i^t \Big / \sum_{j=0}^{t}N_j \Big]\
\end{equation*}

\item For $1\leq l \leq t-1$ and $1\leq i \leq N_l$, update the past importance weights as
\begin{equation*}
\delta_i^l \leftarrow \delta_i^l + N_t q_t(\boldsymbol{\theta}_i^l;\widehat{\boldsymbol\gamma_t}) \qquad\text{and}\qquad w_i^l \leftarrow f(\boldsymbol{\theta}_i^l) \Big / \Big[\delta_i^l \Big / \sum_{j=0}^{t}N_j \Big]\
\end{equation*}

\item Estimate $\widehat{\boldsymbol\gamma_{t+1}}$ using all the weighted samples 
\begin{equation*}
(\{\boldsymbol{\theta}_1^0, w_1^0\}, ..., \{\boldsymbol{\theta}_{N_0}^0, w_{N_0}^0\}, ..., \{\boldsymbol{\theta}_1^t, w_1^t\}, ..., \{\boldsymbol{\theta}_{N_t}^t, w_{N_t}^t\})
\end{equation*}
and the same efficiency criterion for estimation.
\end{enumerate}
\end{itemize}
\rule{\textwidth}{0.5pt} 
\end{table}

In this paper, we use a Gaussian importance density with mean $\boldsymbol{\mu}$ and covariance $\boldsymbol{\Sigma}$, i.e., $\boldsymbol\gamma_t = (\boldsymbol{\mu}^t,\boldsymbol{\Sigma}^t)$. We also choose moment matching as the efficiency criterion to estimate $\widehat{\boldsymbol\gamma_t} = (\hat{\boldsymbol{\mu}}^t, \hat{\boldsymbol{\Sigma}}^t)$ using the self-normalized AMIS estimator:
\begin{equation*}
\hat{\boldsymbol{\mu}}^t = \frac{\sum_{l=0}^{t}\sum_{i=1}^{N_l}w_i^l\boldsymbol{\theta}_i^l}{\sum_{l=0}^{t}\sum_{i=1}^{N_l}w_i^l} \qquad \text{and}\qquad \hat{\boldsymbol{\Sigma}}^t = \frac{\sum_{l=0}^{t}\sum_{i=1}^{N_l}w_i^l(\boldsymbol{\theta}_i^l - \hat{\boldsymbol{\mu}}^t)(\boldsymbol{\theta}_i^l - \hat{\boldsymbol{\mu}}^t)^T}{\sum_{l=0}^{t}\sum_{i=1}^{N_l}w_i^l}
\end{equation*}

Despite the efficiency brought by AMIS compared with other AIS techniques, proving convergence of this algorithm is not straightforward.  
The work in \cite{Marin14} proposed a modified version of AMIS called MAMIS, aiming at obtaining a variant of AMIS where convergence can be more easily established. 
The difference is that the new parameters $ \widehat{\boldsymbol\gamma_t}$ are estimated based only on samples produced at iteration $t$, i.e., $\boldsymbol{\theta}_1^t,...,\boldsymbol{\theta}_{N_t}^t$, with classical weights $f(\boldsymbol{\theta}_i^t)/q(\boldsymbol{\theta}_i^t; \widehat{\boldsymbol\gamma_t} )$. 
Then the weights of all simulations are updated according to \eqref{eq:amisweight} to give the final output. 
The sample size $N_t$ is suggested to grow at each iteration so as to improve the accuracy of $\widehat{\boldsymbol\gamma_t}$.
MAMIS effectively solves any convergence issues of AMIS, but convergence is slower due to the fact that less samples are used to update the importance distribution.

\subsection{Pseudo-Marginal AMIS} \label{sec:pm-amis}

 The above AMIS/MAMIS are designed for the general analytically computable marginal likelihood such as in the case of GP regression. 
In this paper, we proposed using AMIS to sample from the posterior where the likelihood is analytically intractable but can be unbiasedly estimated. 
 In this case, we can attempt employing AMIS replacing the exact calculation of the marginal likelihood by an unbiased estimate, obtaining an unbiased estimate of the posterior up to a normalizing constant:
 \begin{equation} \label{eq:pm_unbiased_posterior}
     \tilde{f}(\boldsymbol\theta) =  \tilde{p}(\mathbf{y} \mid \boldsymbol\theta) p(\boldsymbol\theta)
 \end{equation}
We refer to this as Pseudo-Marginal AMIS (PM-AMIS), inspired by the name Pseudo-Marginal MCMC that was given to the class of MCMC algorithms that replace exact calculations of the likelihood by unbiased estimates \citep{Andrieu09}. 
The pseudo-code of PM-AMIS is similar to that of AMIS described in Algorithm 1 except that the target distribution up to a constant $f(\boldsymbol\theta)= p(\mathbf{y} \mid \boldsymbol\theta)p(\boldsymbol\theta)$ is replaced by the above unbiased estimate $\tilde{f}(\boldsymbol\theta)$. 

It is known that despite the fact that calculations are approximate, Pseudo-Marginal MCMC methods yields samples from the correct posterior distribution over covariance parameters, so a natural question is whether the same arguments hold for our proposal.
In the remainder of this section, we provide an analysis of the properties of Pseudo-Marginal AMIS, discussing the conditions under which it yields expectations with respect to the posterior distribution over covariance parameters that are unbiased.
 As in \cite{Tran14} \cite{Pitt12}, we introduce a random variable $z$ whose distribution (denoted by $p(z \mid \boldsymbol\theta)$ herein) is determined by the randomness occurring when carrying out the unbiased estimation of the likelihood $p(\mathbf{y} \mid \boldsymbol\theta)$. Define:
 \begin{equation} \label{eq:z_for_radomness_estimate}
  z = \log \tilde{p}(\mathbf{y} \mid \boldsymbol\theta) - \log p(\mathbf{y} \mid \boldsymbol\theta)
 \end{equation}
 i.e.
 \begin{equation} \label{eq:z_for_radomness_estimate_transform}
   \tilde{p}(\mathbf{y} \mid \boldsymbol\theta) = p(\mathbf{y} \mid \boldsymbol\theta)e^z
  \end{equation}

Due to the unbiased property ($ E[\tilde{p}(\mathbf{y} \mid \boldsymbol\theta)] = p(\mathbf{y} \mid \boldsymbol\theta)$), we readily verify that $E[e^z] = 1$.
For the sake of clarity, it is useful to define the unnormalized joint density of $z$ and $\boldsymbol\theta$ as:
\begin{equation} \label{eq:f_z_theta}
 f(z, \boldsymbol\theta) =  p(\mathbf{y} \mid \boldsymbol\theta)e^z p(z \mid \boldsymbol\theta) p(\boldsymbol\theta)
\end{equation} 
with a corresponding normalized version
\begin{equation} \label{eq:pi_z_theta}
\pi(z, \boldsymbol\theta) = \frac{ f(z, \boldsymbol\theta)}{Z}
\end{equation}
Marginalizing this joint density with respect to $z$
\begin{equation} \label{eq:integral_pi_theta_z_dz}
\int \pi (z, \boldsymbol\theta)dz = \int \frac{f(z, \boldsymbol\theta)}{Z} dz = \frac{p(\mathbf{y} \mid \boldsymbol\theta)p(\boldsymbol\theta)}{Z} E[e^z] = \frac{f(\boldsymbol\theta)}{Z} 
\end{equation} 
yields the target posterior $\pi(\boldsymbol{\theta})$ of interest defined in Equation \eqref{eq:posterior}.

Recall that our objective is analyzing expectations under the posterior over the parameters $\pi(\boldsymbol{\theta})$ of some function $h(\boldsymbol{\theta})$ 
\begin{equation} \label{eq:expec_of_h(theta)_by_f_theta}
E_{\pi(\boldsymbol\theta)}[h(\boldsymbol\theta)] = \int h(\boldsymbol{\theta}) \pi(\boldsymbol{\theta})d\boldsymbol{\theta}
= \int h(\boldsymbol{\theta}) \frac{f(\boldsymbol{\theta})}{Z}d\boldsymbol{\theta} 
\end{equation}
We begin our analysis by substituting Equation \eqref{eq:integral_pi_theta_z_dz} into Equation \eqref{eq:expec_of_h(theta)_by_f_theta}, obtaining
\begin{equation} \label{eq:expec_of_h(theta)_by_f_z_and_theta}
E_{\pi(\boldsymbol\theta)}[h(\boldsymbol\theta)] = \int h(\boldsymbol{\theta}) \frac{f(z, \boldsymbol\theta)}{Z}d\boldsymbol{\theta}dz 
\end{equation}

In PM-AMIS, let $N_t$ denote the number of samples generated at each iteration $t$, $q_t(\boldsymbol\theta)$ denote the importance density at each iteration for $\pi(\boldsymbol\theta)$. We also define
\begin{equation} \label{eq:importance_density_with_z}
q_t(z,\boldsymbol\theta) =  p(z \mid \boldsymbol\theta)q_t(\boldsymbol\theta)
\end{equation}
 as the joint importance density at each iteration for $\pi(z,\boldsymbol\theta)$, $(z_i^t, {\boldsymbol{\theta}}_i^t)$ as samples drawn from $q_t(z,\boldsymbol\theta)$ with $0 \leq t \leq T, 1 \leq i \leq N_t$. 
 
Since in a practical setting $f(z, \boldsymbol\theta)$ is the only function that we can evaluate, the expectation defined in Equation \eqref{eq:expec_of_h(theta)_by_f_z_and_theta} is estimated by the self-normalized PM-AMIS estimator:
\begin{equation} \label{eq:monte_carlo_estimation_PM_AMIS}
\frac{1}{\sum_{t=0}^{T} \sum_{i=1}^{N_t} {w_i}^t}\sum_{t=0}^{T}\sum_{i=1}^{N_t}{w_i}^t h({\boldsymbol\theta}_i^t)
\end{equation} where the weights of this estimator are computed as
\begin{equation} \label{eq:weight_with_z}
{w_i}^t = \frac{f(z_i^t, {\boldsymbol\theta}_i^t)}{\frac{1}{\sum_{j=0}^{T}N_j}\sum_{l=0}^{T}N_l q_l(z_i^t, {\boldsymbol\theta}_i^t)}
\end{equation}

Expanding the terms in the computations of the weights, namely substituting Equation \eqref{eq:f_z_theta} \eqref{eq:importance_density_with_z} into Equation \eqref{eq:weight_with_z}, we have
\begin{equation} \label{eq:pm-amis-weigh_derive1}
\begin{gathered}
{w_i}^t = \frac{p(\mathbf{y} \mid {\boldsymbol\theta}_i^t)e^{z_i^t} p(z_i^t \mid {\boldsymbol\theta}_i^t) p({\boldsymbol\theta}_i^t)}{\frac{1}{\sum_{j=0}^{T}N_j}\sum_{l=0}^{T}N_l p(z_i^t \mid {\boldsymbol\theta}_i^t) q_l({\boldsymbol\theta}_i^t)}  \\
= \frac{p(\mathbf{y} \mid {\boldsymbol\theta}_i^t)e^{z_i^t}  p({\boldsymbol\theta}_i^t)}{\frac{1}{\sum_{j=0}^{T}N_j}\sum_{l=0}^{T}N_l q_l({\boldsymbol\theta}_i^t)} 
\end{gathered} 
\end{equation}
which can be rewritten in terms of the unbiased estimate of the marginal likelihood as

\begin{equation} \label{eq:pm-amis-weight_with_unbiased_estimate}
{w_i}^t = \frac{\tilde{p}(\mathbf{y} \mid {\boldsymbol\theta}_i^t) p({\boldsymbol\theta}_i^t)}{\frac{1}{\sum_{j=0}^{T}N_j}\sum_{l=0}^{T}N_l q_l({\boldsymbol\theta}_i^t)} =
\frac{\tilde{f}({\boldsymbol\theta}_i^t)}{\frac{1}{\sum_{j=0}^{T}N_j}\sum_{l=0}^{T}N_l q_l({\boldsymbol\theta}_i^t)}
\end{equation}
Equation \eqref{eq:pm-amis-weight_with_unbiased_estimate} shows how the importance weights can be computed by the unbiased estimator of the marginal likelihood.

We now appeal to Lemma 1 in \cite{Cornuet12}, which gives the conditions under which the self-normalized estimator of AMIS will converge to Equation \eqref{eq:expec_of_h(theta)_by_f_theta}. 
Following the conditions in Lemma 1 in \cite{Cornuet12}, when $T$ and $N_0$, ..., $N_{T-1}$ are fixed, and when $N_T$ goes to infinity, ${w_i}^t$ (Equation \eqref{eq:weight_with_z}) becomes:
\begin{equation} \label{pm-amis-weight3}
  {w_i}^t \simeq \frac{f(z_i^t, {\boldsymbol\theta}_i^t)}{q_T({z_i^t, \boldsymbol\theta}_i^t)} 
\end{equation}
Then we have 
\begin{eqnarray*}
  E_{q_t(z,\boldsymbol\theta)}\left[\frac{1}{\sum_{t=0}^{T} \sum_{i=1}^{N_t} {w_i}^t}\sum_{t=0}^{T}\sum_{i=1}^{N_t}{w_i}^t h({\boldsymbol\theta}_i^t)\right] & = &
  \frac{1}{Z\sum_{t=0}^{T}N_t}\sum_{t=0}^{T} N_t \int h(\boldsymbol\theta) \frac{f(z,\boldsymbol\theta)}{q_T(z,\boldsymbol\theta)} q_T(z,\boldsymbol\theta) d\boldsymbol\theta dz\\ \nonumber
  & = & \frac{1}{\sum_{t=0}^{T}N_t}\sum_{t=0}^{T} N_t \int 
  h(\boldsymbol\theta) \frac{f(z,\boldsymbol\theta)}{Z} d\boldsymbol\theta dz\\ \nonumber
  & = & \frac{1}{\sum_{t=0}^{T}N_t}\sum_{t=0}^{T} N_t \int h(\boldsymbol\theta) \frac{f(\boldsymbol\theta)}{Z} d\boldsymbol\theta\\ \nonumber
  & = & \int h(\boldsymbol\theta) \frac{f(\boldsymbol\theta)}{Z} d\boldsymbol\theta = E_{\pi(\boldsymbol\theta)}[h(\boldsymbol\theta)] \nonumber
\end{eqnarray*}
where the normalizing constant $Z$ is estimated by $\frac{\sum_{t=0}^{T} \sum_{i=1}^{N_t} {w_i}^t}{\sum_{t=0}^{T}N_t}$ 

Therefore, under the conditions that $T$ and $N_0$, ..., $N_{T-1}$ are fixed and that $N_T$ goes to infinity, which are the same conditions mentioned in Lemma 1 in \cite{Cornuet12}, the estimator of Equation \eqref{eq:monte_carlo_estimation_PM_AMIS} proves to be an unbiased estimator of $E_{\pi(\boldsymbol\theta)}[h(\boldsymbol\theta)]$.
As noted in \cite{Cornuet12}, we remark that these conditions might prove restrictive in practice; however, these conditions provide some solid grounds onto which convergence can be established for AMIS.
Furthermore, we note that in a practical setting, when in doubt as to whether convergence might be an issue, it is always possible to switch to the modified version of AMIS \citep{Marin14} during execution.

\section{Experiments}

\subsection{Competing sampling methods}
In this section, we present the state-of-the-art MCMC and AIS sampling methods considered in this work. The aim is to find out whether adaptive imporatnce sampling (AMIS/MAMIS) can improve speed of convergence
 with respect to computational complexity compared to MCMC approaches (MH \citep{Metropolis53,Hastings70}, HMC \citep{Duane87,Neal93}, NUTS and NUTSDA \citep{Hoffman11}, SS (Slice Sampling) \citep{Neal03}). 
The competing sampling algorithms considered in this work are given in Table \ref{tab:competing:samplers}.

\begin{table}[th]
 \begin{center}
  \begin{tabular}{|l|l|}
  \hline
   {\bf Sampler} & \multicolumn{1}{C{6cm}|}{{\bf Tuning parameters}}\\
  \hline
 Metropolis-Hastings (MH) & \multicolumn{1}{C{6cm}|}{Covariance matrix $\boldsymbol{\Sigma}$ }\\
 \hline
  Hybrid Monte Carlo (HMC) &\multicolumn{1}{C{6cm}|}{Mass matrix $\boldsymbol{\Sigma}$, Leapfrog stepsize $\epsilon$, Number of leapfrog steps $L$} \\
 \hline
  No-U-Turn Sampler (NUTS) &\multicolumn{1}{C{6cm}|}{Mass matrix $\boldsymbol{\Sigma}$, Leapfrog stepsize $\epsilon$} \\
 \hline
NUTS with Dual Averaging (NUTSDA) &\multicolumn{1}{C{6cm}|}{ Mass matrix $\boldsymbol{\Sigma}$} \\
  \hline 
 Slice Sampling (SS) &\multicolumn{1}{C{6cm}|}{Width of the initial bracket} \\
  \hline 
\end{tabular} 
   \captionsetup{justification=centering}   
   \caption{Competing sampling algorithms considered in this work.\label{tab:competing:samplers}}
 \end{center}  
\end{table}
\FloatBarrier

\subsection{Datasets}
The sampling methods considered in this work are tested on six UCI datasets \citep{Asuncion07}. The Concrete, Housing and Parkinsons datasets are used for GP regression, whereas the Glass, Thyroid and Breast datasets are used for GP classification. The number of data points and the dimension of the features for each data point are given in Table \ref{tab:dataset}. 
For the original Parkinsons dataset 
we randomly sampled $4$ records for each of the $42$ patients, resulting in $168$ data points in total.

\begin{table}[ht]
\begin{center}
\begin{tabular}{|c|c|c|c|c|c|c|}
 \hline  
 \multirow{1}{2em}{} &\multicolumn{3}{c|}{Datasets for regression} &\multicolumn{3}{c|}{Datasets for classification} \\ 
 \cline{2-7}
 & Concete & Housing & Parkinsons & Glass & Thyroid & Breast \\
 \hline
 n & 1030 & 506  & 168 & 214 & 215 & 682\\
 \hline
 d & 8 & 13 & 20 & 9& 5 & 9\\
 \hline   
\end{tabular}
 \caption{Datasets used in this paper; $n$ denotes the number of data points, $d$ denotes the dimension of the features.\label{tab:dataset}} 
\end{center}  
\end{table} 
\FloatBarrier

\subsection{Experimental setup}
\subsubsection{Settings for GP regression} \label{Experi_set_GP_regression}
We compare three different covariances for the proposals 
of the MH algorithm.
The first is proportional to the identity matrix.
The second and third covariances are proportional to the inverse of the negative Hessian (denoted by $\mathbf{H}$) evaluated at the mode (denoted by $\mathbf{m}$); one uses the full matrix, whereas the other uses its diagonal only, namely $\mathrm{diag}((\mathbf{- H})^{-1})$. 
 The mode $\mathbf{m}$ is found by the maximum likelihood optimization using the ``BFGS'' method. 

Thus the proposals that we compare in this work take the form of ${\cal N}(\boldsymbol\theta\mid \mathbf{m},\alpha\mathbf{I})$, ${\cal N}(\boldsymbol\theta\mid \mathbf{m},\alpha(\mathbf{- H})^{-1})$, and ${\cal N}(\boldsymbol\theta\mid \mathbf{m},\alpha$ $\mathrm{diag}((\mathbf{- H})^{-1}))$, where $\alpha$ is a tuning parameter. 
We tune $\alpha$ in pilot runs until we get the desired acceptance rate (around $25\%$), as suggested by \cite{Roberts97b}. 

The approximate distribution ${\cal N}(\boldsymbol\theta\mid \mathbf{m},(\mathbf{- H})^{-1})$ is used to be the initial importance density for AMIS/MAMIS. 
This approximation is also used to initialize several independent sequences of samples from other samplers considered in this work (listed in Table \ref{tab:competing:samplers}). In this way, valid summary inference from multiple independent sequences can be obtained \citep{Gelman92}.
 For AMIS/MAMIS, we explored two kinds of update of the covariance of the importance density. One updates the full covariance, whereas the other updates only the diagonal of the covariance.
The first two rows of Table \ref{tab:amis_mamis} show the experimental settings for AMIS/MAMIS.
 
Motivated by the fact that using knowledge of scales and correlation of the position variables can improve the performance of HMC \citep{Neal11}, we also chose three kinds of mass matrix for HMC, namely the identity matrix, the inverse of the approximate covariance, and the inverse of the diagonal of the approximate covariance. We set the maximum leapfrog steps to be $10$. We then tune the stepsize $\epsilon$ until a suggested acceptance rate (around $65\%$) is reached \citep{Beskos13,Neal11}. The three forms of mass matrix apply to NUTS, NUTSDA as well; a full description of the pseudo codes of these algorithms can be found in Algorithm 3 and 6 respectively in \cite{Hoffman11}. 
NUTS requires the tuning of a stepsize $\epsilon$. After a few trials, we set the stepsize of NUTS to $0.1$. Although tuning leapfrog steps and stepsize is not an issue in NUTSDA, the parameters ($\gamma,t_0,\kappa$) for the dual averaging scheme therein have to be tuned by hand to produce reasonable results. By trying a few settings for each parameter, finally the values $\gamma = 0.05, t_0 = 30, \kappa = 0.75$ were used in both the RBF and ARD kernel case.
 
 The slice sampling algorithm adopted in this paper makes component-wise updates of the parameters, where a new sample is drawn according to the ``stepping out'' and ``shrinkage'' procedures as described in \cite{Neal03}.
In our implementation, we set the estimate of the typical size of a slice $w$ to $1.5$.

\begin{table}[ht]
\begin{center}
\begin{tabular}{|c|c|c|c|c|}
 \hline  
 \multirow{2}{2em}{} &\multicolumn{2}{c|}{RBF kernel} &\multicolumn{2}{c|}{ARD kernel} \\ 
 \cline{2-5}
 & \multicolumn{1}{C{0.5cm}|}{ $T$} & \multicolumn{1}{C{0.5cm}|}{$N_t$} & \multicolumn{1}{C{1cm}|}{$T$} &\multicolumn{1}{C{1cm}|}{$N_t$} \\
 \hline
 AMIS & 1120 & 25  & 280 & 100\\
 \hline
 MAMIS & 46 & $26t$ & 5 & $3000 + 1000t$ \\
 \hline
 PM-AMIS & 60 & 400 & 60 & 400 \\
  \hline   
\end{tabular}
\caption{Experimental settings for AMIS/MAMIS/PM-AMIS. $T$ is the total number of iterations, $N_t$ is the sample size at each iteration $t$.
   \label{tab:amis_mamis}} 
\end{center}  
\end{table} 
\subsubsection{Settings for GP classification} 

We compared only PM-AMIS and Pseudo-Marginal MH (PM-MH) in the case of GP classification. Since the likelihood is analytically intractable and thus is unbiasedly estimated, the critical property of reversibility and preservation of volume of HMC, NUTS, NUTSDA is no longer satisfied. Therefore, these algorithms are not considered in the GP classification case. It can be proved that slice sampling with the noisy estimate $\tilde{f}(\boldsymbol\theta)$ is still valid, but the adaptation of the slice estimate $w$ such as the "stepping out" and "doubling" procedure in \cite{Neal03} is not proper for this use case and naively running standard SS with the noisy estimate $\tilde{f}(\boldsymbol\theta)$ worked very poorly \citep{Murray15}. Consequently, SS is also not compared in this case.

Both the EP and LA approximations are used as the importance densities to estimate the marginal likelihood. The last row of Table \ref{tab:amis_mamis} shows the settings of PM-AMIS in both the RBF and ARD cases except for the Breast dataset in the ARD case using LA approximation, where the total number of iterations $T$ is set to $240$ for the sake of presentation. The initial importance density is obtained by the same optimization method as described in Section \ref{Experi_set_GP_regression} except that the gradient required to perform the optimization cannot be computed analytically but is estimated from the EP or LA approximations. We update the full covariance of the importance density during the adaptation process. The proposal of PM-MH also takes the form of ${\cal N}(\boldsymbol\theta\mid \mathbf{m},\alpha(\mathbf{- H})^{-1})$ where $H$ is the Hessian matrix obtained again from the approximate  marginal likelihood obtained by the EP or LA algorithms. We tuned $\alpha$ until we reached the recommended acceptance rate around $25\%$ and use this tuned proposal to generate new samples.

\subsection{Convergence analysis} 
 
Since the classic $\hat{R}$ score is for MCMC convergence analysis and not suitable for importance sampling, convergence analysis here is performed based on the IQR (interquartile range) of the expectation of norm of parameters ($E_{p(\boldsymbol\theta\mid \mathbf{y},\mathbf{X})}[\,\norm{\boldsymbol\theta}]$) over several repetitions against the number of $O(n^3)$ operations; This is reported to be a more reliable measure of complexity than running time, as many other factors such as implementation details that do not relate directly to the actual computing complexity of the algorithms can affect the running time \citep{FilipponeML13}. 
For GP regression the IQR is computed over $100$ replicates, whereas for GP classification it is based on $20$ repetitions. 

For AMIS/MAMIS/SS/MH, the computing complexity lies in the computation of the function of
$f(\mathbf{\boldsymbol\theta})$, where one $O(n^3)$ operation is required to perform the Cholesky decomposition of the covariance matrix $\mathbf{C}$. Whereas for HMC/NUTS/NUTSDA where computing the gradient is necessary, two extra $O(n^3)$ operations are needed for the computation of the inverse of the covariance matrix $\mathbf{C}$. 

For PM-AMIS/PM-MH, the computing complexity largely comes from the EP or LA approximation of the posterior of the latent variables in order to compute the unbiased estimate $\tilde{f}(\boldsymbol{\theta})$. Both EP and LA approximations take two $O(n^3)$ operations to perform the Cholesky decomposition. One is for the Cholesky decomposition of the covariance matrix $\mathbf{K}$ of the GP prior, the other is for the Cholesky decomposition of the covariance of the approximating Gaussian. Each iteration of EP and LA requires three $O(n^3)$ operations and one $O(n^3)$ operation respectively. In LA approximation, two extra $O(n^3)$ operations are needed to compute the covariance of the Gaussian approximation.

\subsection{Results}
\subsubsection{Notation of samplers}
Table \ref{tab:notation_GP_samplers} shows the notation for the different samplers considered in this paper. 
\begin{table*}[th]
 \begin{center}
  \begin{tabular}{|c|c|}
  \hline
 AMIS/MAMIS & \multicolumn{1}{C{9cm}|}{AMIS/MAMIS for GP regression where the full covariance matrix of the proposal  distribution is updated at each iteration}\\
 \hline
 AMIS-D/MAMIS-D &\multicolumn{1}{C{9cm}|}{AMIS/MAMIS for GP regression where only the diagonal of the covariance matrix of the proposal distribution is updated at each iteration} \\
 \hline
 MH-I &\multicolumn{1}{C{9cm}|}{MH for GP regression where the covariance of the starting proposal distribution for tuning is the identity matrix} \\
 \hline
 MH-D &\multicolumn{1}{C{9cm}|}{MH for GP regression where the covariance of the starting proposal distribution for tuning is the diagonal of the approximate covariance from the optimization} \\
  \hline 
 MH-H &\multicolumn{1}{C{9cm}|}{MH for GP regression where the covariance of the starting proposal distribution for tuning is the approximate covariance from the optimization} \\
  \hline 
 HMC-I/NUTS-I/NUTSDA-I &\multicolumn{1}{C{9cm}|}{HMC family for GP regression where the mass matrix is the identity matrix} \\
  \hline 
HMC-D/NUTS-D/NUTSDA-D &\multicolumn{1}{C{9cm}|}{HMC family for GP regression where the mass matrix is the inverse of the diagonal of the approximate covariance from the optimization} \\
  \hline 
HMC-H/NUTS-H/NUTSDA-H &\multicolumn{1}{C{9cm}|}{HMC family for GP regression where the mass matrix is the inverse of the approximate covariance from the optimization} \\
\hline
 PM-AMIS &\multicolumn{1}{C{9cm}|}{AMIS for GP classification where the full covariance matrix of the proposal distribution is updated at each iteration} \\
 \hline
  PM-MH &\multicolumn{1}{C{9cm}|}{MH for GP classification where the covariance of the starting proposal distribution for tuning is the approximate covariance from the optimization} \\
  \hline
\end{tabular} 
   \captionsetup{justification=centering}   
   \caption{Notation for the samplers.  \label{tab:notation_GP_samplers}}
 \end{center}  
\end{table*}

\subsubsection{Convergence of samplers for GP regression} \label{sec:convergence_samplers_GP_regression}
In this section, we present the comparison of convergence of samplers for GP regression considered in this paper.

\begin{figure*}[th]
  \begin{center}
\begin{tabular}{ccc}
\includegraphics[width=\figwidth,angle=0]{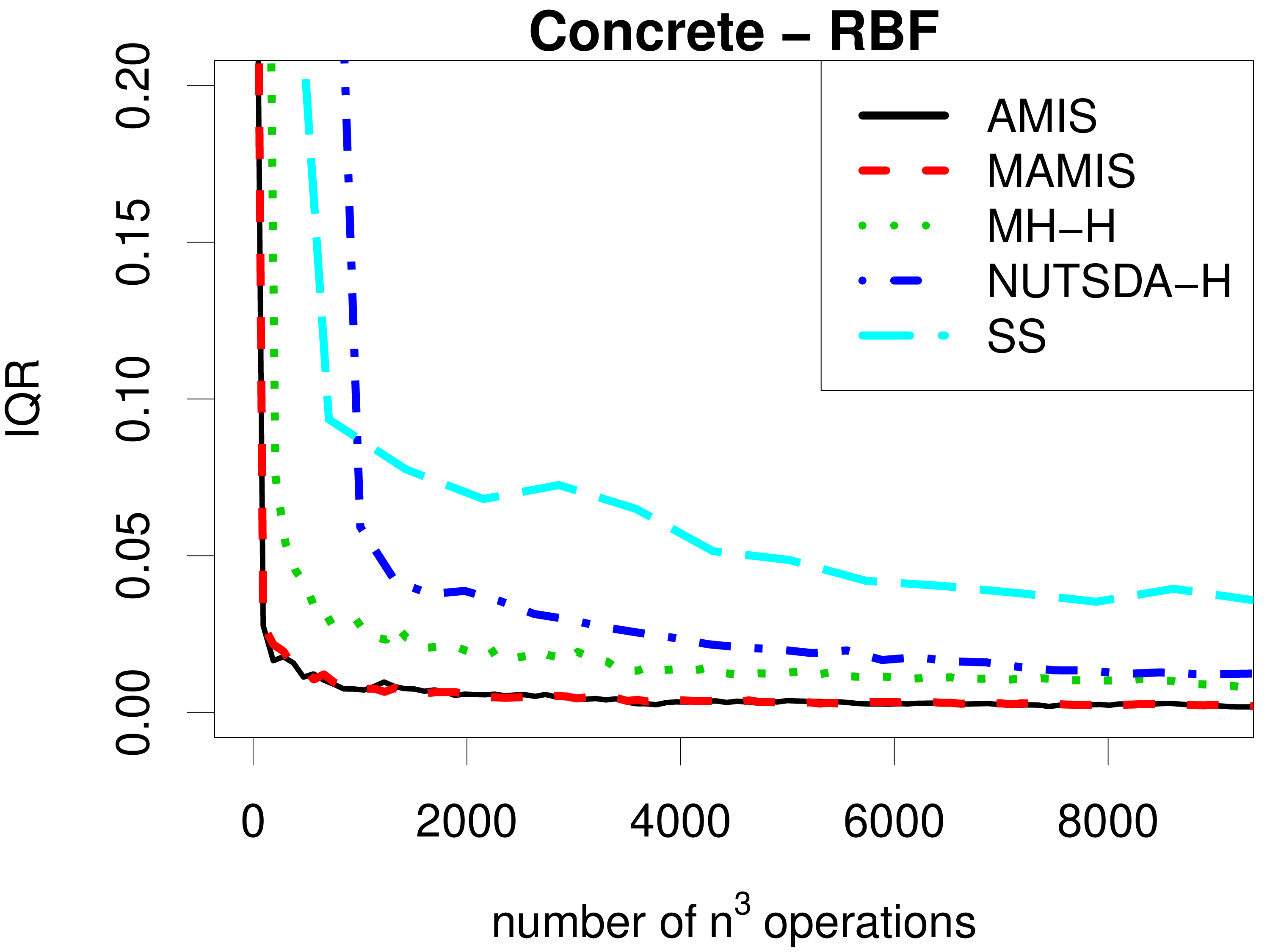} & \includegraphics[width=\figwidth,angle=0]{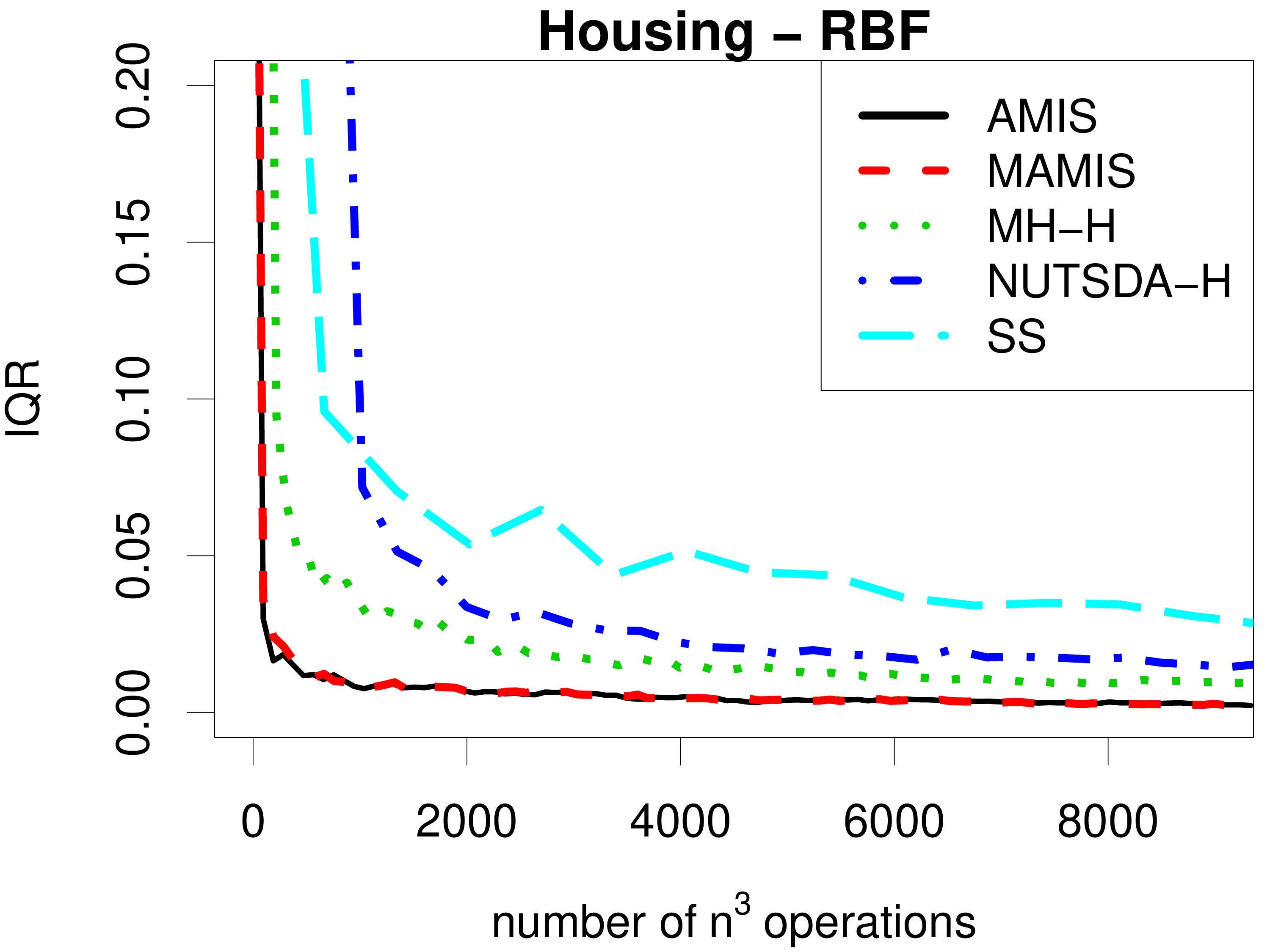} & \includegraphics[width=\figwidth,angle=0]{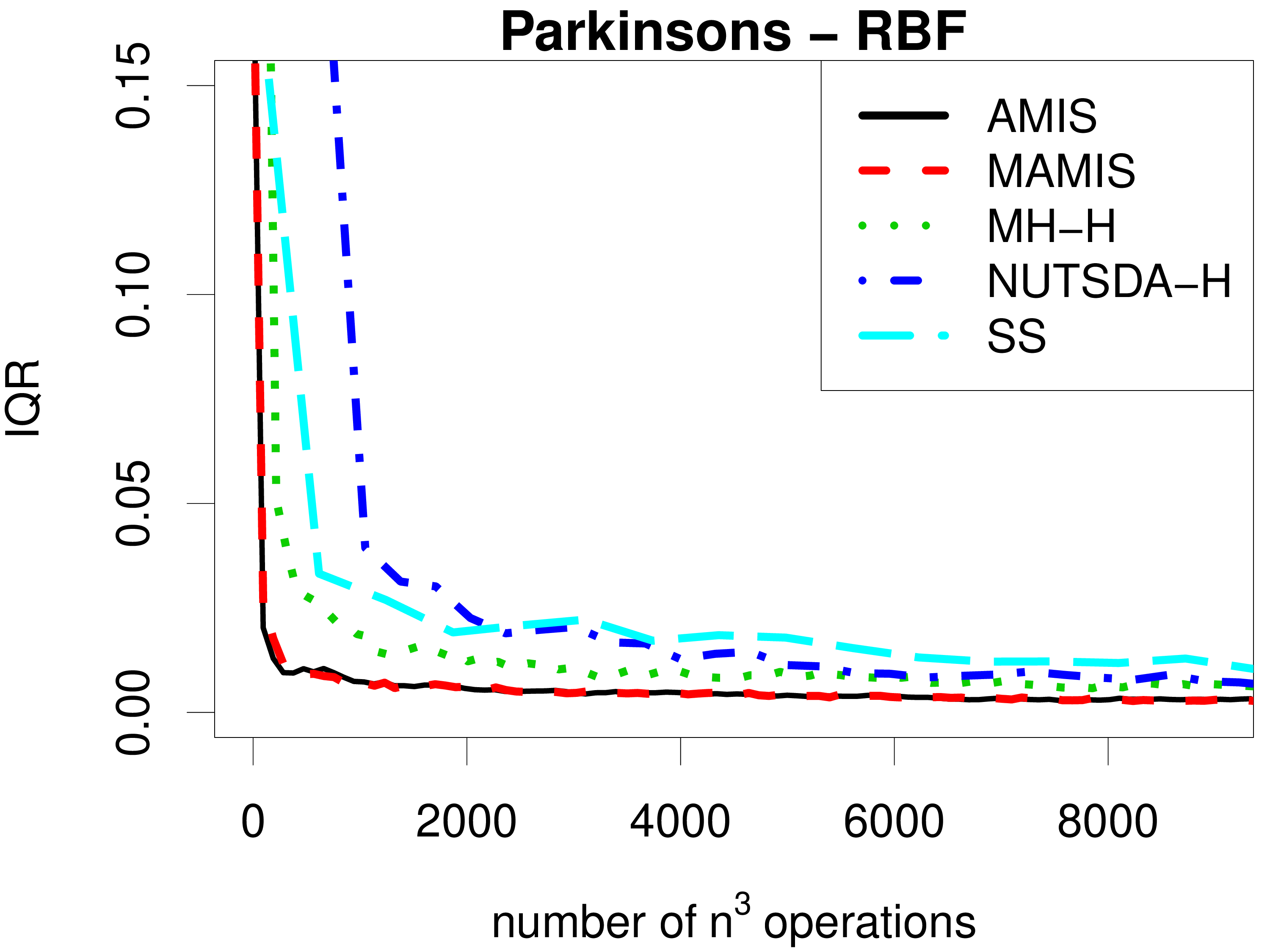} \\
\includegraphics[width=\figwidth,angle=0]{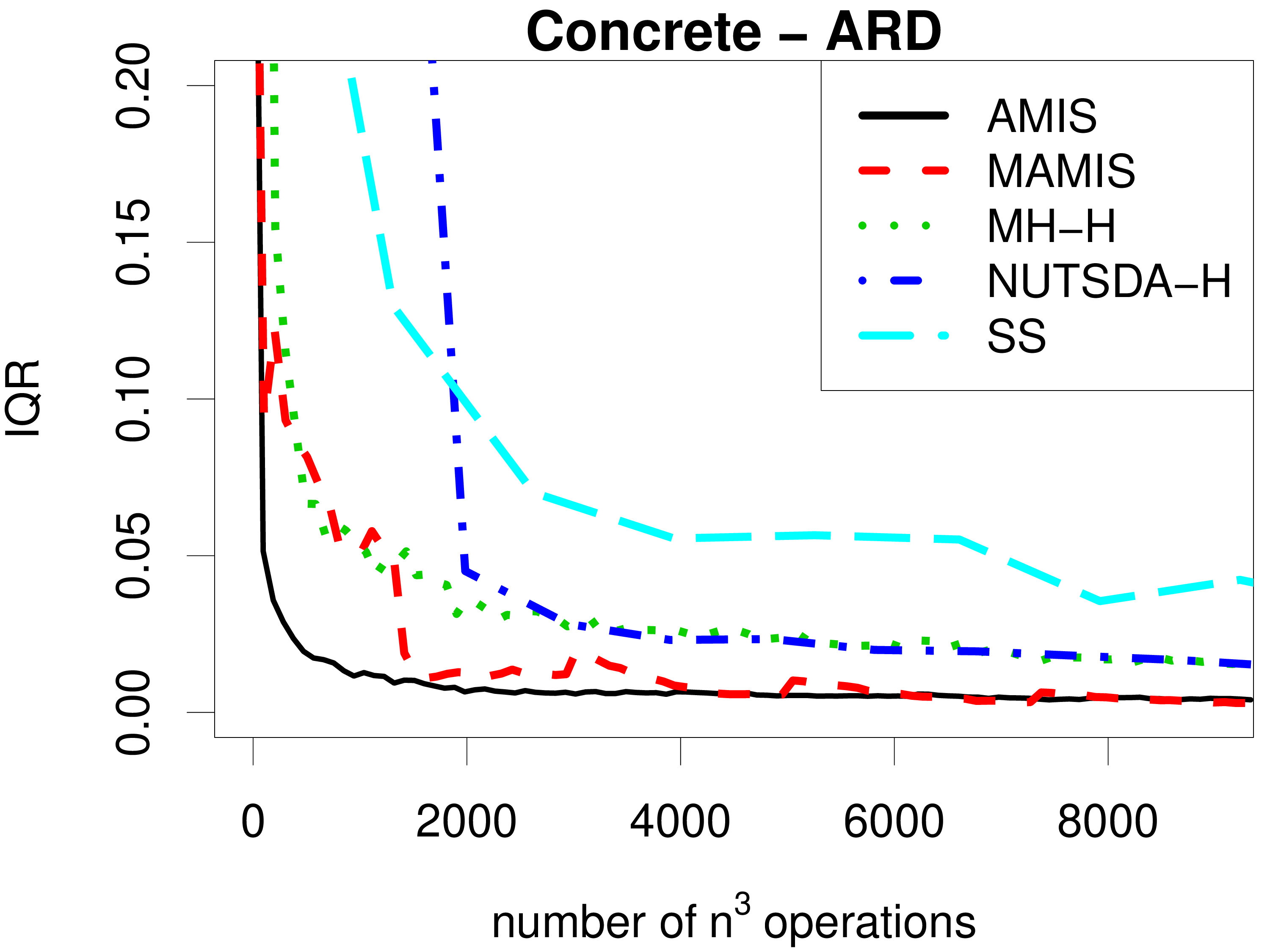} & \includegraphics[width=\figwidth,angle=0]{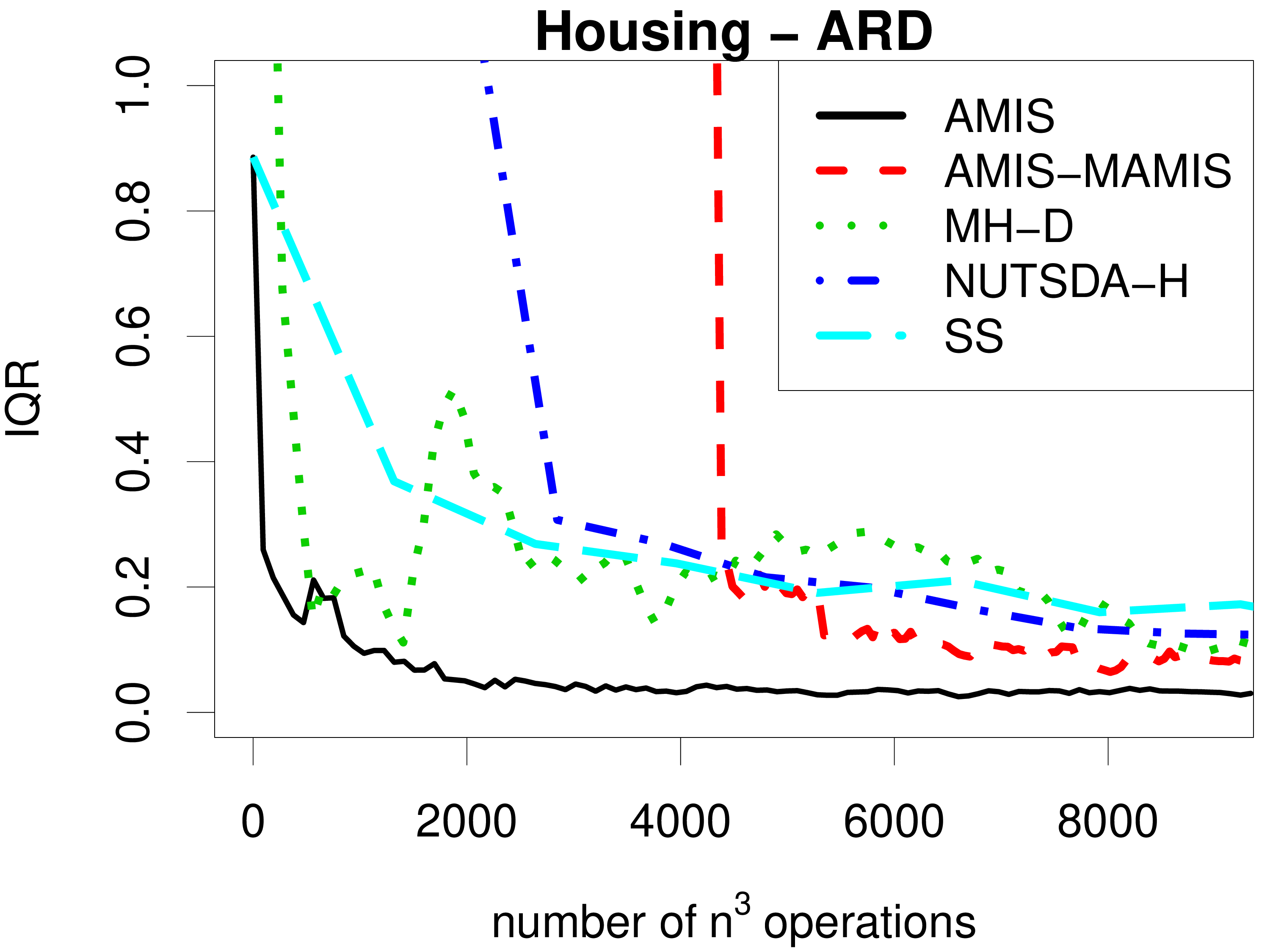} & \includegraphics[width=\figwidth,angle=0]{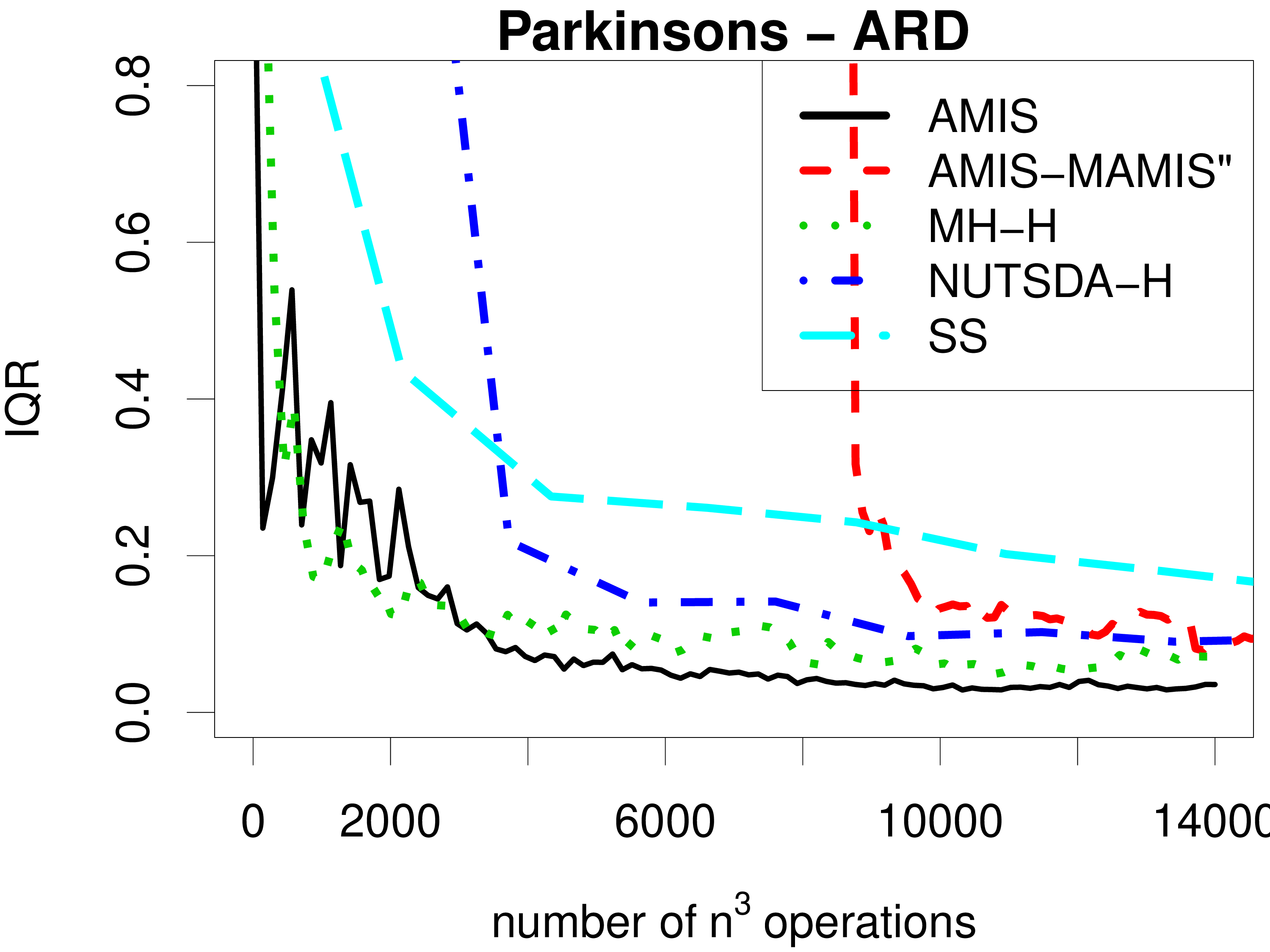} 
\end{tabular}          
  \caption{Convergence of AMIS, Best of MAMIS, Best of MH family, Best of HMC family, SS for GP regression. \label{fig:convergence_gp_regress}}
  \end{center}
\end{figure*}
 Details of convergence results of AMIS family (AMIS/MAMIS), MH family (MH-I/MH-D/MH-H) and HMC family (standard HMC , NUTS, NUTSDA) can be found in \ref{App:AppendixA} and \ref{App:AppendixB}. 
 Fig. \ref{fig:convergence_gp_regress} shows the final result of AMIS, best of MAMIS, best of MH family, best of HMC family and SS for the three datasets in both the RBF and ARD kernel case. It is interesting to see that AMIS/MAIMS performs best among all methods in terms of convergence speed in the RBF kernel case. In the ARD kernel case, AMIS also converges much faster than the other approaches. However, our experiments show that in the ARD kernel case, although MAMIS converges faster than the other approaches in the Concrete dataset, it converges slowly in the Housing and Parkinsons datasets, which is probably due to the higher dimensionality compared to the previous cases.

 In cases where MAMIS converges slowly, we can exploit the fact that AMIS converges faster than MAMIS by running AMIS for a fixed number of iterations and then switch to MAMIS. In this way, we can ensure fast convergence of the adaptive scheme while ensuring that the scheme converges without issues. 
 In the experiments, we tested this AMIS-MAMIS combination in cases where MAMIS converges slowly. We treated samples from AMIS as tuning cost for MAMIS to get an accurate initial importance density as is shown in bottom-right of Fig. \ref{fig:MH_HMC_NUT_NUTDA_HOU_ARD} and Fig. \ref{fig:MH_HMC_NUT_NUTDA_PARK_ARD} with EOT (end of tuning) indicated by the vertical dotted line. Three settings (Table \ref{tab:AMIS-MAMIS}) of AMIS-MAMIS were tested for the Parkinsons dataset.
 \begin{table}[th]
  \begin{center}
   \begin{tabular}{|c|c|c|c|}
  \hline  
    & \multicolumn{1}{C{1cm}|}{$N_t$ for MAMIS}& \multicolumn{1}{C{2cm}|}{$\star$ number of tuning samples for MAMIS} & \multicolumn{1}{C{1.5cm}|}{the corresponding tuning cost}\\
      \hline
  AMIS-MAMIS & \multicolumn{1}{C{1cm}|}{ $1000t$}& \multicolumn{1}{C{2cm}|}{ 13000} &\multicolumn{1}{C{1.5cm}|}{ 4333}\\
  \hline
   AMIS-MAMIS' & \multicolumn{1}{C{1cm}|}{ $5000t$}& \multicolumn{1}{C{2cm}|}{ 13000}&\multicolumn{1}{C{1.5cm}|}{ 4333}\\
    \hline
  AMIS-MAMIS" & \multicolumn{1}{C{1cm}|}{ $5000t$}& \multicolumn{1}{C{2cm}|}{ 26000}&\multicolumn{1}{C{1.5cm}|}{ 8667}\\
   \hline
 \end{tabular} 
    \caption{Settings for AMIS-MAMIS. $N_t$ is the sample size at each iteration $t$. $\star$ refers to the number of samples generated from AMIS for tuning the initial importance density of MAMIS. Unit of the tuning cost: number of $n^3$ operations. \label{tab:AMIS-MAMIS}}
  \end{center}  
 \end{table}
 
  For the Housing dataset, we tested only AMIS-MAMIS in Table \ref{tab:AMIS-MAMIS}. The results for the Housing and Parkinsons datasets in the ARD kernel case prove the convergence of AMIS-MAMIS. In particular, AMIS-MAMIS and AMIS-MAMIS" seem to compete well with the other MCMC approaches in terms of convergence for the Housing dataset and the Parkinsons dataset respectively. As is shown in the bottom-right of Fig. \ref{fig:MH_HMC_NUT_NUTDA_PARK_ARD}, the best performance of AMIS-MAMIS" for the Parkinsons dataset suggests that for higher dimensional problem, a more accurate initialization and a larger sample size at each iteration for MAMIS are necessary to achieve faster convergence.

 Another attempt that we made in this paper to improve convergence speed of the adaptive importance sampling schemes is to regularize the estimation of the parameters of the importance distribution as illustrated in \cite{Smidl14}.
 The regularization stems from the use of an informative prior on $\boldsymbol\gamma$ of the importance distribution $q_t(\boldsymbol\gamma)$ of MAMIS and treat the update of these parameters in a Bayesian fashion \citep{Kulhavy96}. This construction makes it possible to avoid situations where the importance distribution degenerates to low rank due to few importance weights dominating all the others. In this work, we use an informative prior based on a Gaussian approximation to the posterior over covariance parameters. We denote this method by MAMIS-P and in the ARD kernel case it was tested only in the Housing dataset. The result indicates that even though MAMIS-P improves on MAMIS, its convergence is slower than AMIS-MAMIS (bottom-right of Fig. \ref{fig:MH_HMC_NUT_NUTDA_HOU_ARD}).
 
\subsubsection{Convergence of samplers for GP classification} \label{sec:convergence_of_samplers_GP_classification}
The comparison of convergence of samplers for GP classification (PM-AMIS and PM-MH) is presented in this section.

\begin{figure*}[th]
  \begin{center}
\begin{tabular}{ccc}
\includegraphics[width=\figwidth,angle=0]{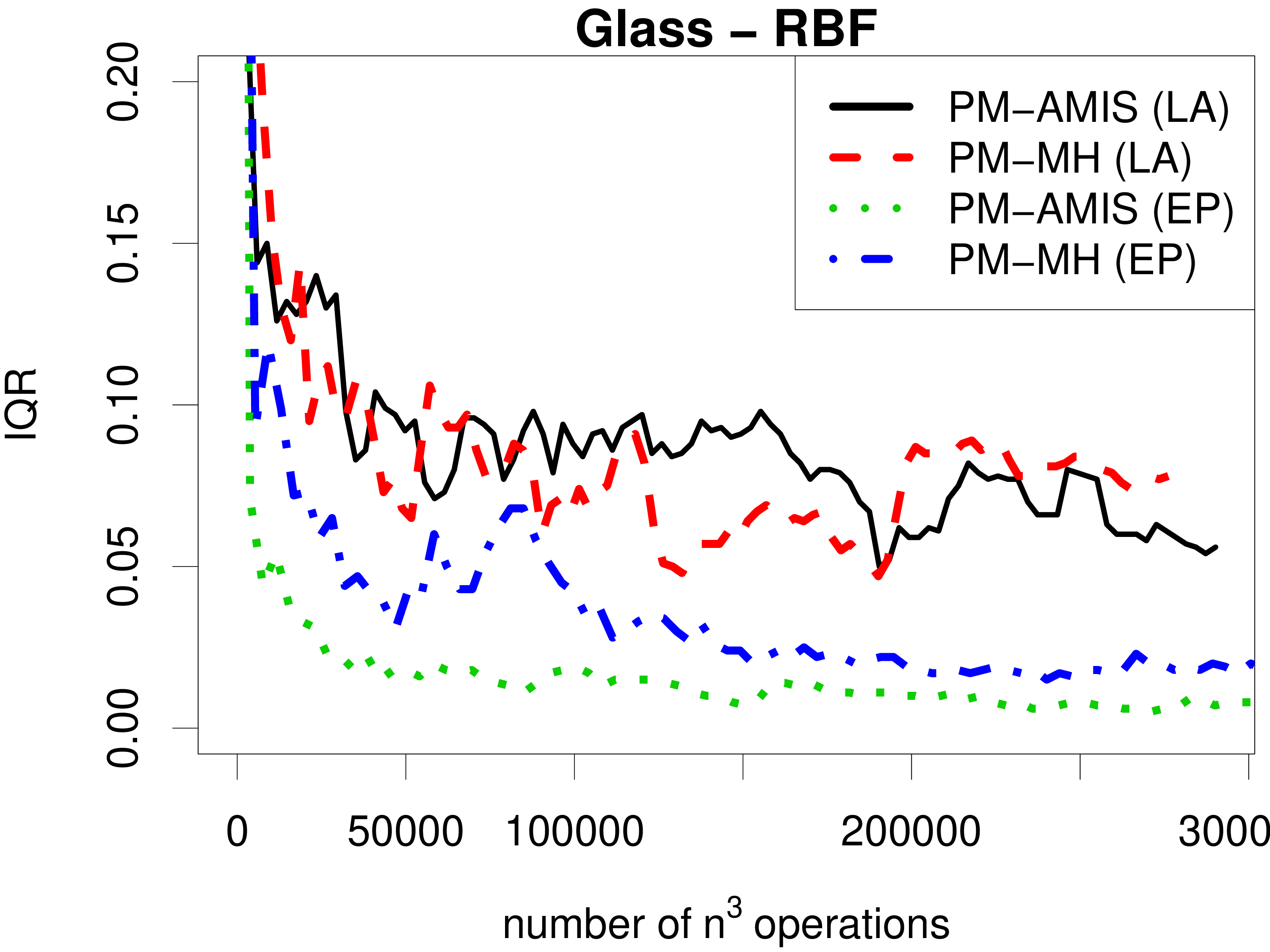} & \includegraphics[width=\figwidth,angle=0]{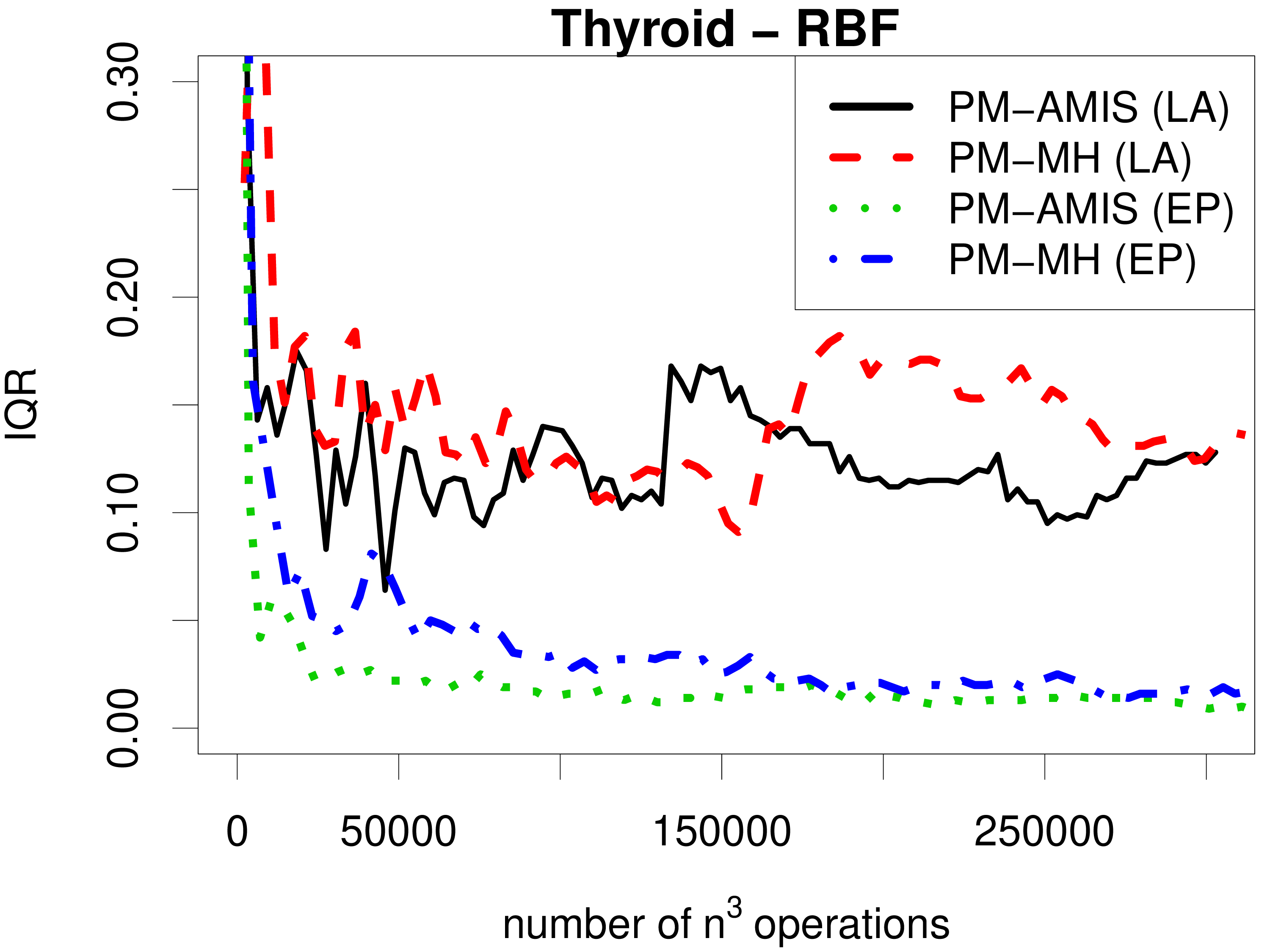} & \includegraphics[width=\figwidth,angle=0]{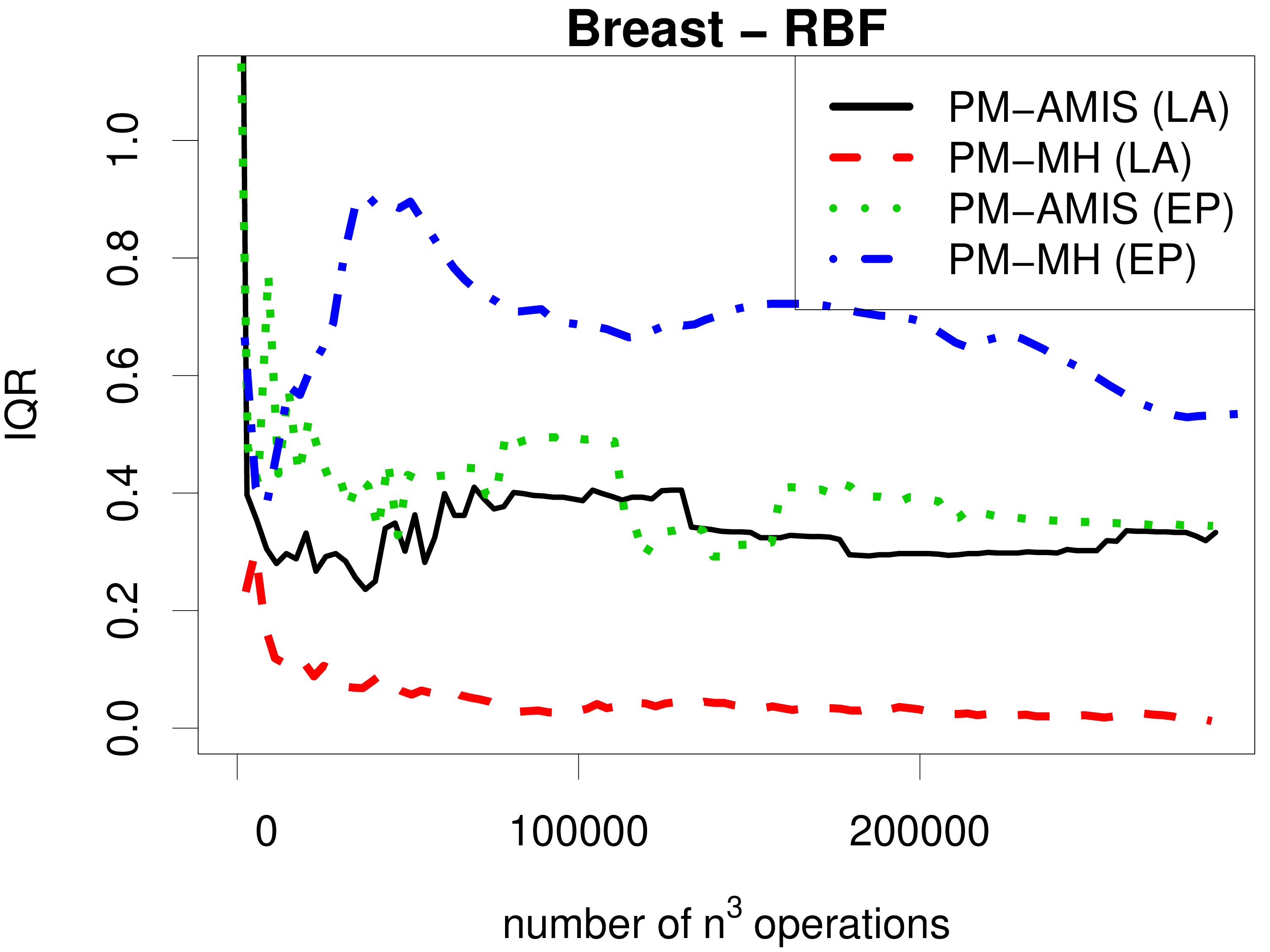} \\
\includegraphics[width=\figwidth,angle=0]{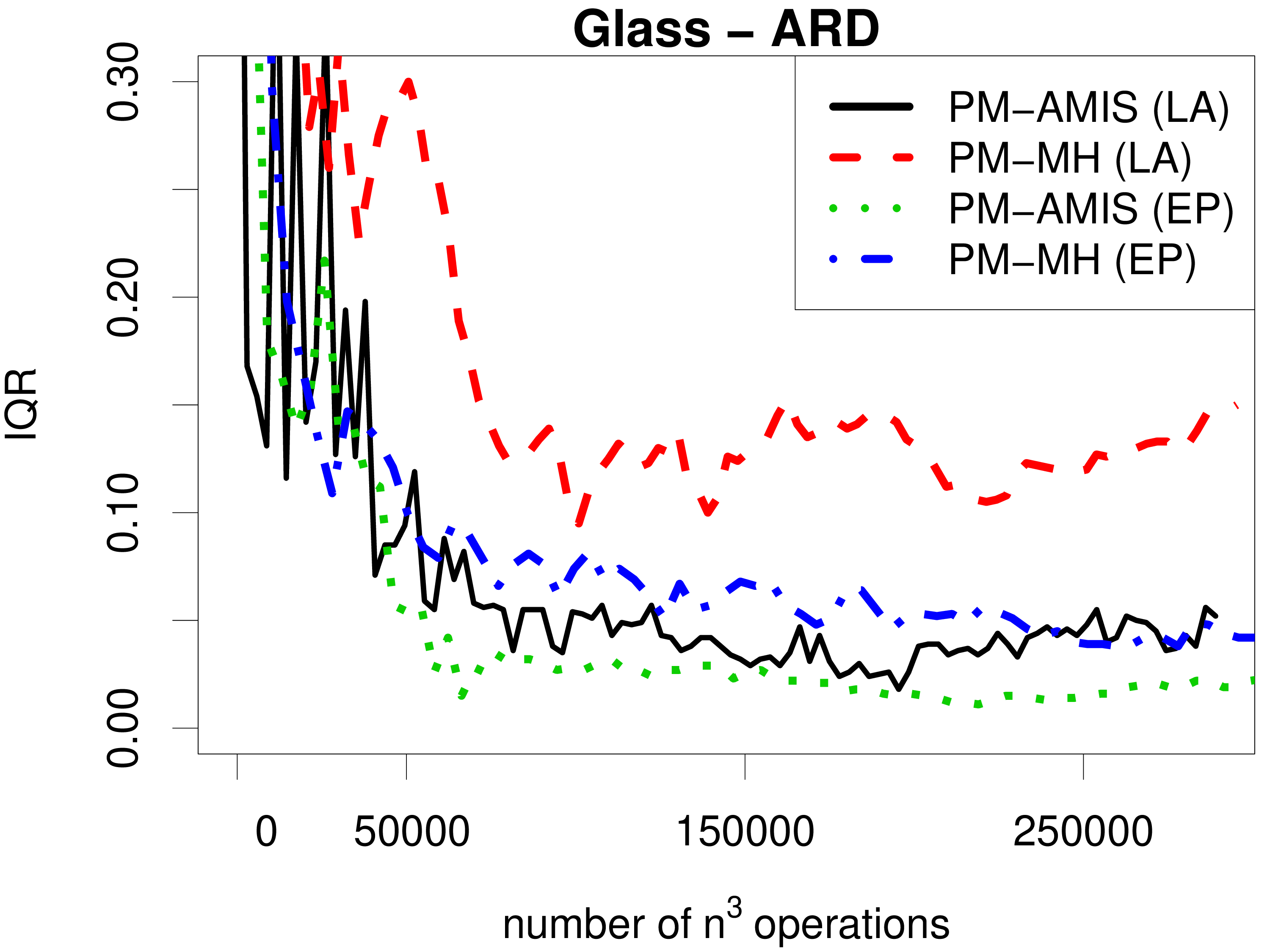} & \includegraphics[width=\figwidth,angle=0]{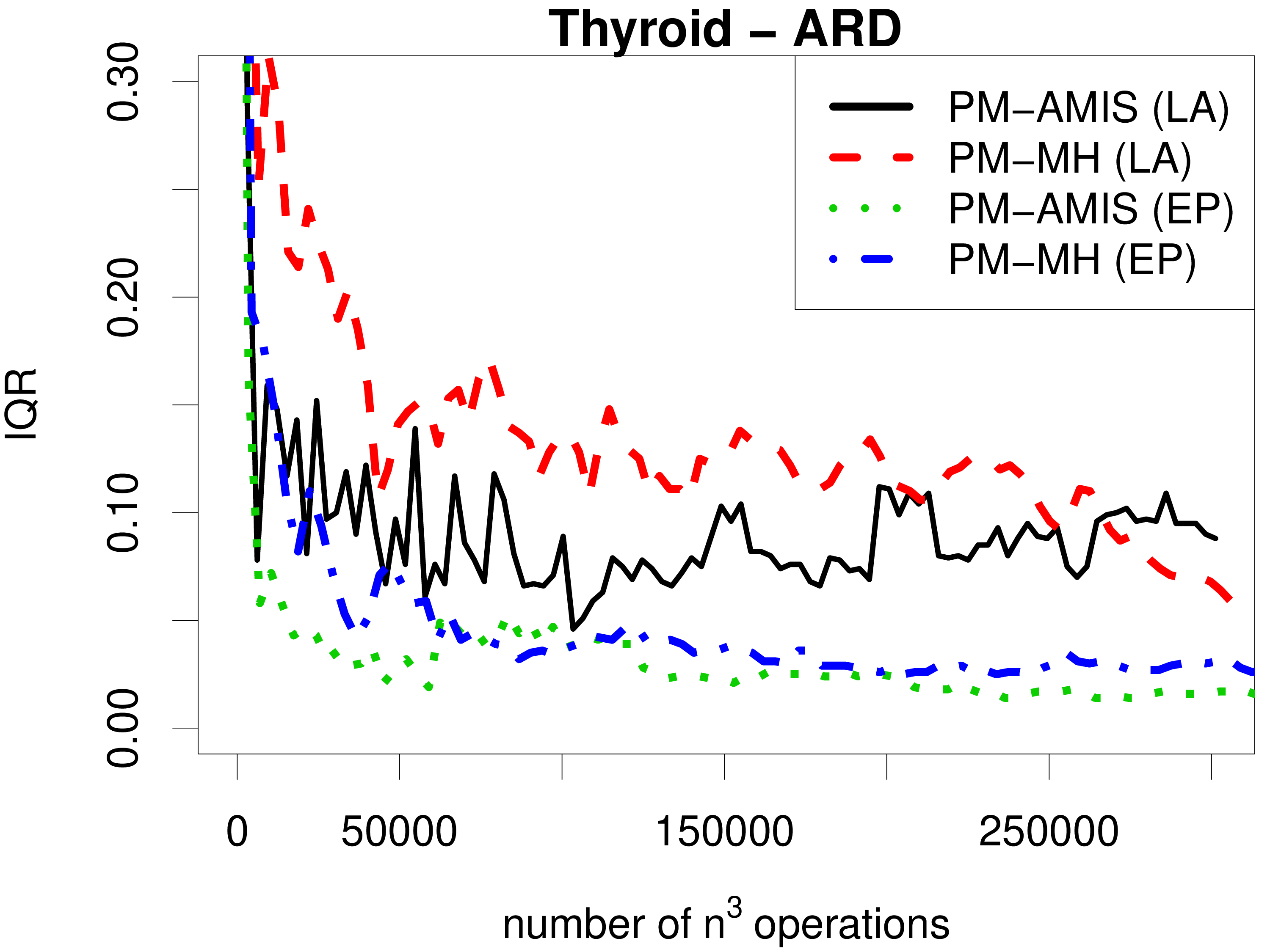} & \includegraphics[width=\figwidth,angle=0]{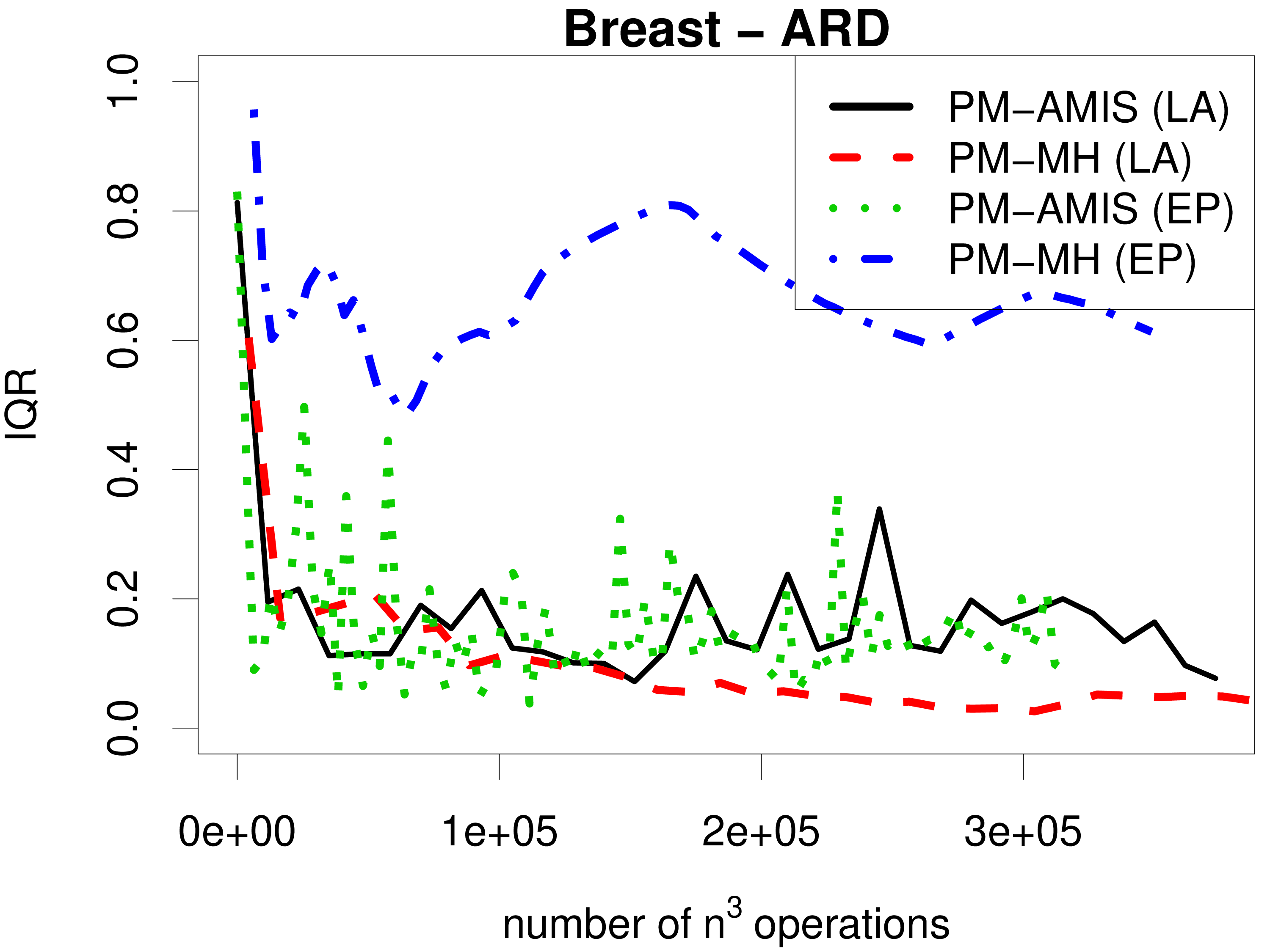}
\end{tabular}          
  \caption{Convergence of Best of PM-AMIS, Best of PM-MH using EP optimization for GP classification.  LA in the brackets indicates the case where the Gaussian approximation to the posterior of the latent variables used in the corresponding method is obtained by LA approximation, whereas EP in the brackets indicates the case where the Gaussian approximation is obtained by EP approximation. \label{fig:convergence_gp_classif_ep_la}}
  \end{center}
\end{figure*}

Fig. \ref{fig:convergence_gp_classif_ep_la} shows the results of PM-AMIS and PM-MH using EP and LA approximation with $N_{imp} = 64$, where $N_{imp}$ denotes the number of importance samples of latent variables $\mathbf{f}$ to estimate the marginal likelihood $p(\mathbf{y} \mid \boldsymbol{\theta})$. The results indicate that PM-AMIS is competitive with PM-MH in terms of convergence speed in all the EP approximation cases and in most of the LA approximation cases. The results also seem to suggest that PM-AMIS/PM-MH converge faster with EP approximation than with LA approximation in most cases, which is probably because EP yields a more accurate approximation than LA as reported in \cite{Kuss05}, \cite{Nickisch08}.
We also tested the performance of PM-AMIS and PM-MH with $N_{imp}= 1$, the results of which are shown in \ref{App:AppendixC}. As is seen from the figures, both PM-AMIS and PM-MH algorithms with higher number of importance samples converge much faster than those with lower number of importance samples as expected. 

\section{Conclusions}

In this paper we proposed the use of adaptive importance sampling techniques to carry out expectations under the posterior distribution of covariance parameters in Gaussian processes. 
The motivation for our proposal is based on a number of observations related to the complexity of dealing with the calculation of the marginal likelihood.
In GPs with a Gaussian likelihood, calculating the marginal likelihood and its gradient with respect to covariance parameters is expensive and standard MCMC algorithms reject proposals leading to a waste of computations.
In GPs with a non-Gaussian likelihood, pseudo marginal MCMC approaches bypass the need to compute the marginal likelihood but may suffer from inefficiencies due to the fact that when a proposal is accepted and the estimated marginal likelihood is largely overestimated, it becomes difficult for the chain to accept any other proposal.
A further motivation behind our proposal is that importance sampling-based algorithms are generally easy to implement and tune, and can be massively parallelized.

The extensive set of results reported in this paper supports our intuition that importance sampling-based inference of covariance parameters is competitive with MCMC algorithms.
In particular, the results indicate that it is possible to achieve convergence of expectations under the posterior distribution of covariance parameters faster than employing MCMC methods in a wide range of scenarios.
Even in the case of around twenty parameters, where importance sampling based methods start to degrade in performance, our proposal is still competitive with MCMC approaches.



\clearpage

\onecolumn{%
\centering
\vspace{0.5cm}
{\bf Appendices}
\vspace{1cm}
}

Appendices \ref{App:AppendixA} and \ref{App:AppendixB} show the convergence results of the samplers for GP regression with the RBF covariance (RBF kernel case) and ARD covariance (ARD kernel case) respectively. The top-left of figures in \ref{App:AppendixA} and \ref{App:AppendixB} demonstrate the result of AMIS/MAMIS. It can be seen that AMIS/MAMIS that exploits the full covariance structure of the proposal distribution performs better than the one that only updates the diagonal of the covariance matrix of the proposal density. For the MH family (MH-I/MH-D/MH-H) and HMC family (standard HMC , NUTS, NUTSDA), figures in \ref{App:AppendixA} and \ref{App:AppendixB} show that, the methods that make use of the scales and correlation of the parameters, perform better than the one that does not in most cases. Also, NUTS/NUTSDA turns out to converge much faster than the standard HMC due to the fact that standard HMC has to be tuned costly in pilot runs. For MH and standard HMC, the computational cost of tuning is counted when comparing the convergence, as is shown in top-center and top-right of figures in \ref{App:AppendixA} and \ref{App:AppendixB} where the end of tuning (EOT) is indicated by three vertical dotted lines, corresponding to the three variants respectively from left to right. For NUTSDA, the computational cost of tuning the parameters of the dual averaging scheme is also counted when determining the convergence, as is displayed in bottom-center of figures in \ref{App:AppendixA} and \ref{App:AppendixB} with EOT indicated by three vertical dotted lines, relating to the three variants respectively from left to right. Table \ref{tab:tuning_cost} shows the corresponding computational cost of tuning: 
 \begin{table}[ht]
 \begin{center}
 \begin{tabular}{|c|c|c|c|c|c|c|}
  \hline  
  \multirow{2}{2em}{} &\multicolumn{2}{c|}{Concrete} &\multicolumn{2}{c|}{Housing} &\multicolumn{2}{c|}{Parkinsons} \\ 
  \cline{2-7}
  & \multicolumn{1}{C{0.5cm}|}{ RBF } & \multicolumn{1}{C{0.5cm}|}{ARD} & \multicolumn{1}{C{0.5cm}|}{RBF} &\multicolumn{1}{C{0.5cm}|}{ARD}& \multicolumn{1}{C{0.5cm}|}{RBF} &\multicolumn{1}{C{0.5cm}|}{ARD} \\
  \hline
  HMC-I & 6747 & 5910 & 4779 & 3924 & 1561& 1340 \\
  \hline
  HMC-D & 6042 & 7316 & 7281 & 7726 & 8883 & 8469 \\
  \hline
  HMC-H & 10851 & 9451 & 10987 & 8860 & 10871& 8736\\
  \hline 
  NUTDA-I & 1402 & 3528  & 1193 & 7433 & 1338 & 6488\\
    \hline
  NUTDA-D & 1357 & 1582 & 1124 & 2424 & 975& 1951 \\
    \hline
  NUTDA-H & 682 & 1023 & 670 & 1866 & 728 & 1794 \\
    \hline  
 \end{tabular}
 \caption{Computational cost of tuning for HMC/NUTSDA. Unit: number of $n^3$ operations.
    \label{tab:tuning_cost}} 
 \end{center}  
 \end{table}
\FloatBarrier 
\ref{App:AppendixC} shows the convergence results of PM-AMIS/PM-MH for the RBF (Fig. \ref{fig:PM_RBF}) and ARD (Fig. \ref{fig:PM_ARD}) case respectively. As can be seen from the figures, both PM-AMIS and PM-MH algorithms with higher number of importance samples (Nimp=64) converge much faster than those with lower number of importance samples (Nimp=1) in both EP and LA approximation cases as expected. The results also indicate that  PM-AMIS is competitive with PM-MH in terms of convergence speed in most of the EP and LA approximation cases. Moreover, PM-AMIS/PM-MH seem to converge faster with EP approximation than with LA approximation in most cases which is probably because EP yields a more accurate approximation than LA as reported in \cite{Kuss05}, \cite{Nickisch08}.

\clearpage 
\onecolumn
\appendix

\section{Convergence of samplers for GP regression with the RBF covariance} \label{App:AppendixA}

\begin{figure*}[th]
  \begin{center}
  {\scriptsize  {\bf Concrete dataset - RBF covariance}}\\
  \begin{tabular}{ccc}  
          \includegraphics[width=\figwidth,angle=0]{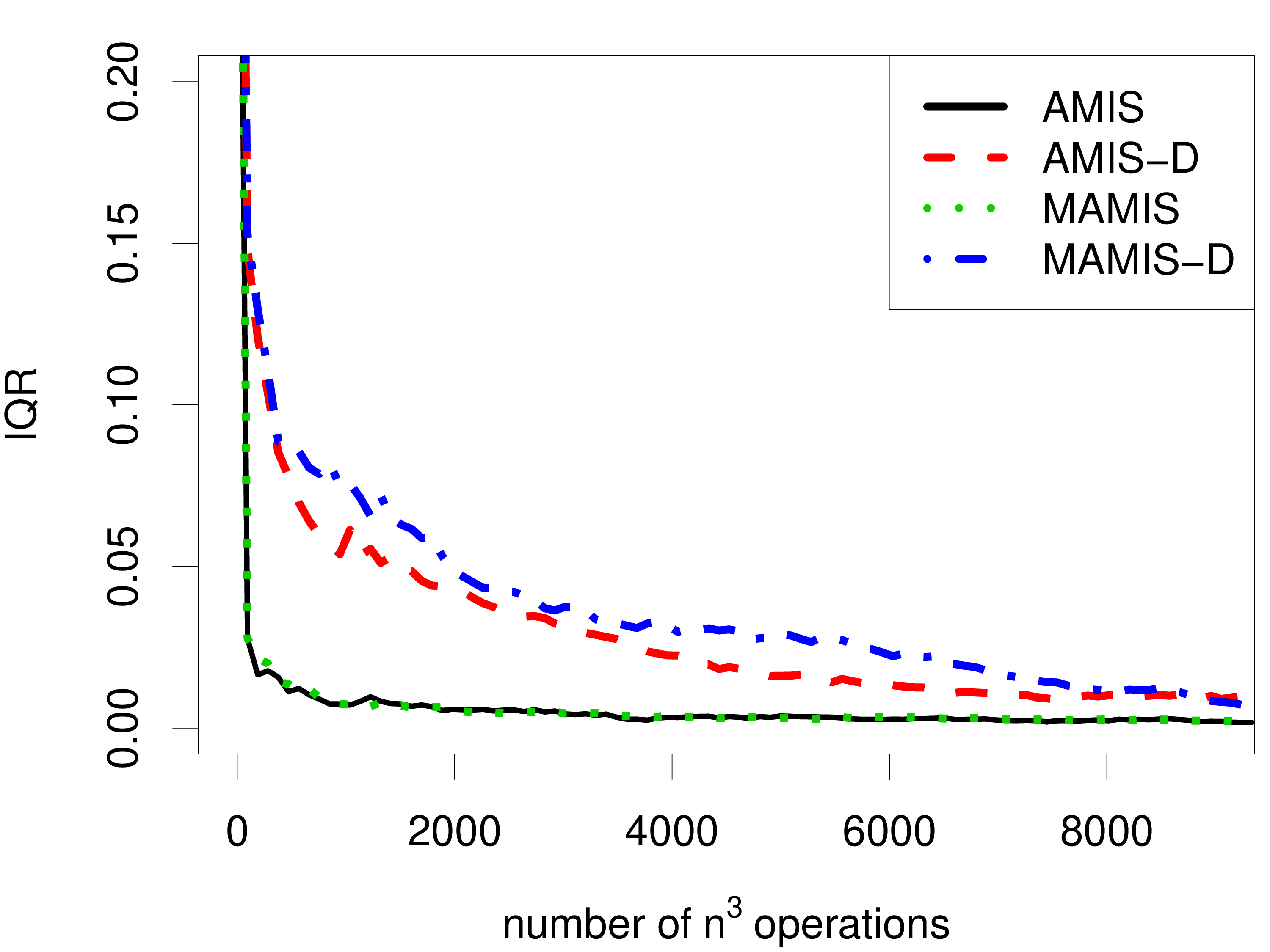}&
          \includegraphics[width=\figwidth,angle=0]{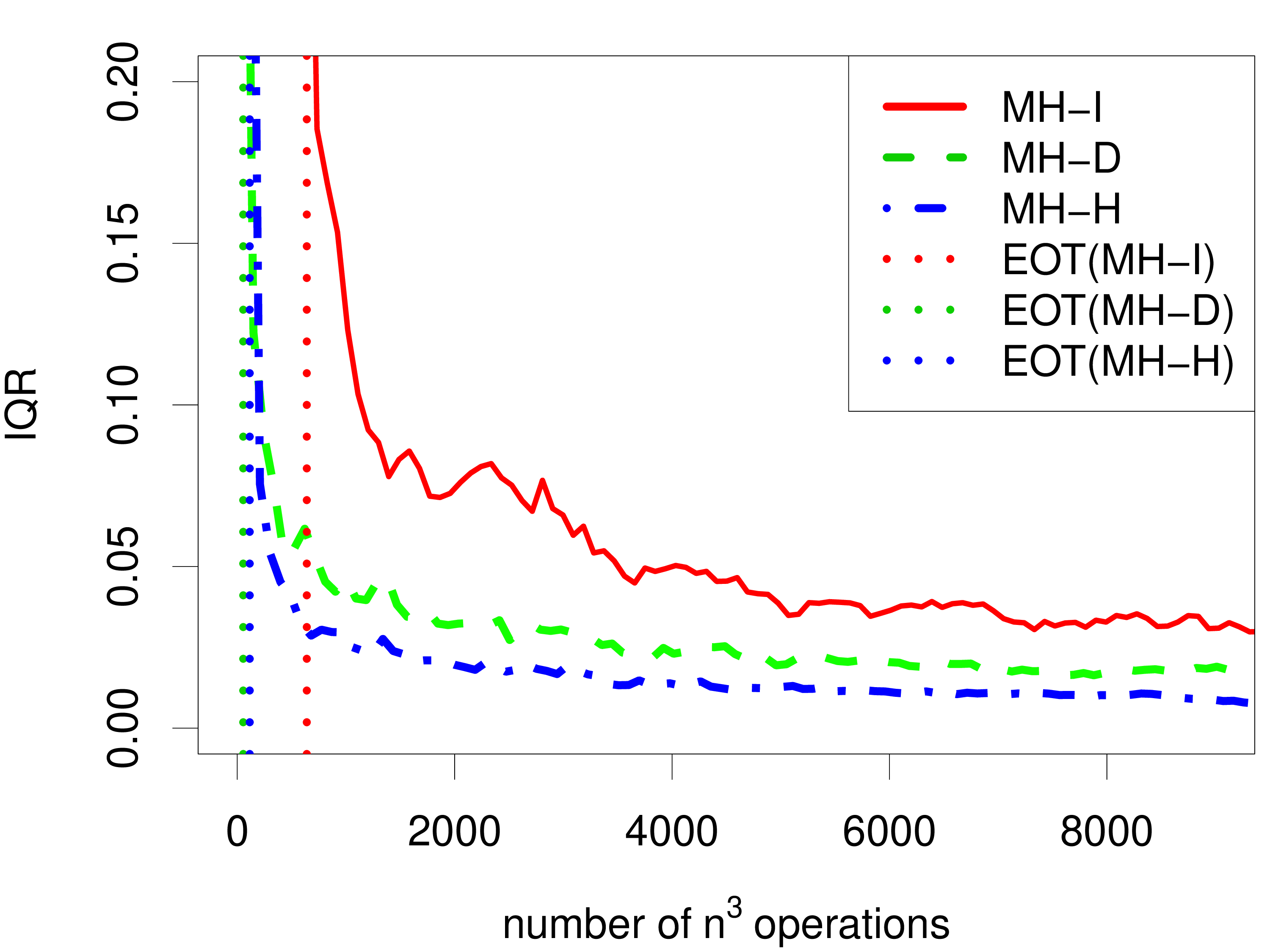}&
           \includegraphics[width=\figwidth,angle=0]{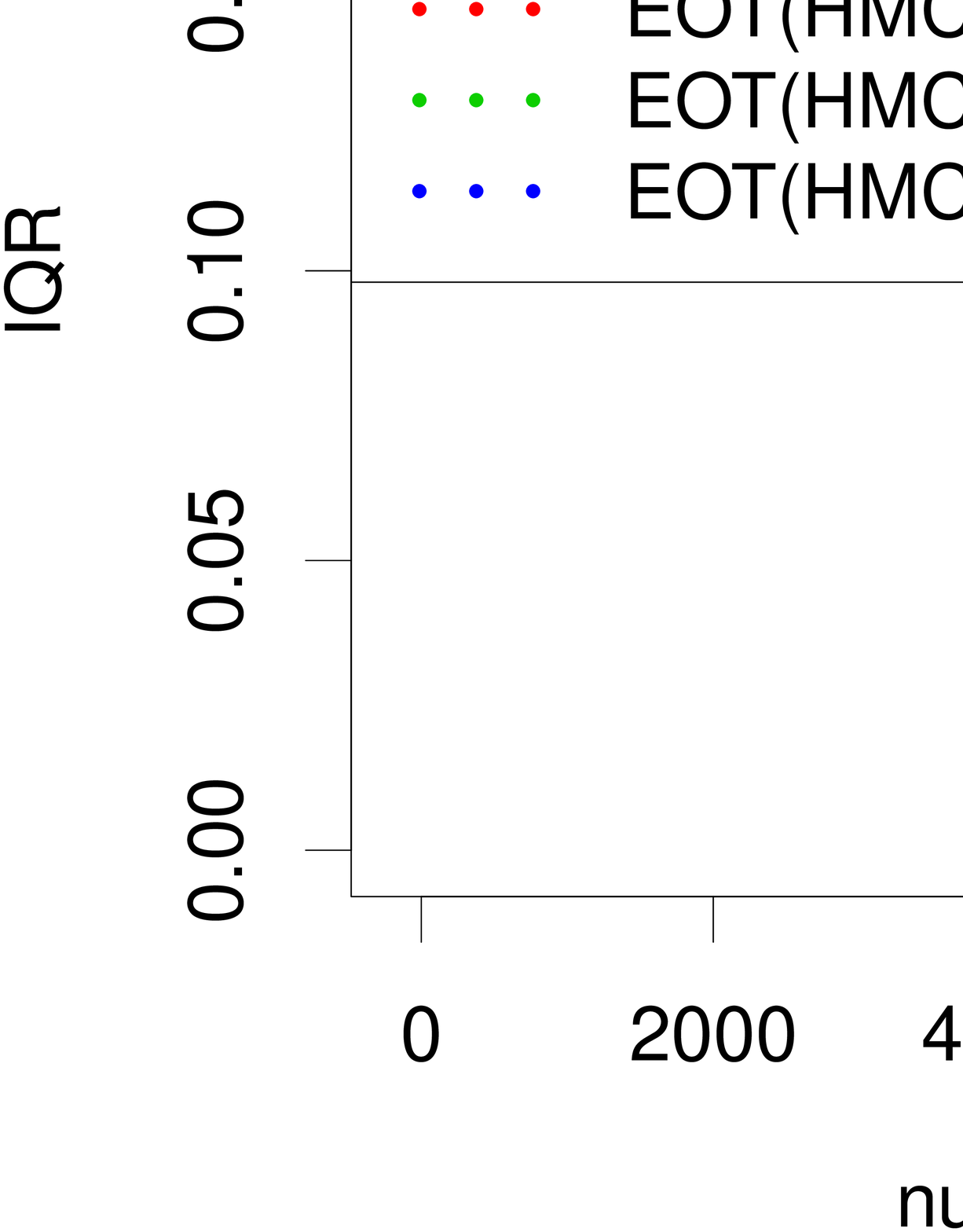}\\
          \includegraphics[width=\figwidth,angle=0]{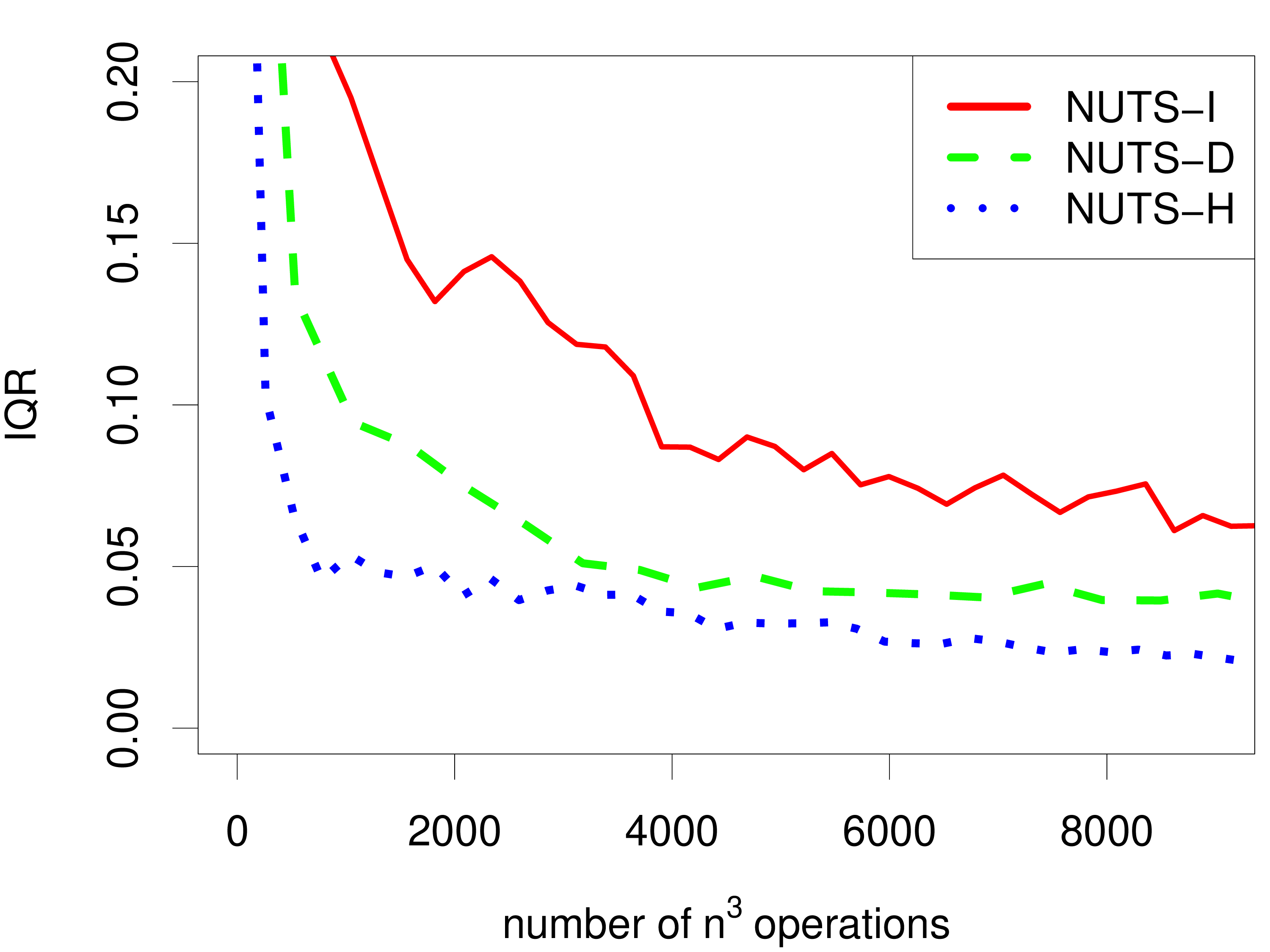}&
          \includegraphics[width=\figwidth,angle=0]{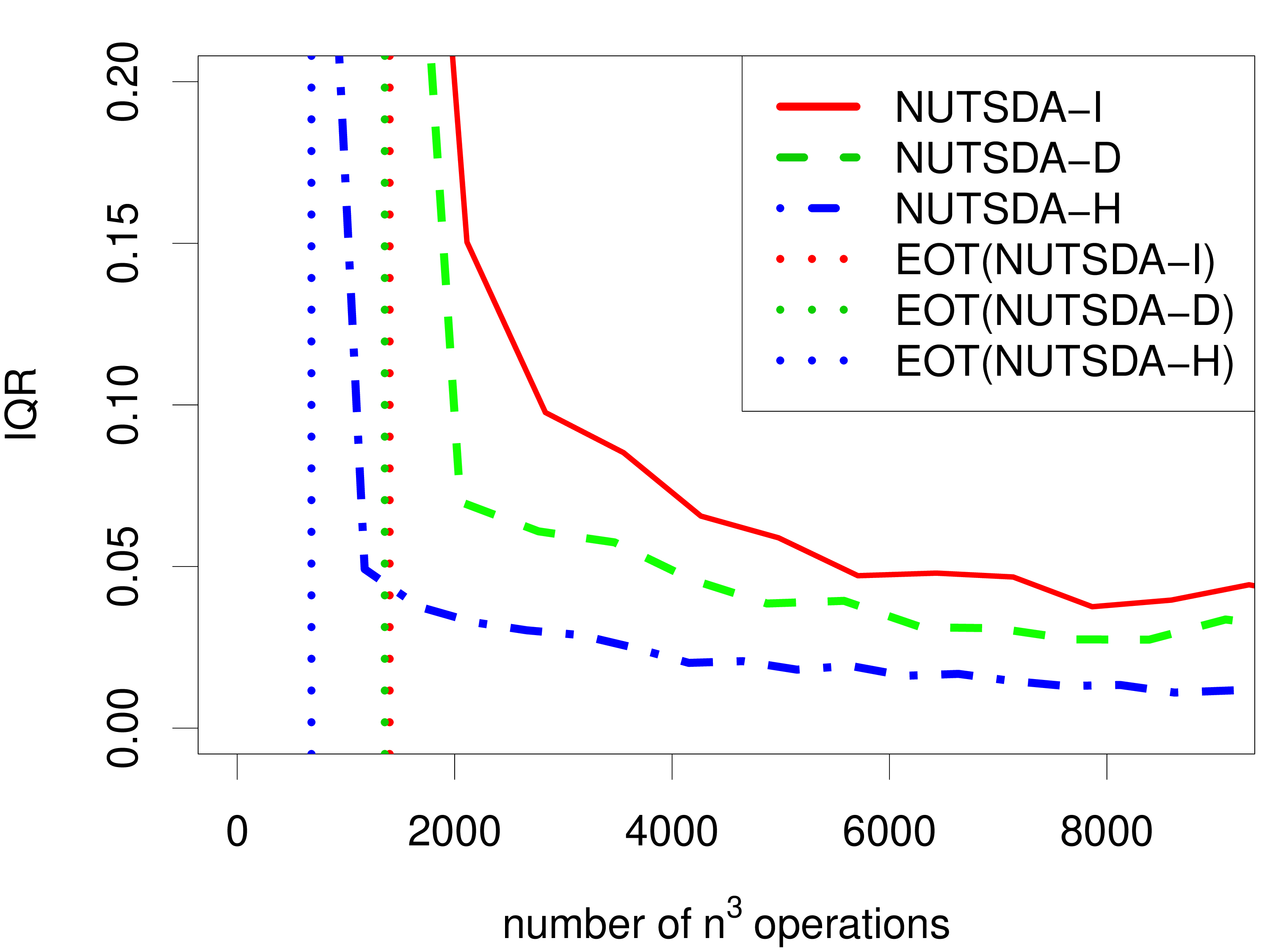}
          
   \end{tabular} 
   \caption{Convergence of AMIS/MAMIS, MH, HMC, NUTS, NUTSDA for the Concrete dataset. EOT stands for "end of tuning". \label{fig:MH_HMC_NUT_NUTDA_CON}}
  \end{center} 
\end{figure*}

\begin{figure*}[th]
  \begin{center}
{\scriptsize  {\bf Housing dataset - RBF covariance}}\\
\begin{tabular}{ccc}
          \includegraphics[width=\figwidth,angle=0]{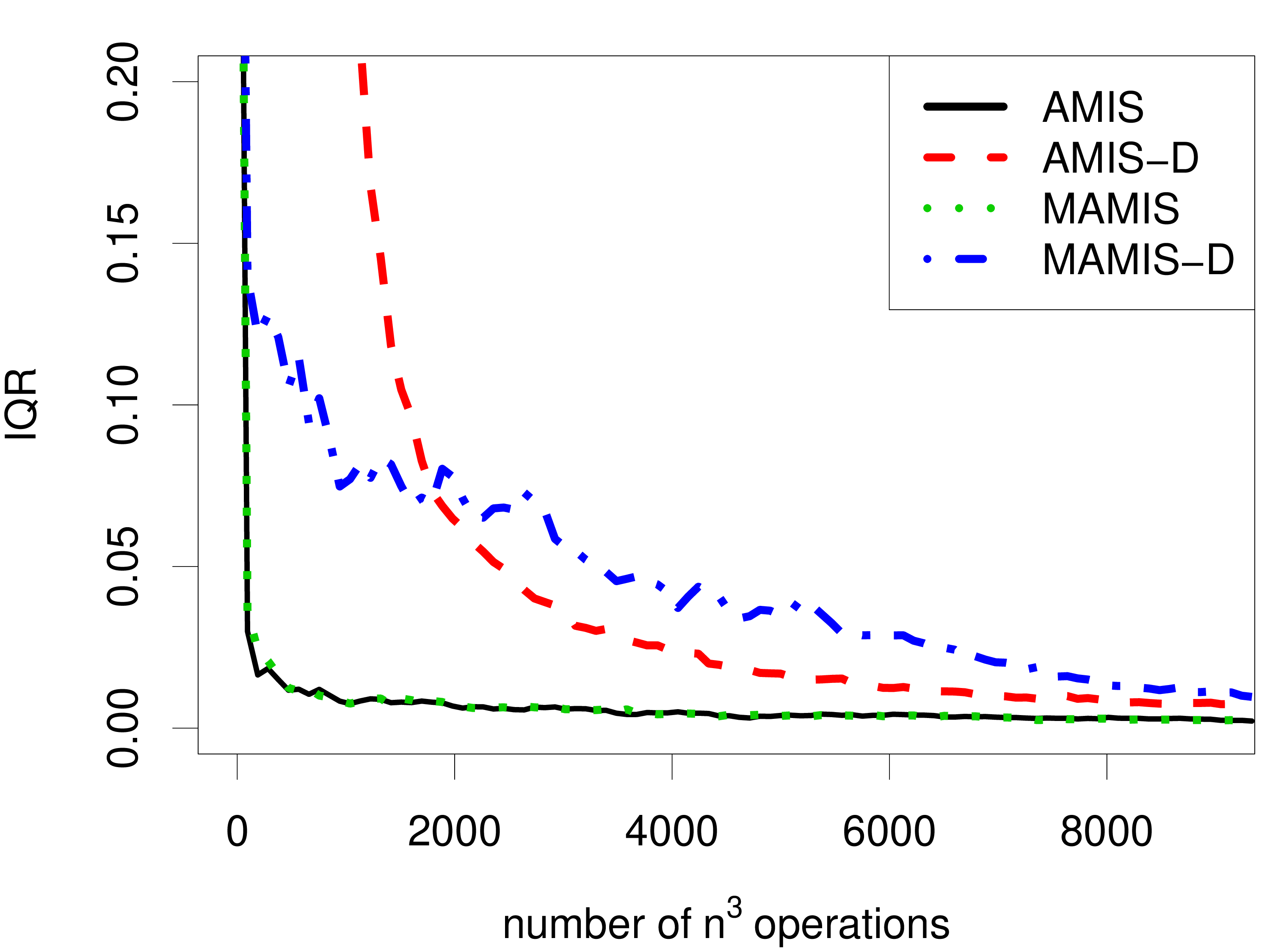} &  \includegraphics[width=\figwidth,angle=0]{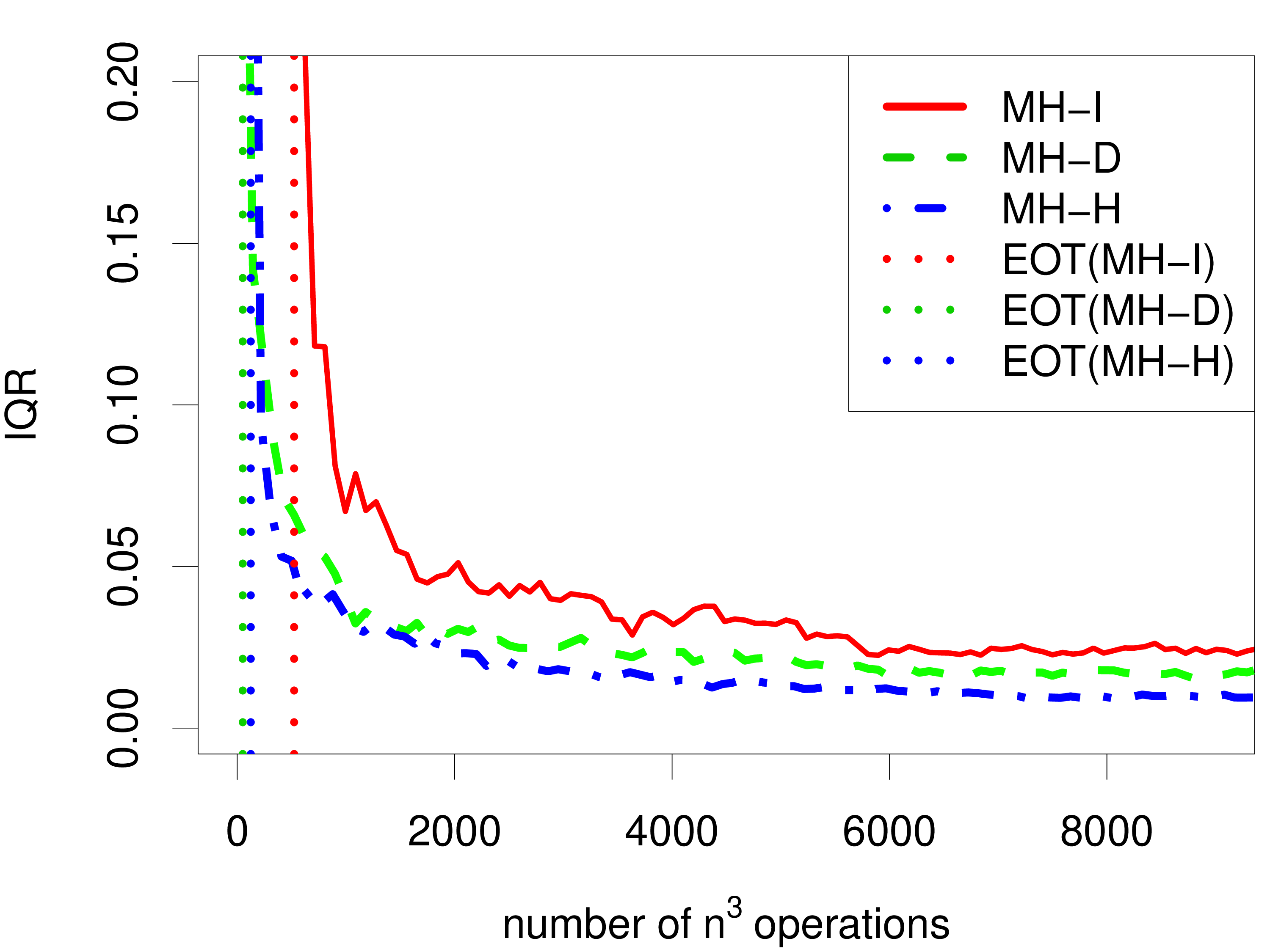} &
          \includegraphics[width=\figwidth,angle=0]{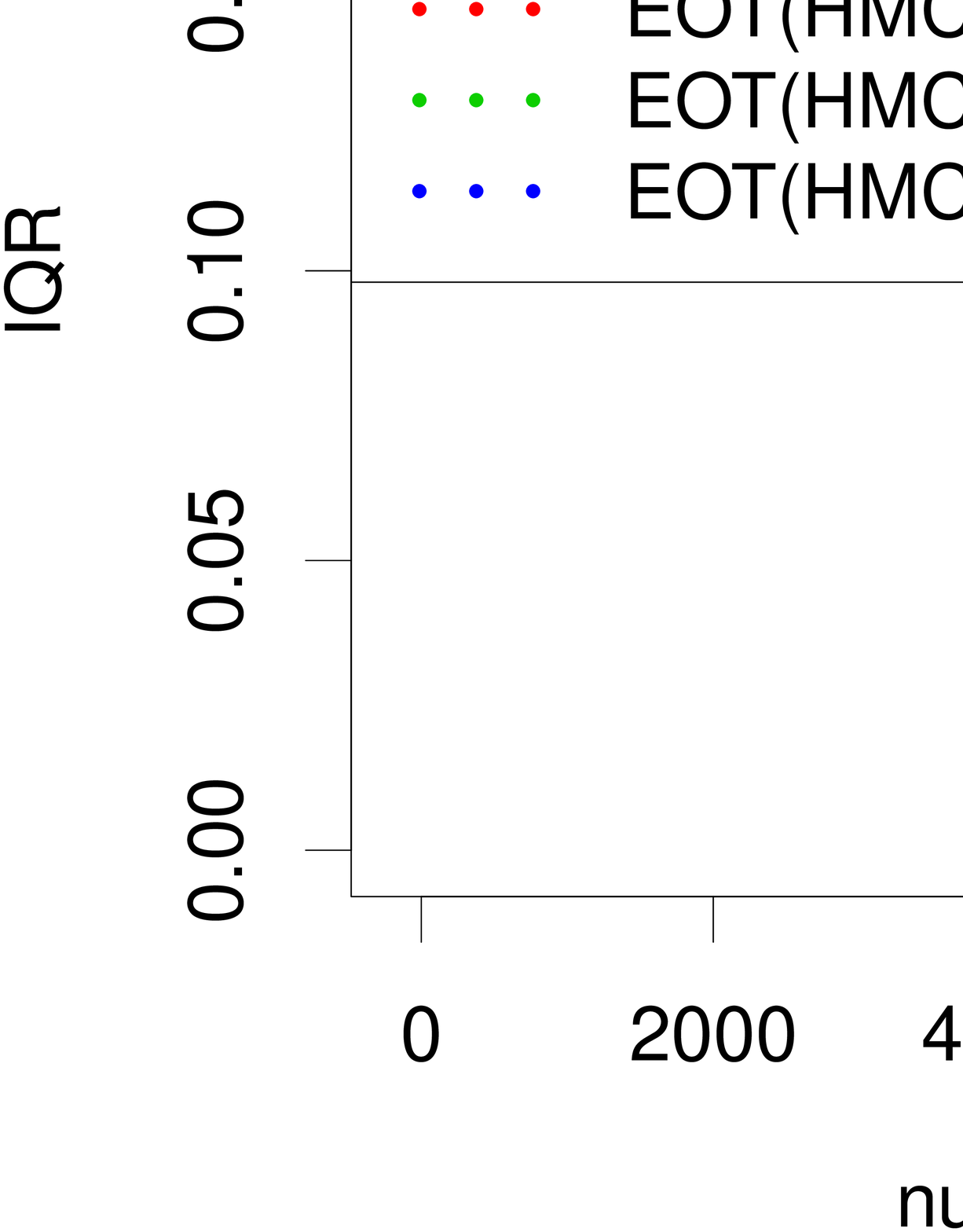} \\ \includegraphics[width=\figwidth,angle=0]{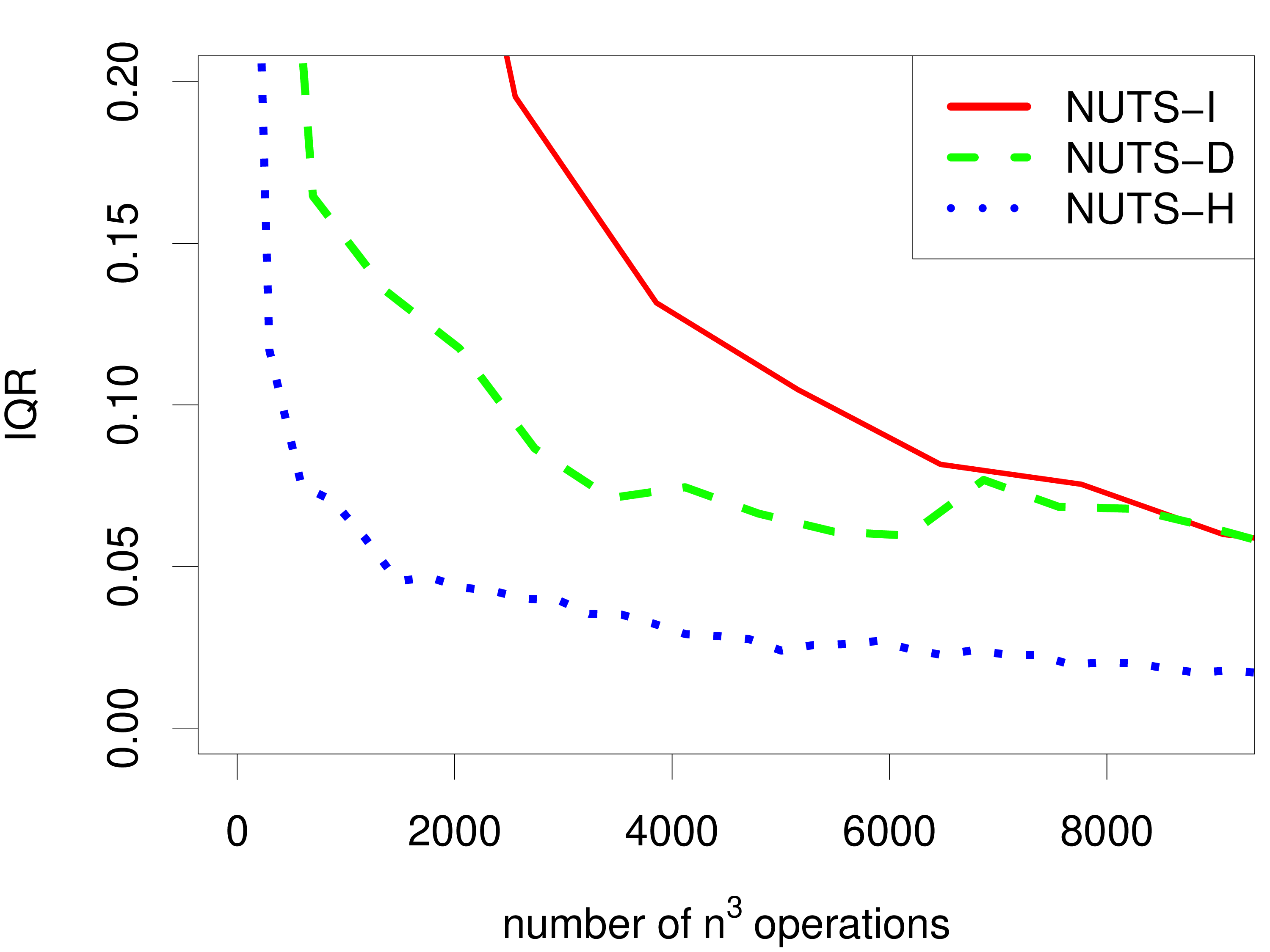} &
           \includegraphics[width=\figwidth,angle=0]{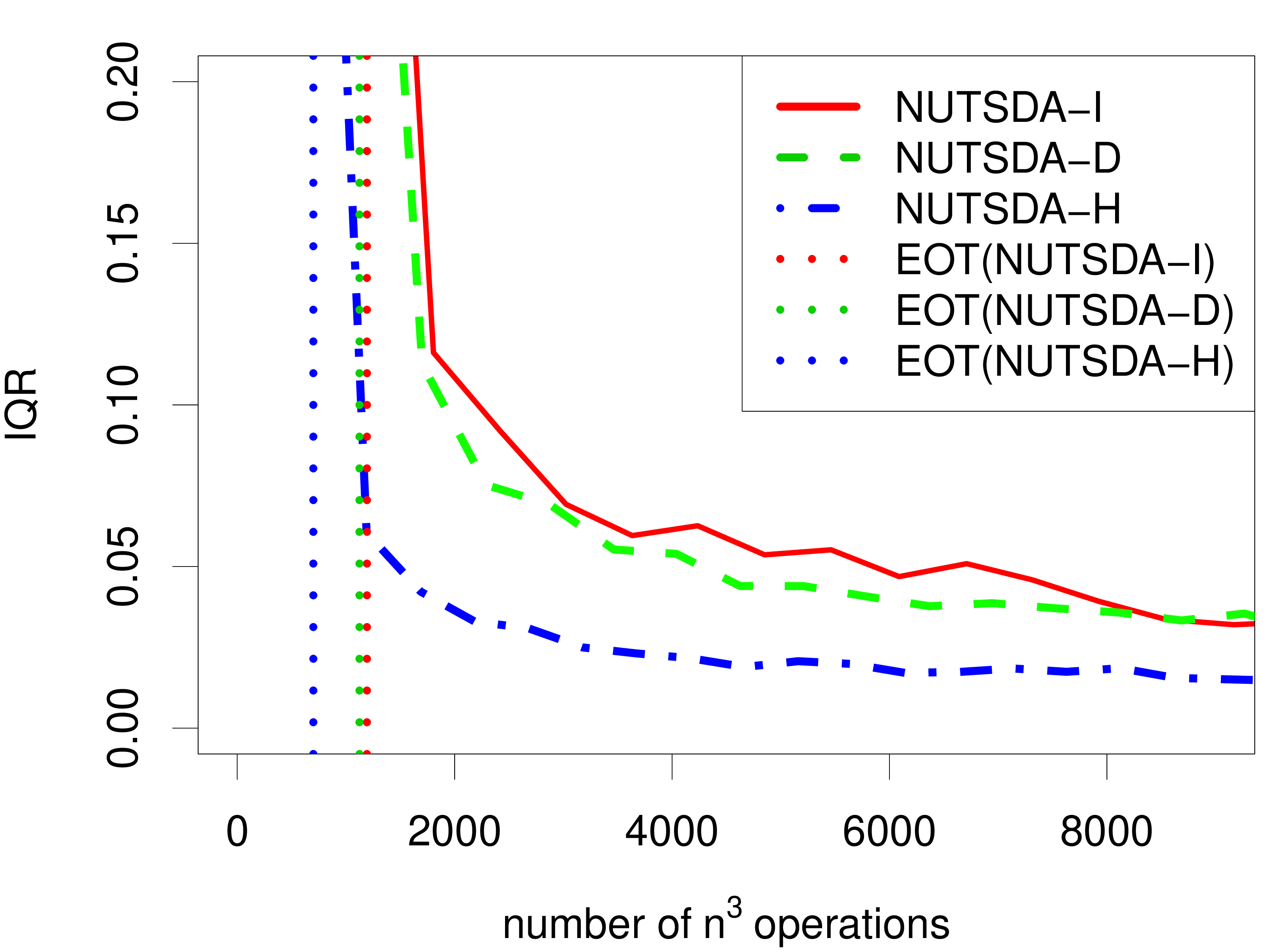} 
\end{tabular}          
  \caption{Convergence of AMIS/MAMIS, MH, HMC, NUTS, NUTSDA for the Housing dataset. EOT stands for "end of tuning". \label{fig:MH_HMC_NUT_NUTDA_HOU}}
  \end{center}
\end{figure*}

\begin{figure*}[th]
  \begin{center}
  {\scriptsize  {\bf Parkinsons dataset - RBF covariance}}\\
  \begin{tabular}{ccc}  
          \includegraphics[width=\figwidth,angle=0]{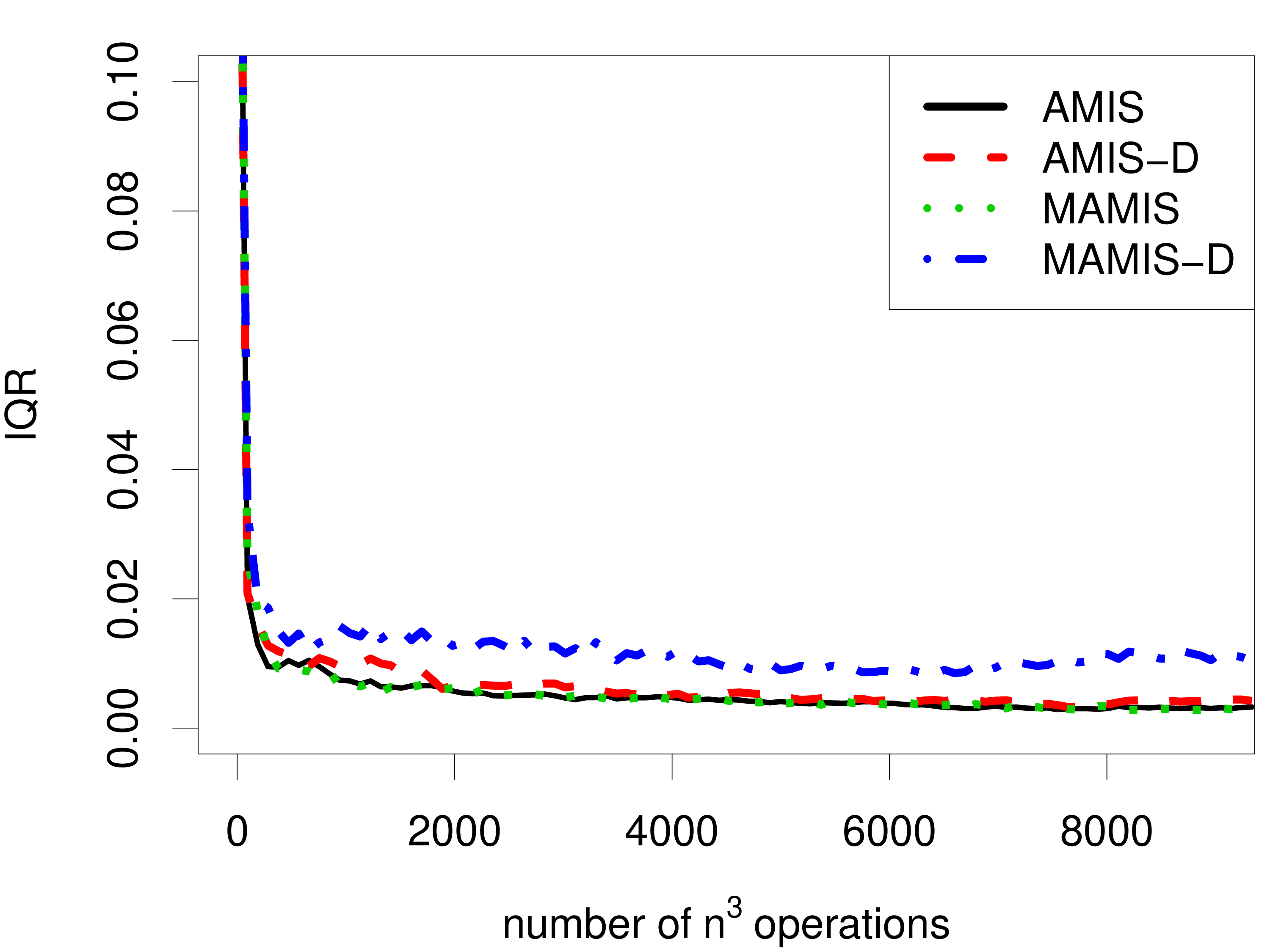}&
          \includegraphics[width=\figwidth,angle=0]{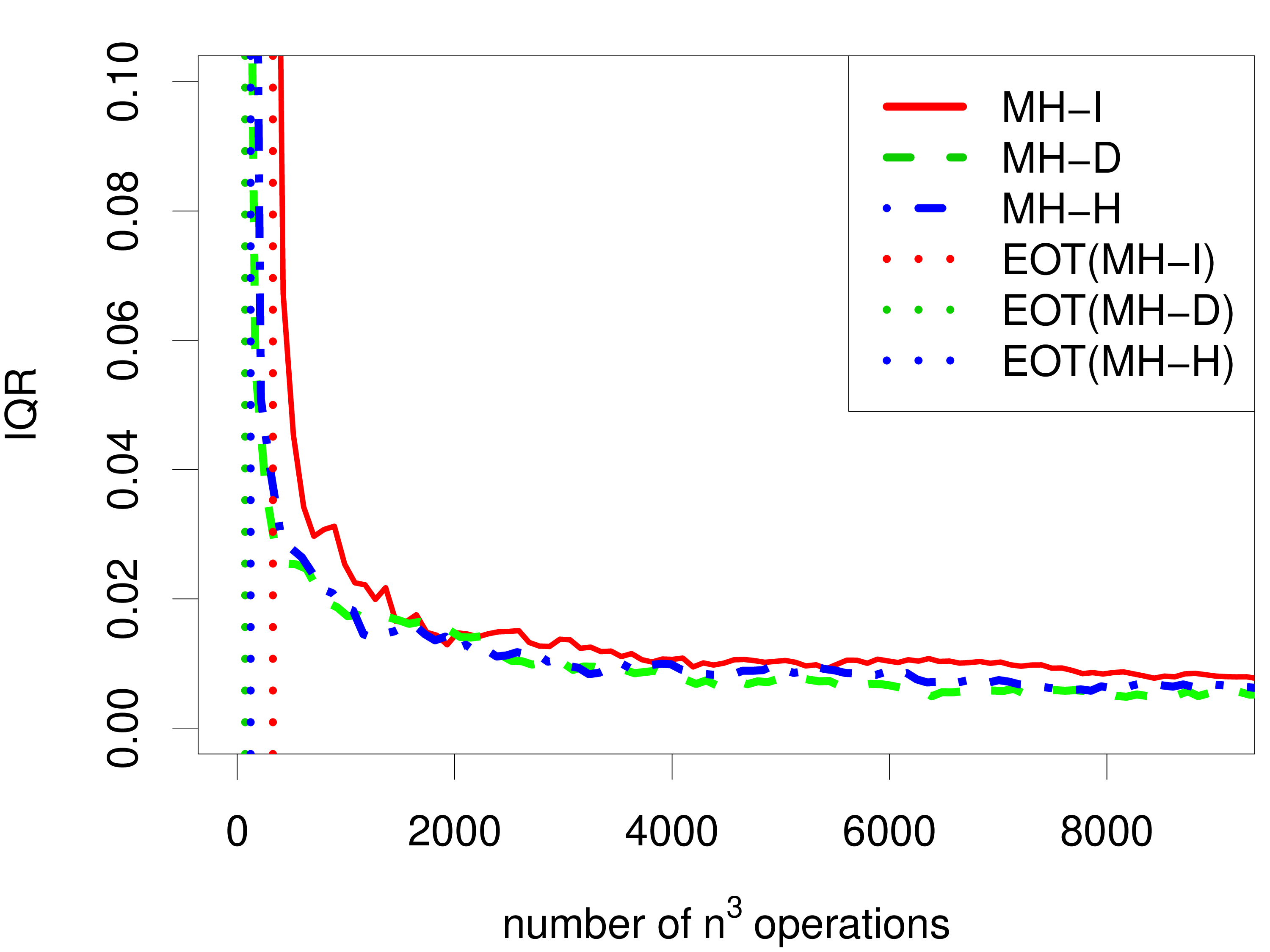}&
           \includegraphics[width=\figwidth,angle=0]{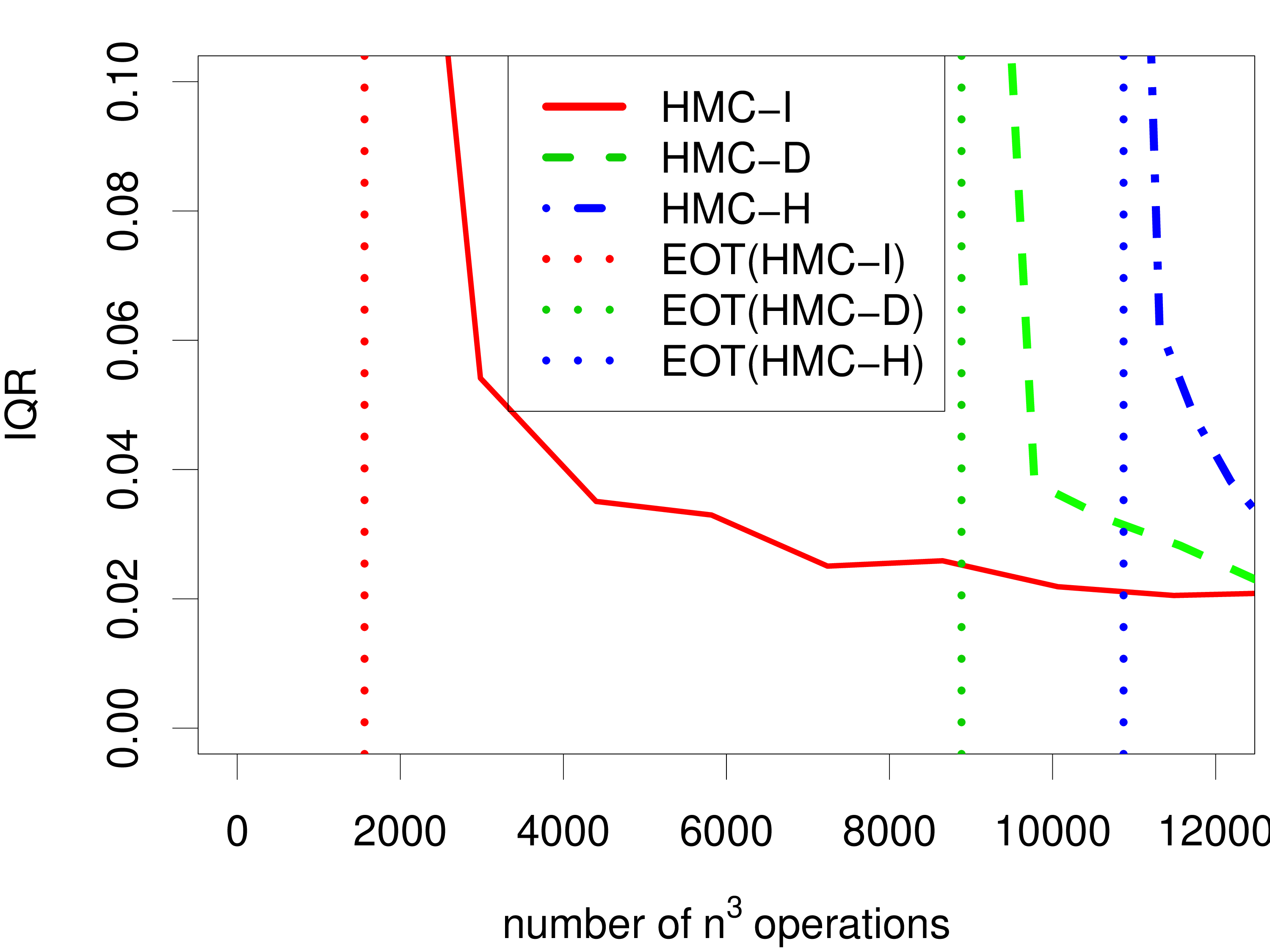}\\
          \includegraphics[width=\figwidth,angle=0]{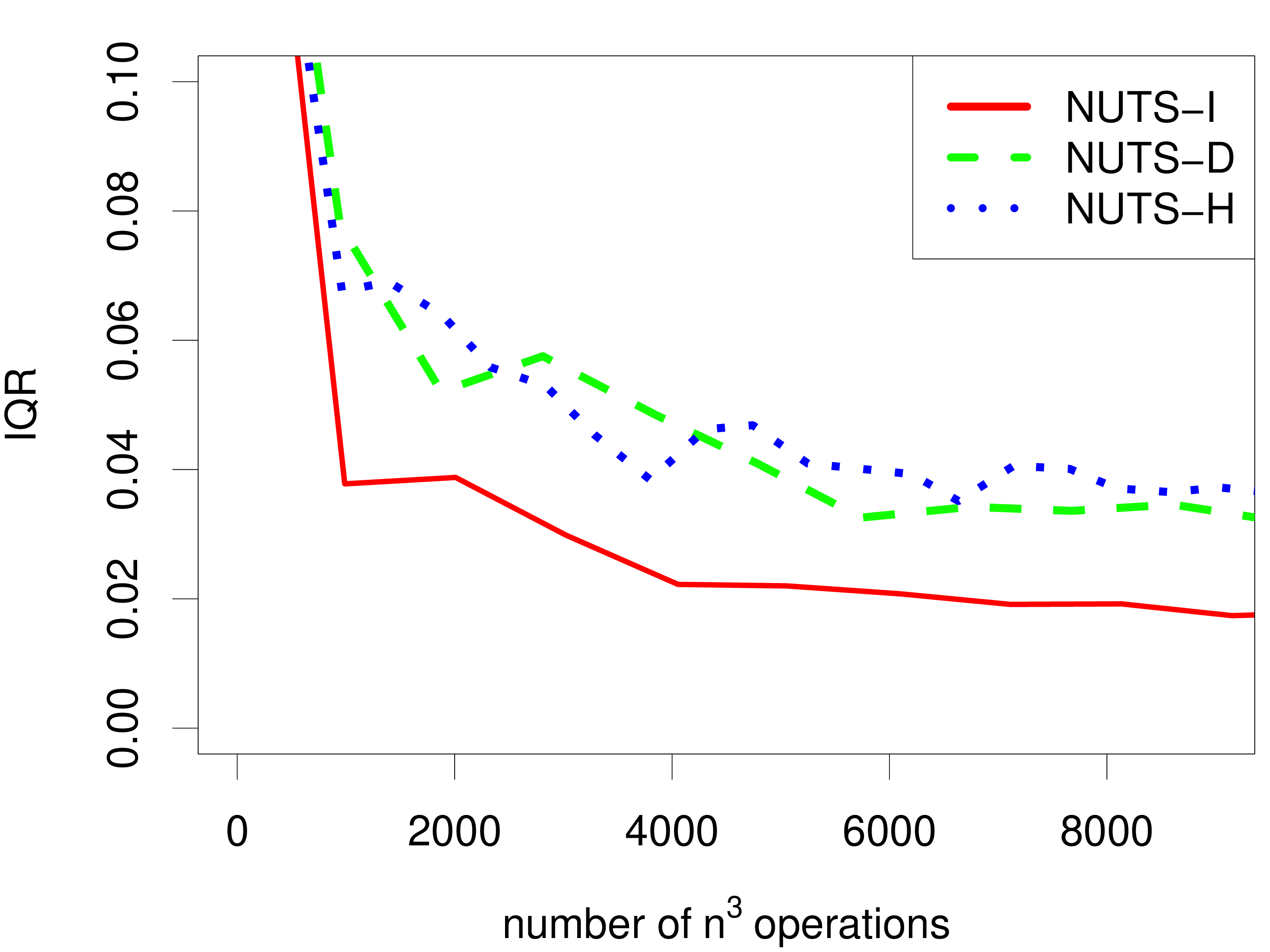}&
          \includegraphics[width=\figwidth,angle=0]{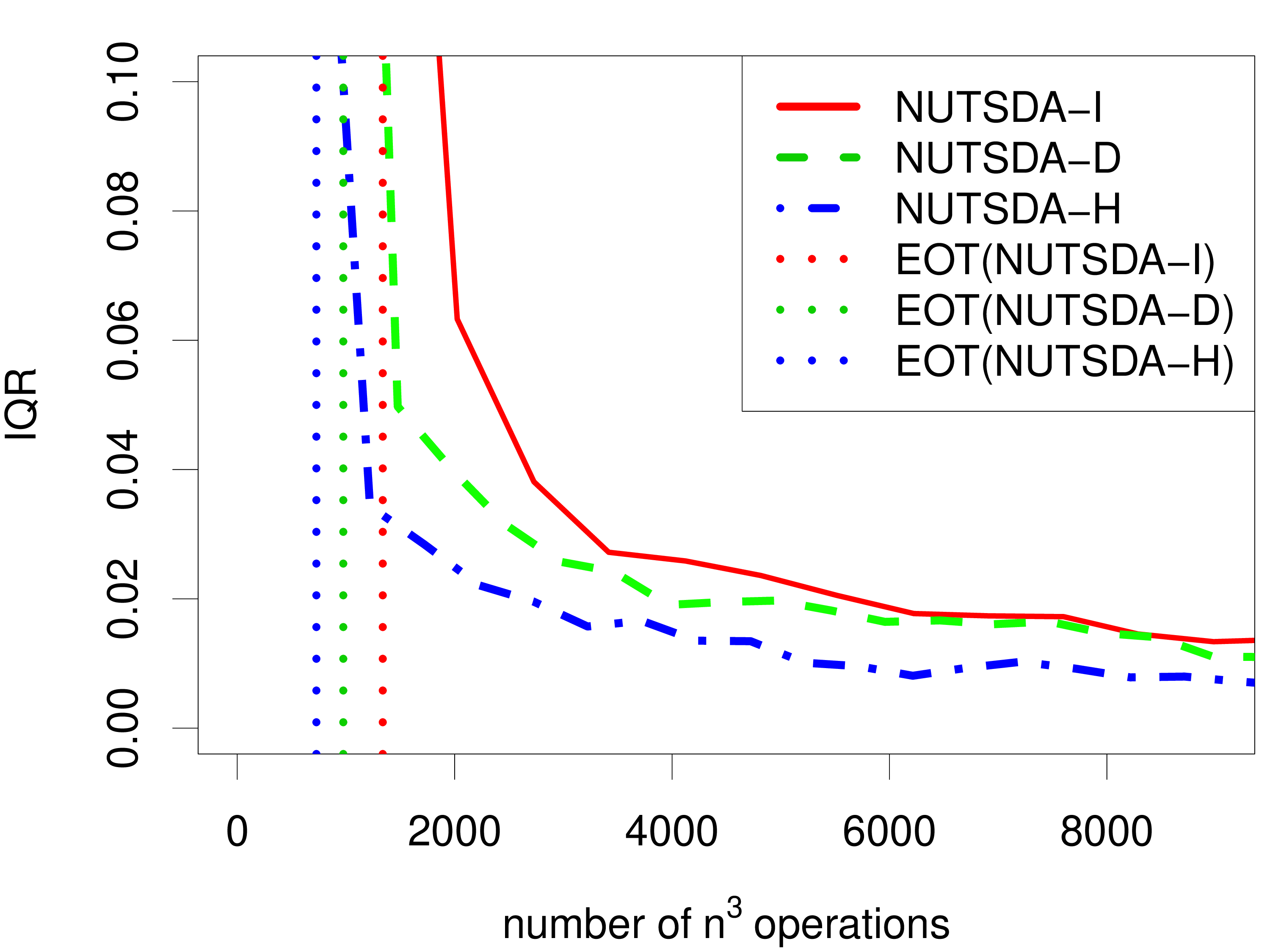}
          
   \end{tabular} 
   \caption{Convergence of AMIS/MAMIS, MH, HMC, NUTS, NUTSDA for the Parkinsons dataset. EOT stands for "end of tuning". \label{fig:MH_HMC_NUT_NUTDA_PARK}}
  \end{center}
 
\end{figure*}

\FloatBarrier

\newpage
\section{Convergence of samplers for GP regression with the ARD covariance} \label{App:AppendixB}

\begin{figure*}[th]
  \begin{center}
  {\scriptsize  {\bf Concrete dataset - ARD covariance}}\\
      \begin{tabular}{ccc}
           \includegraphics[width=\figwidth,angle=0]{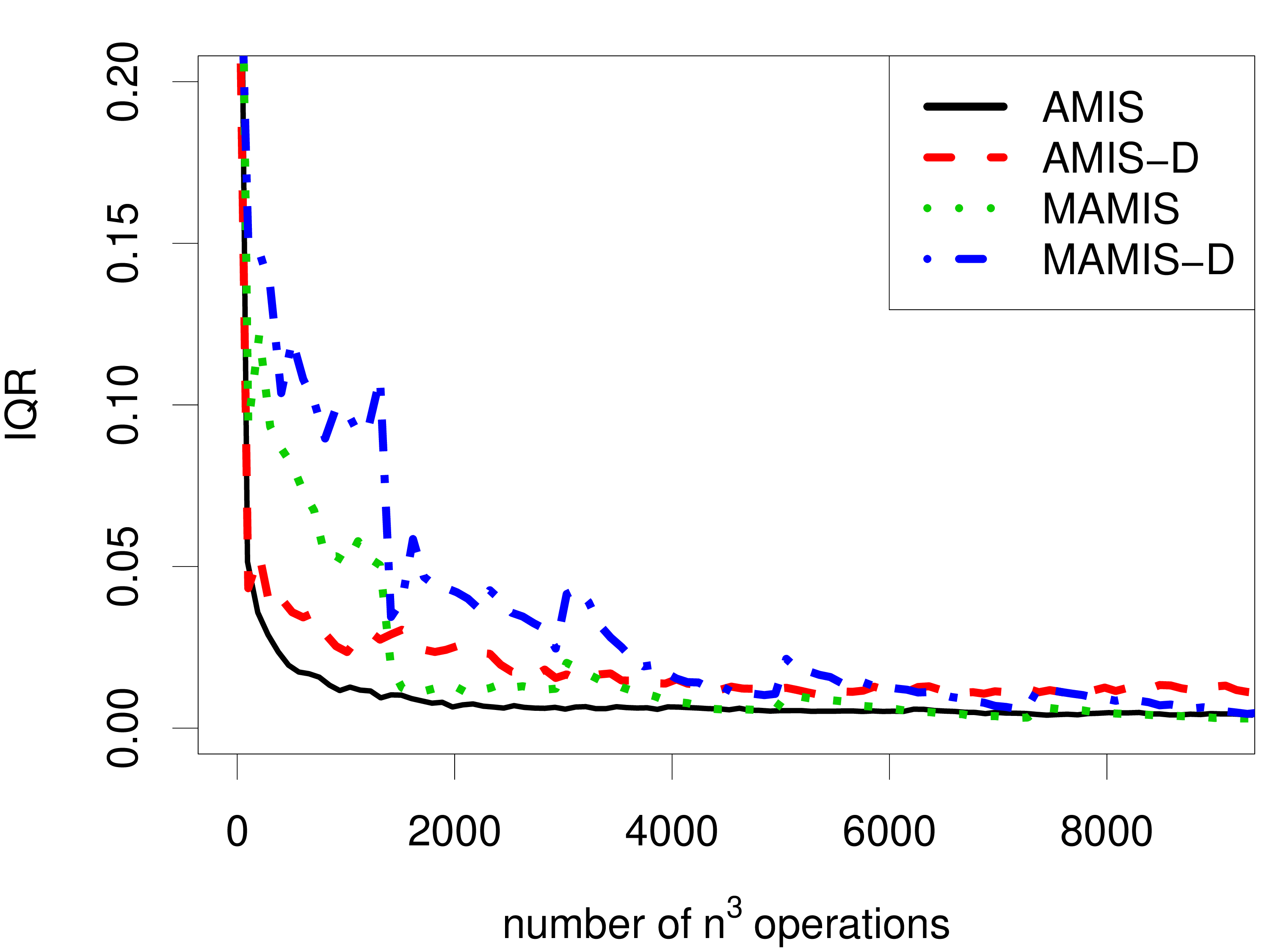}& 
           \includegraphics[width=\figwidth,angle=0]{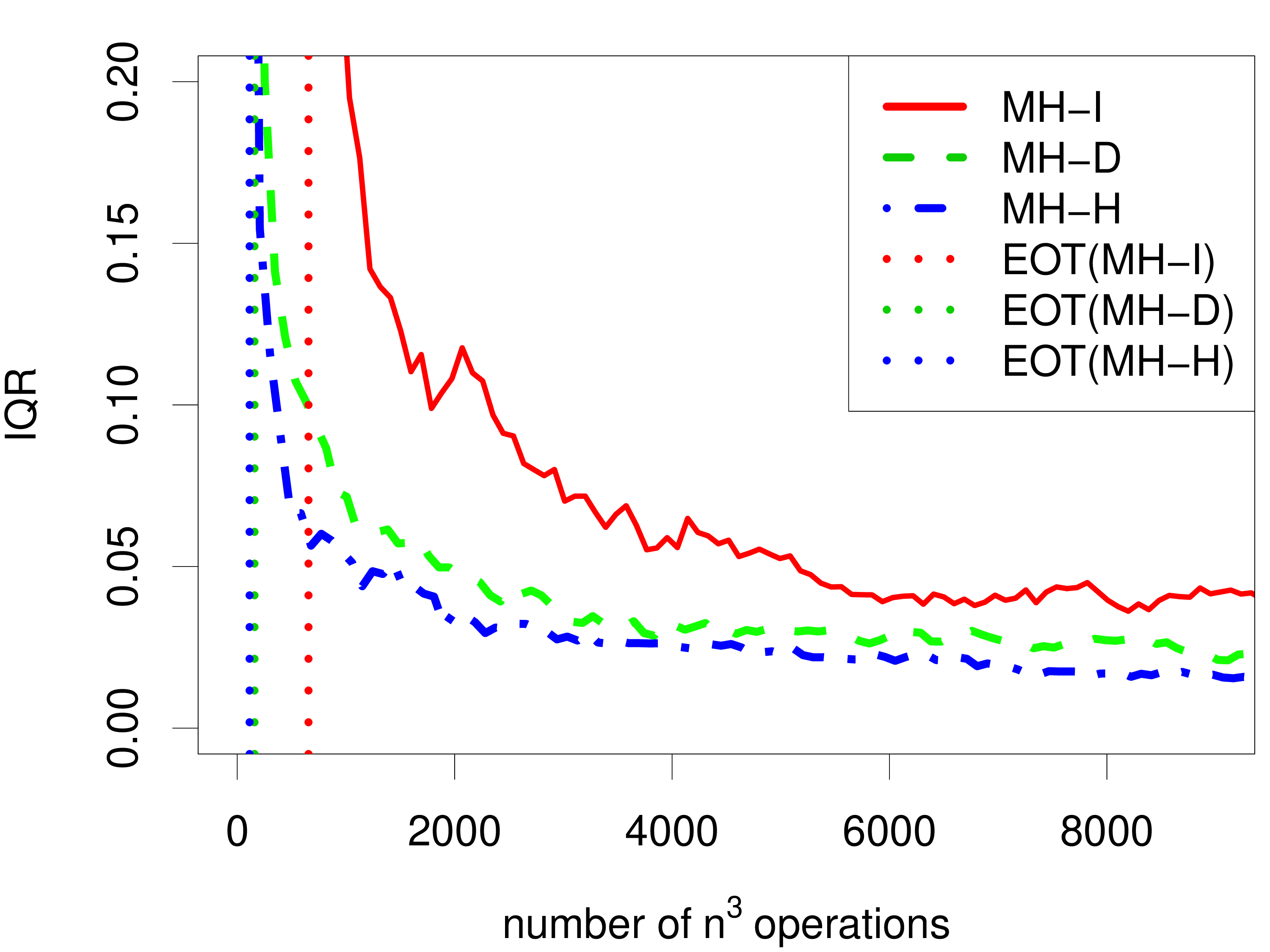}&
          \includegraphics[width=\figwidth,angle=0]{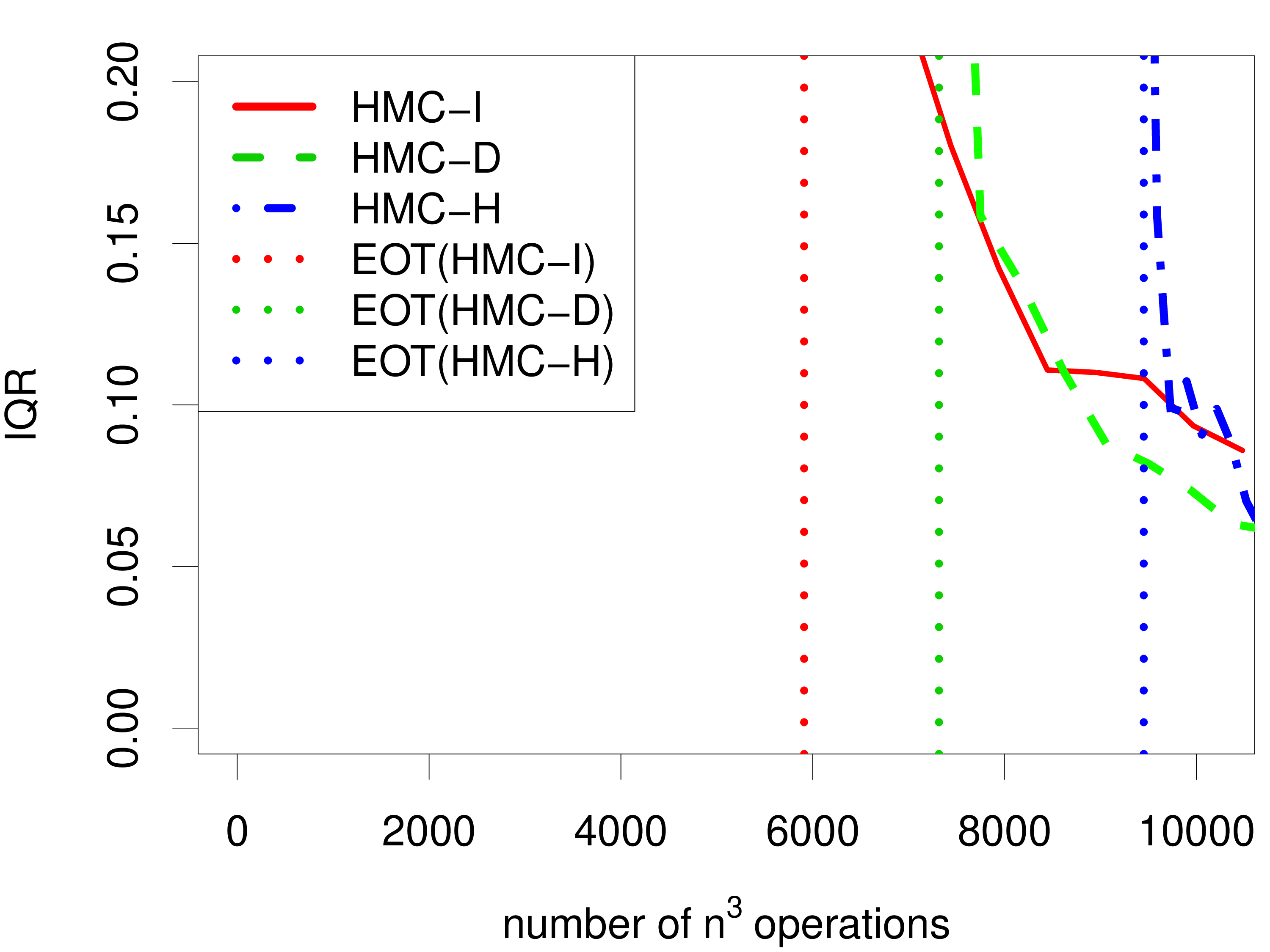}\\
          \includegraphics[width=\figwidth,angle=0]{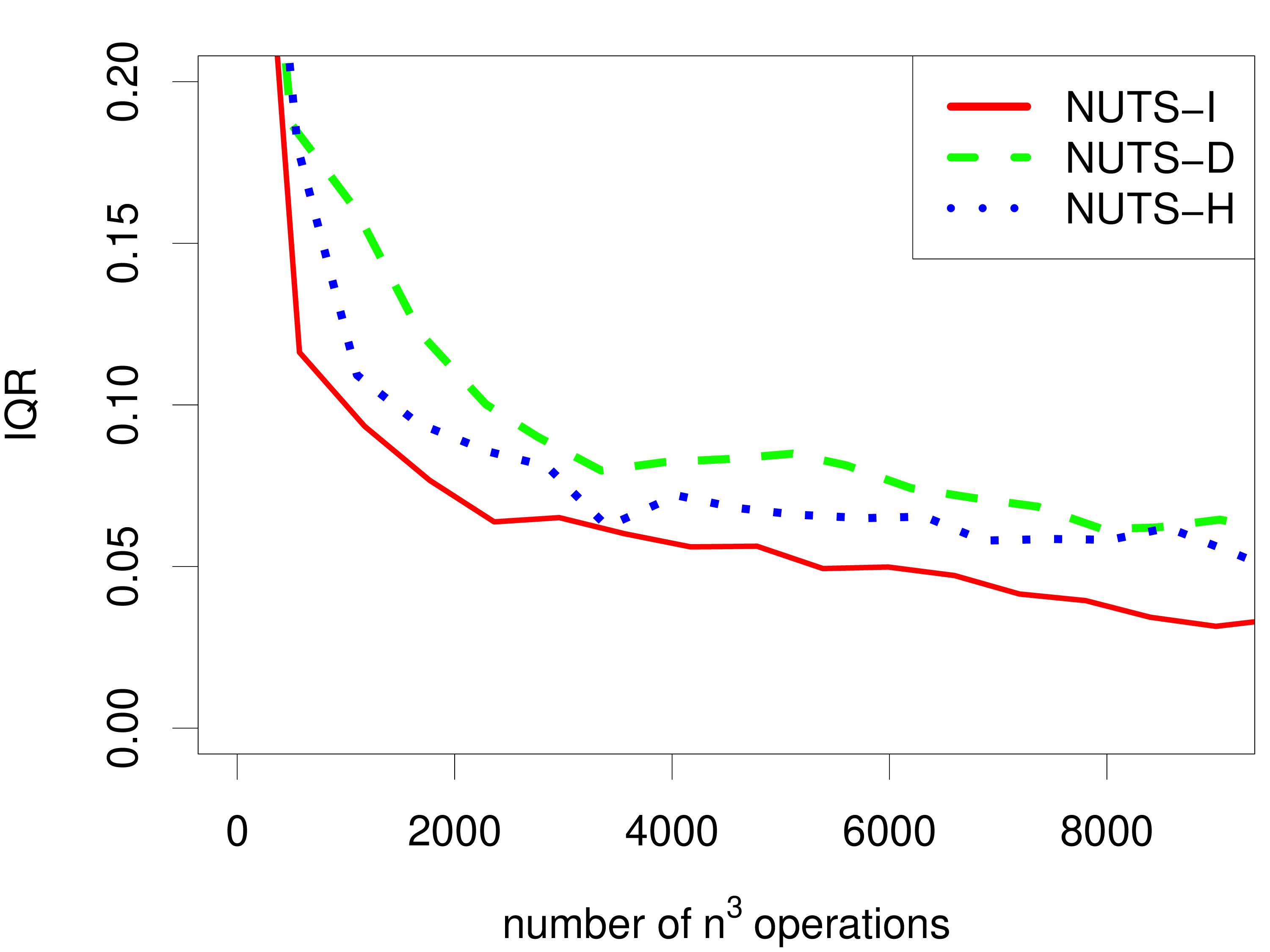}&
          \includegraphics[width=\figwidth,angle=0]{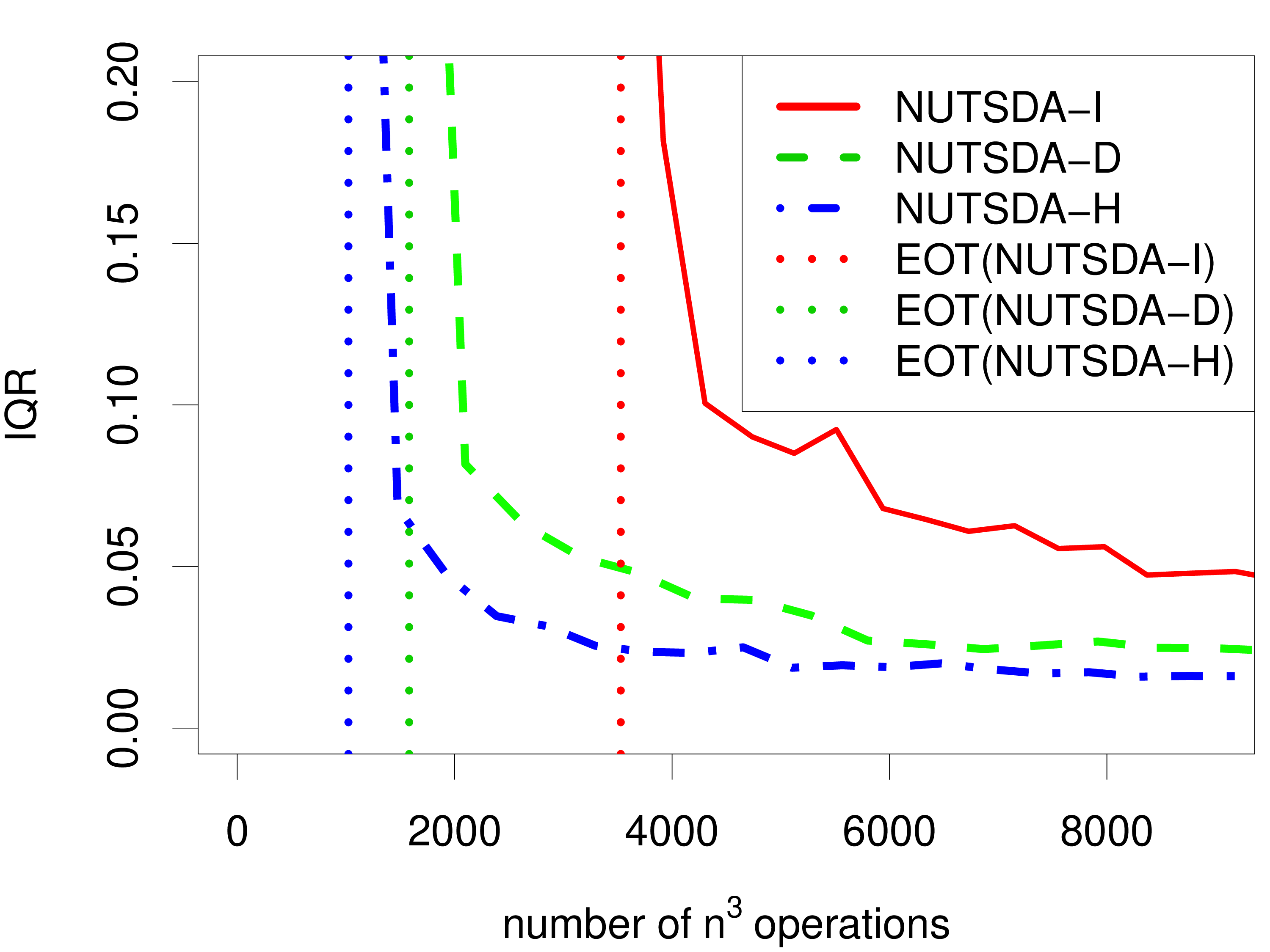}
          
       \end{tabular}
   \caption{Convergence of AMIS/MAMIS, MH, HMC, NUTS, NUTSDA for the Concrete dataset. EOT stands for "end of tuning". \label{fig:MH_HMC_NUT_NUTDA_CON_ARD}}    
  \end{center}  
\end{figure*}

\begin{figure*}[th]
  \begin{center}
  {\scriptsize  {\bf Housing dataset - ARD covariance}}\\
    \begin{tabular}{ccc}  
          \includegraphics[width=\figwidth,angle=0]{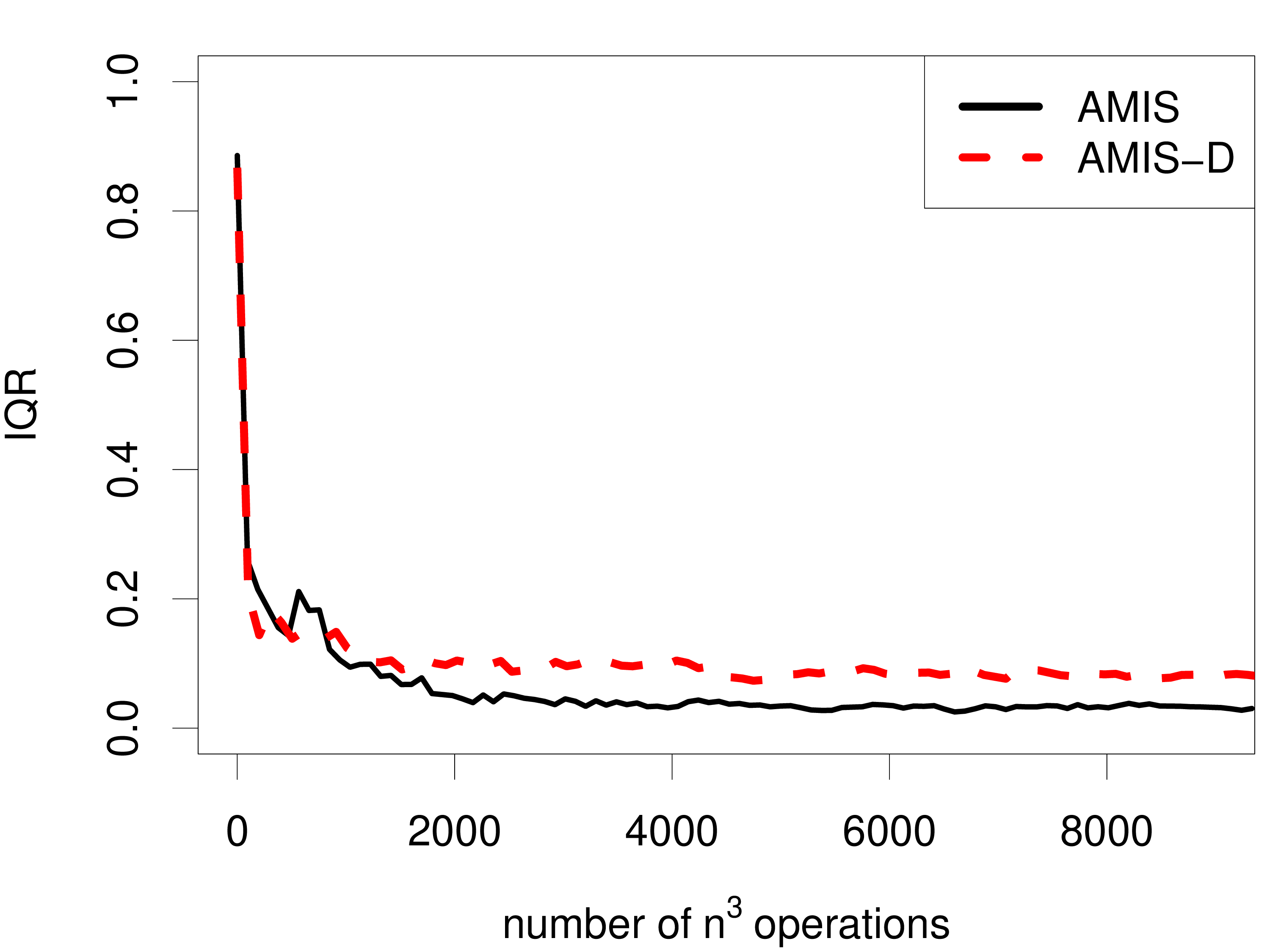}& 
          \includegraphics[width=\figwidth,angle=0]{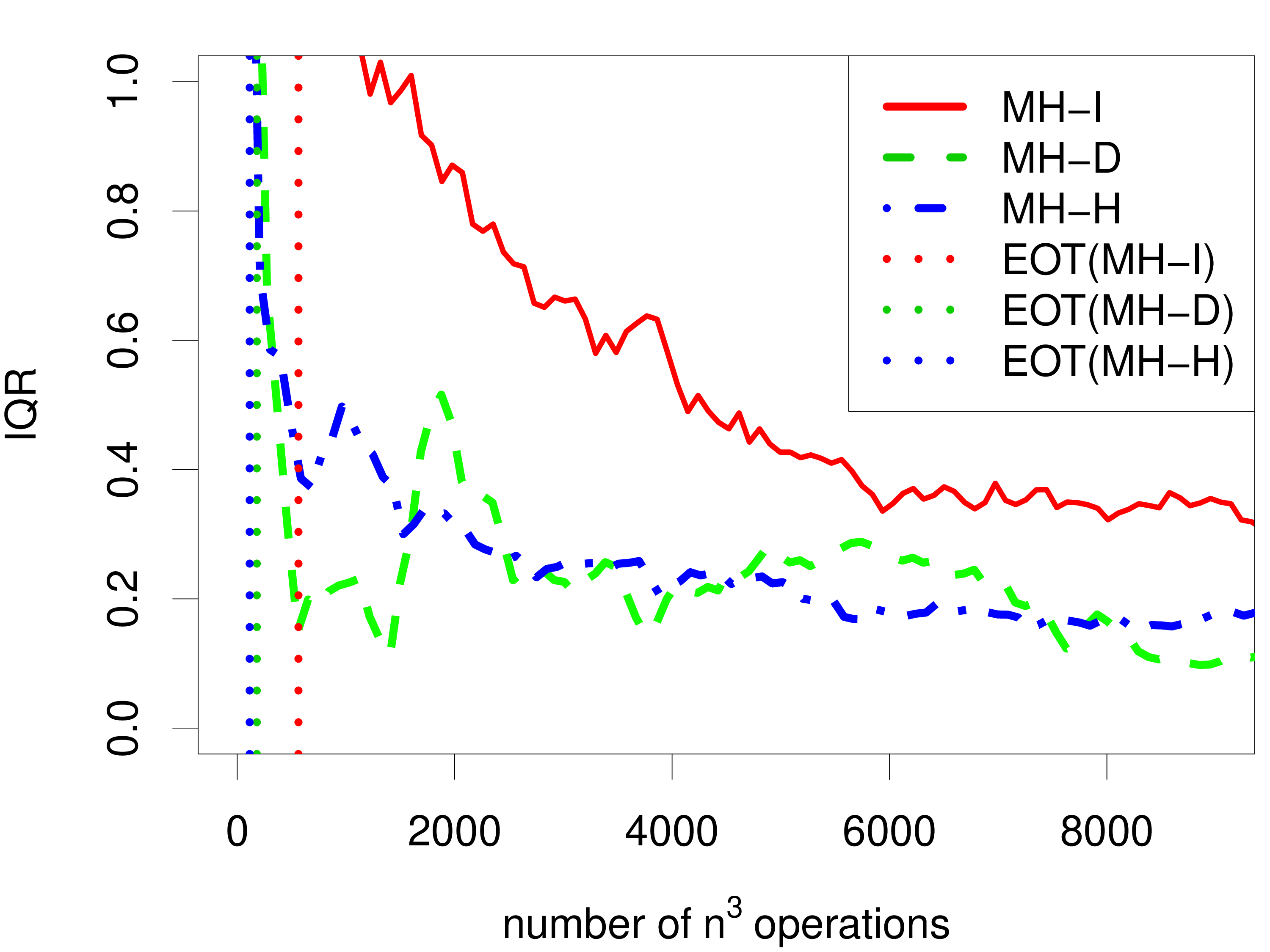}&
          \includegraphics[width=\figwidth,angle=0]{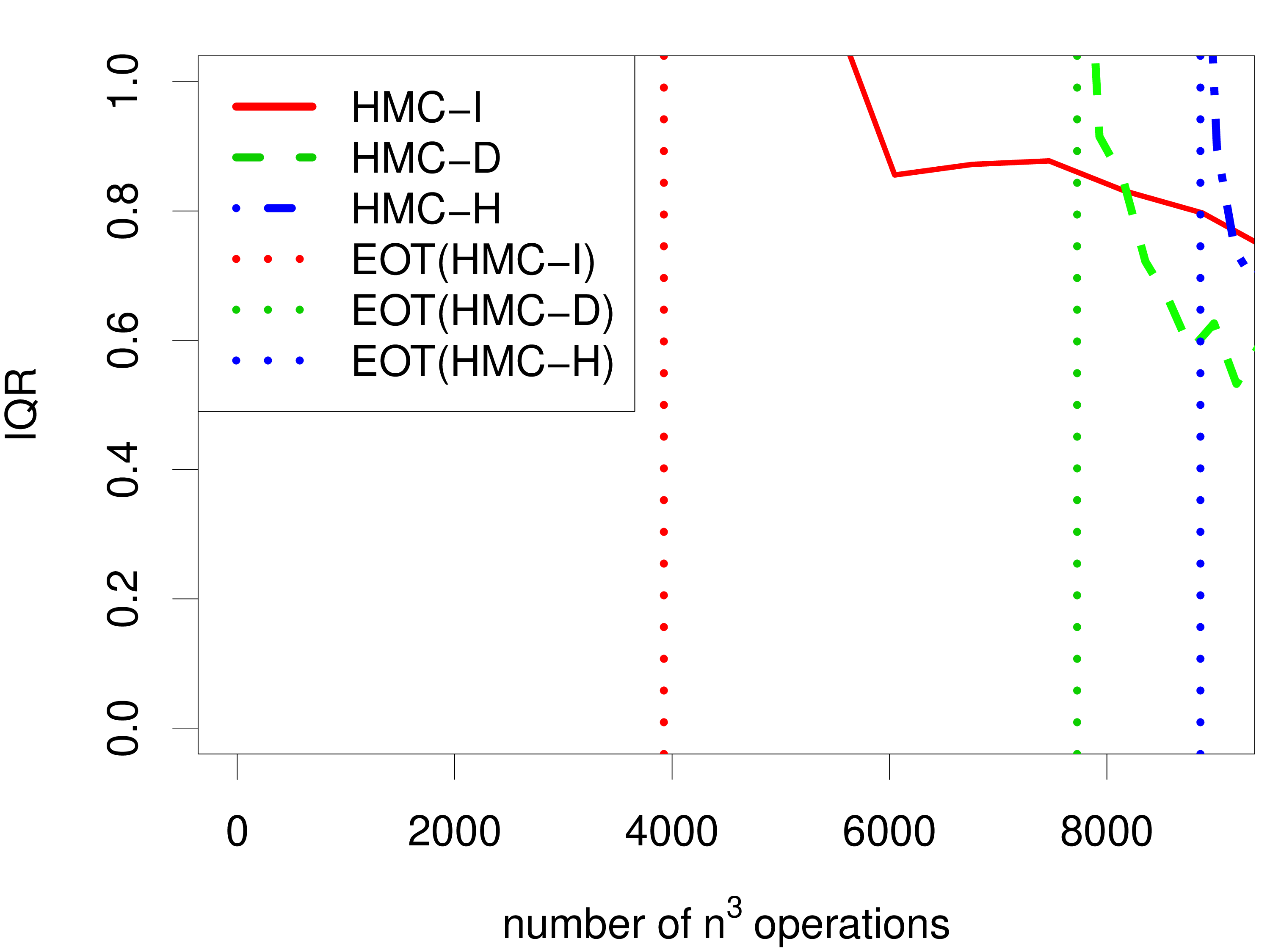}\\                     
          \includegraphics[width=\figwidth,angle=0]{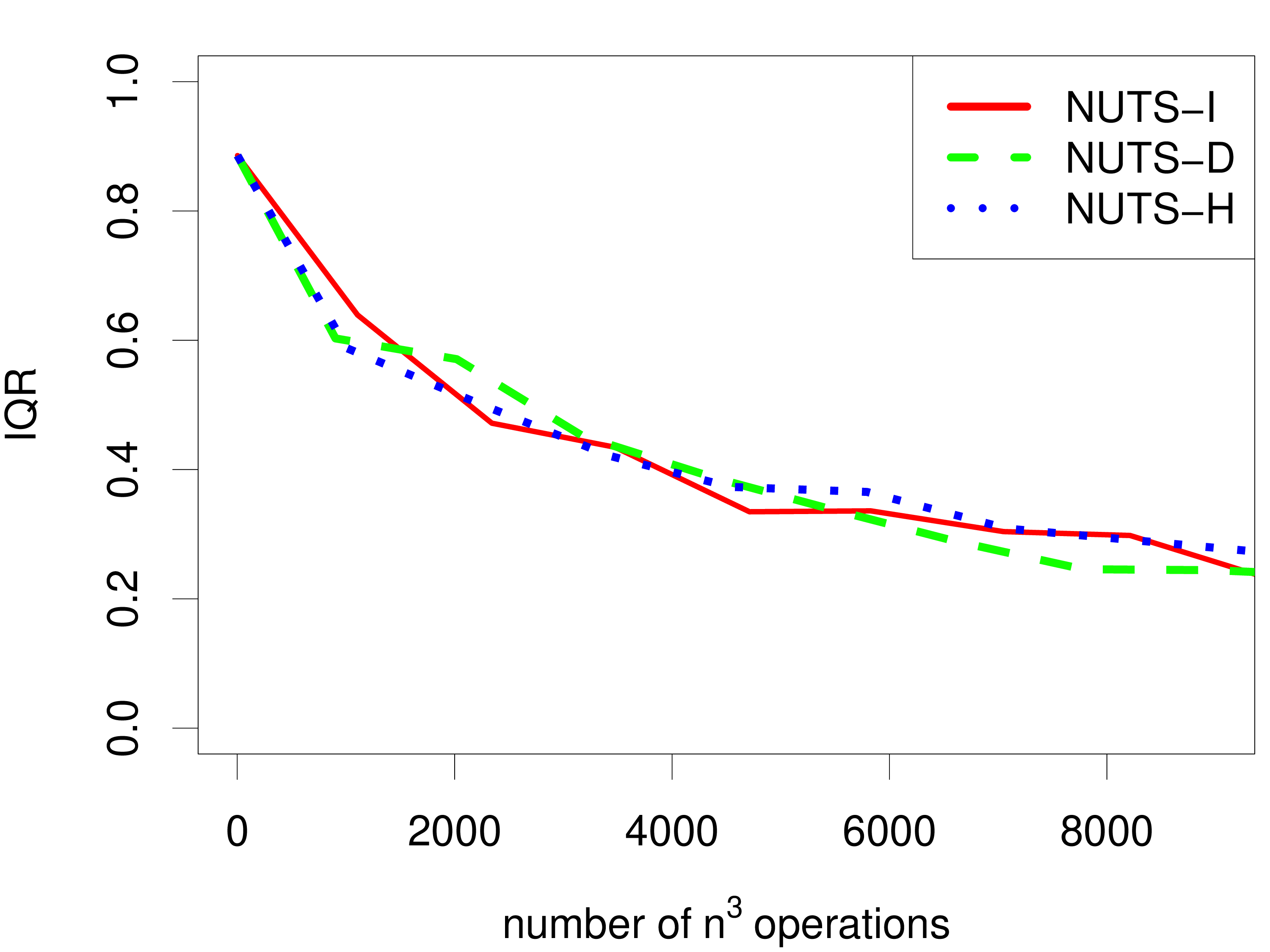}&
          \includegraphics[width=\figwidth,angle=0]{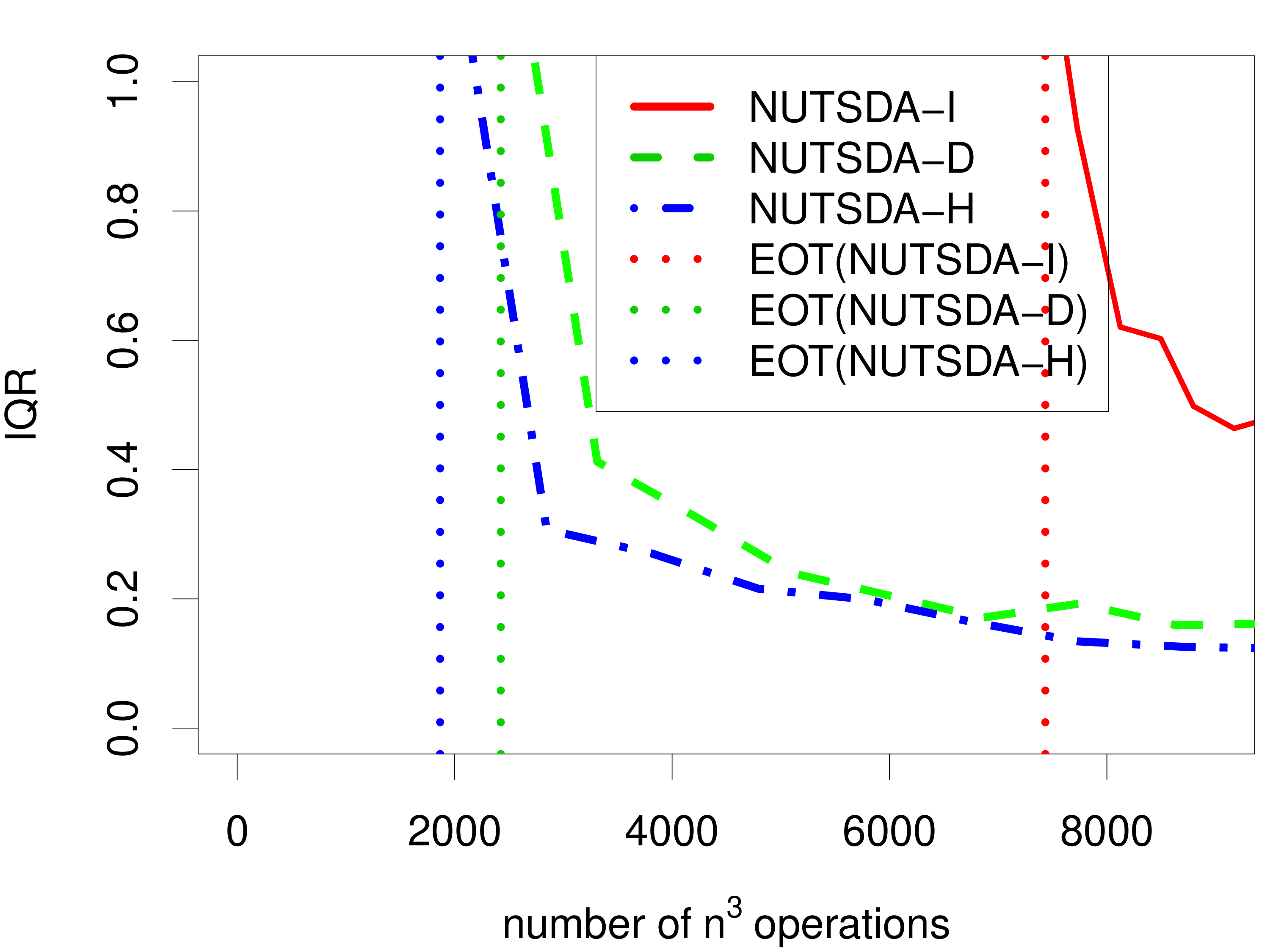}&    \includegraphics[width=\figwidth,angle=0]{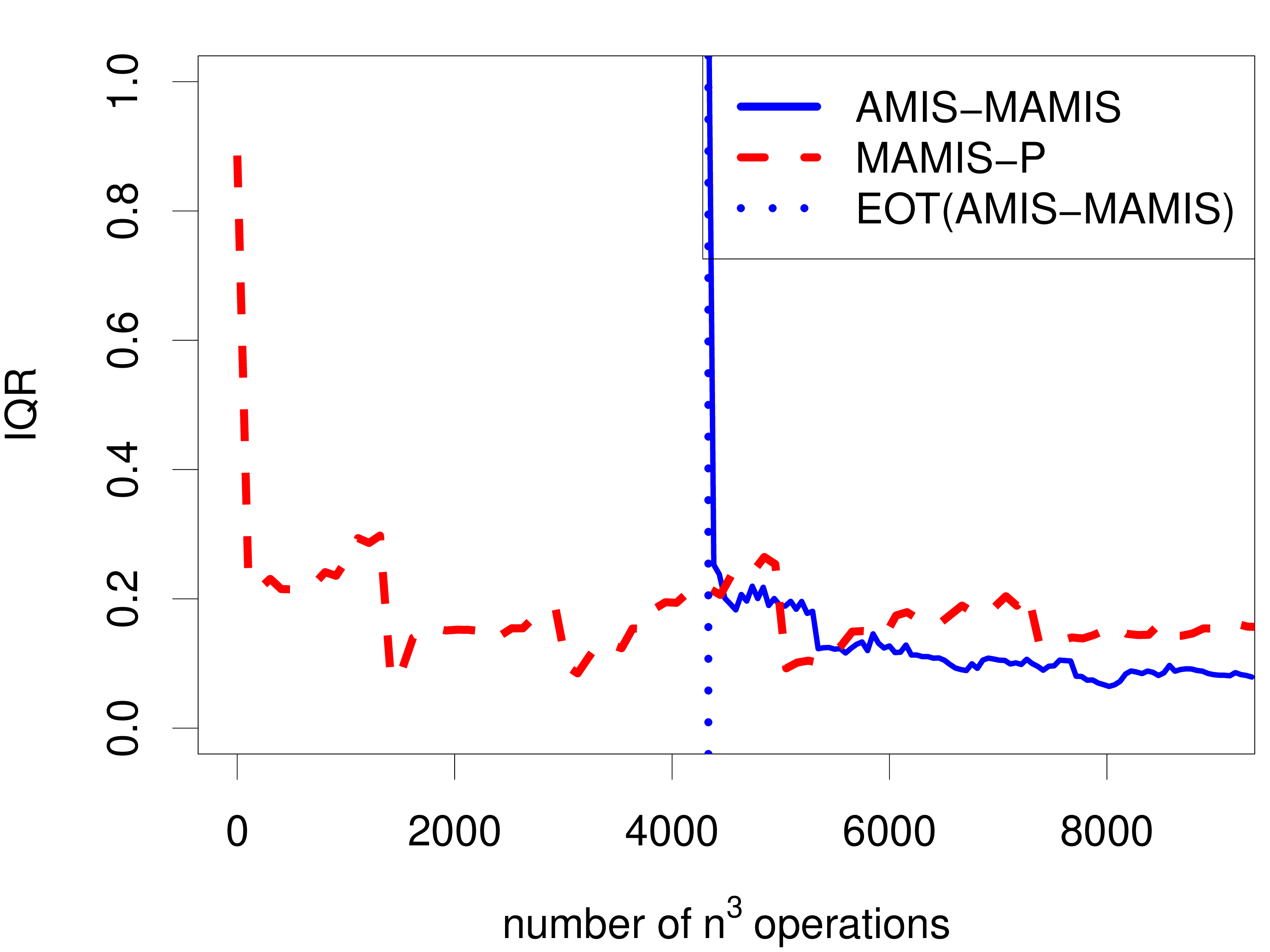}\\     
          
    \end{tabular}
\caption{Convergence of AMIS/MAMIS, MH, HMC, NUTS, NUTSDA for the Housing dataset. EOT stands for "end of tuning". \label{fig:MH_HMC_NUT_NUTDA_HOU_ARD}}
  \end{center}  
\end{figure*}

\begin{figure*}[th]
  \begin{center}
  {\scriptsize  {\bf Parkinsons dataset - ARD covariance}}\\
        \begin{tabular}{ccc}
         \includegraphics[width=\figwidth,angle=0]{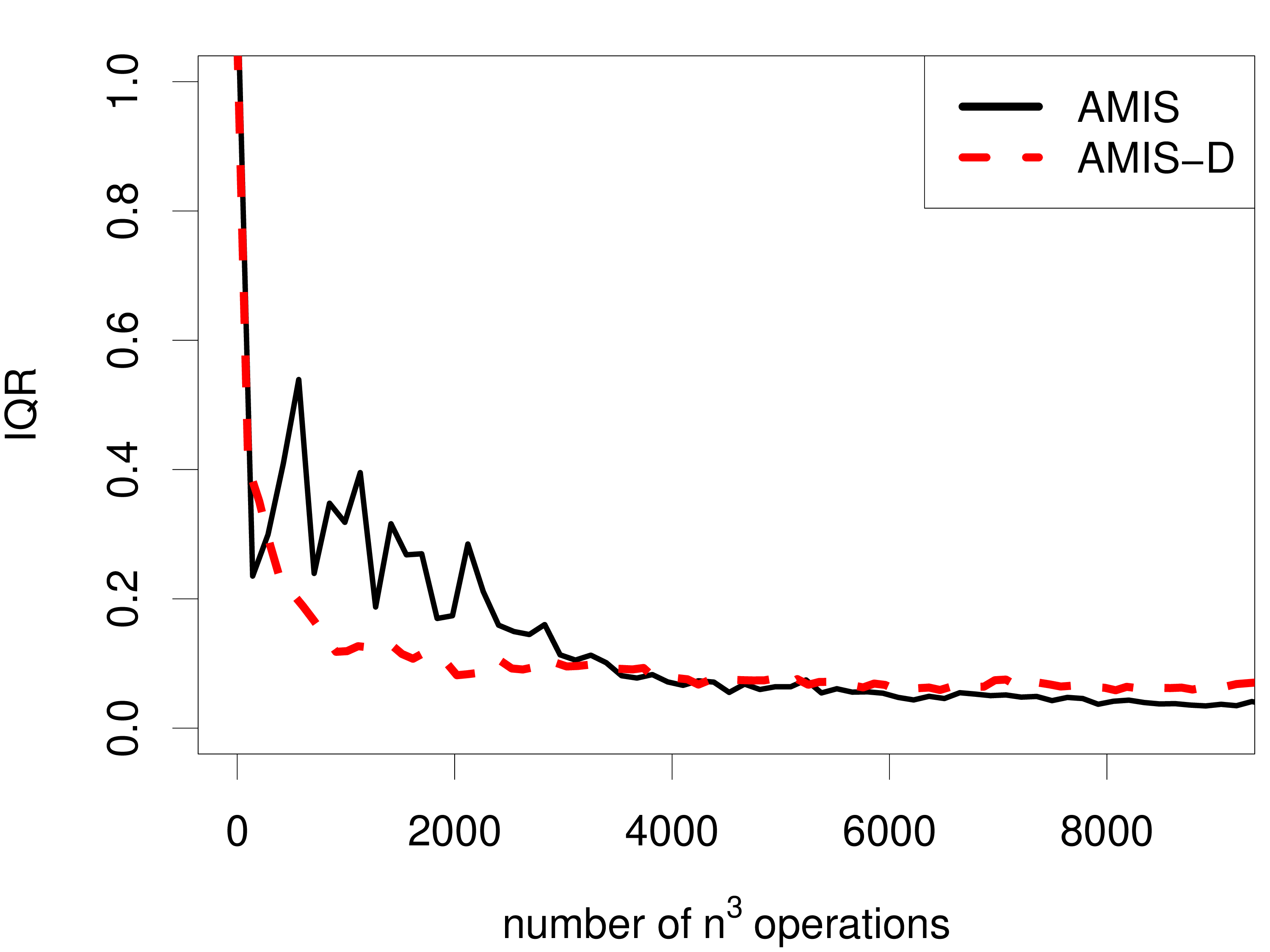}& 
          \includegraphics[width=\figwidth,angle=0]{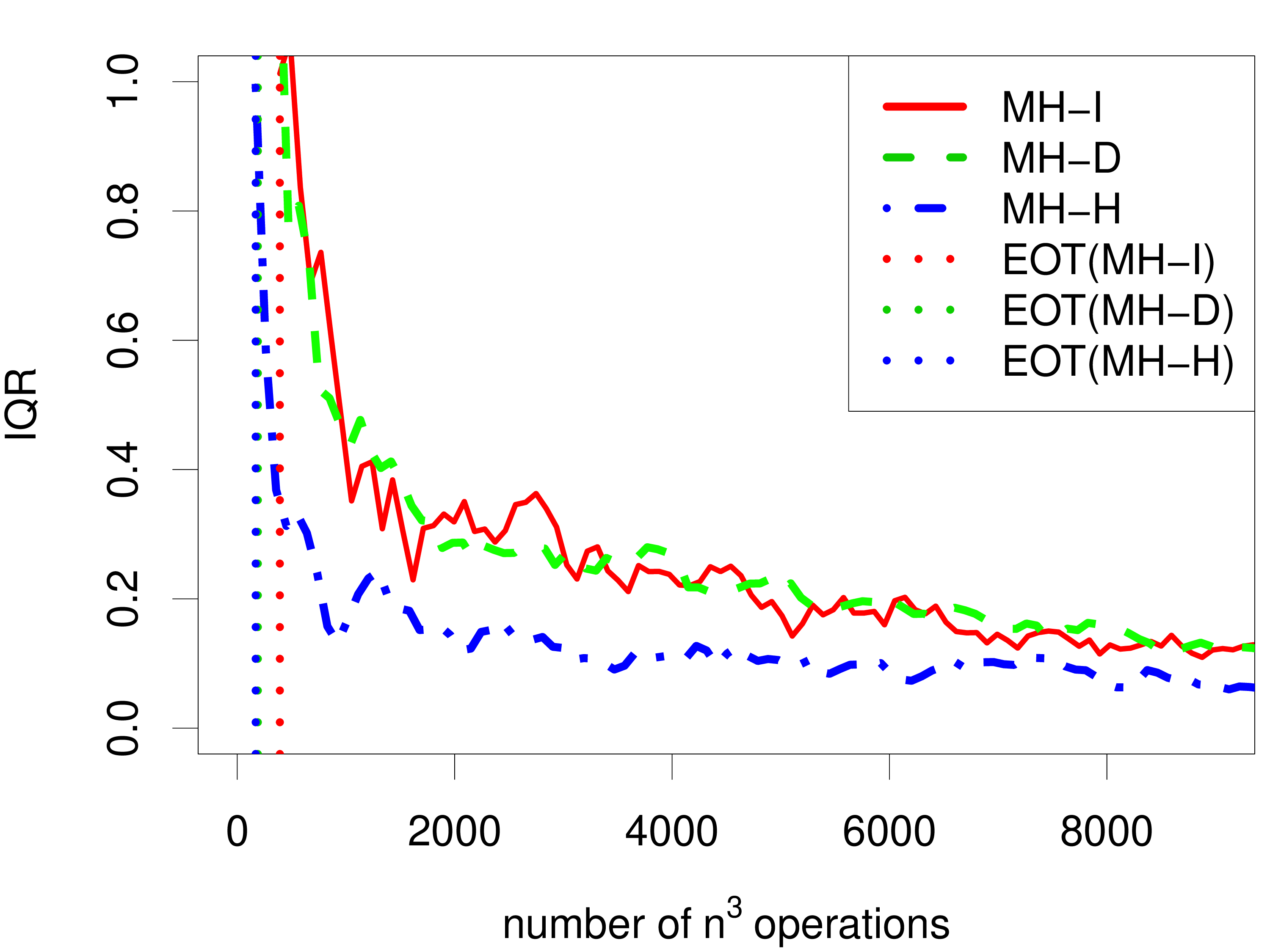}&
          \includegraphics[width=\figwidth,angle=0]{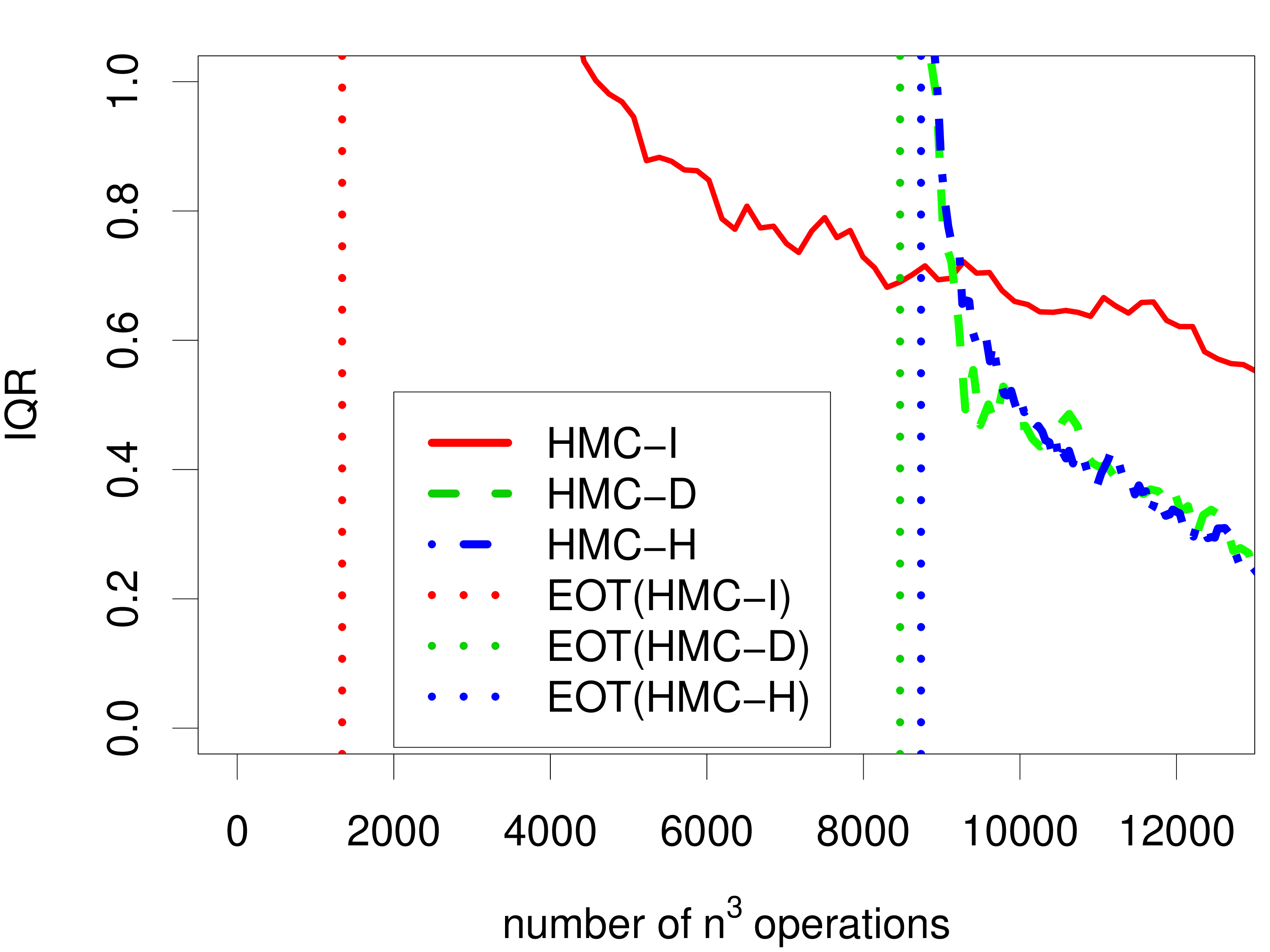}\\
          \includegraphics[width=\figwidth,angle=0]{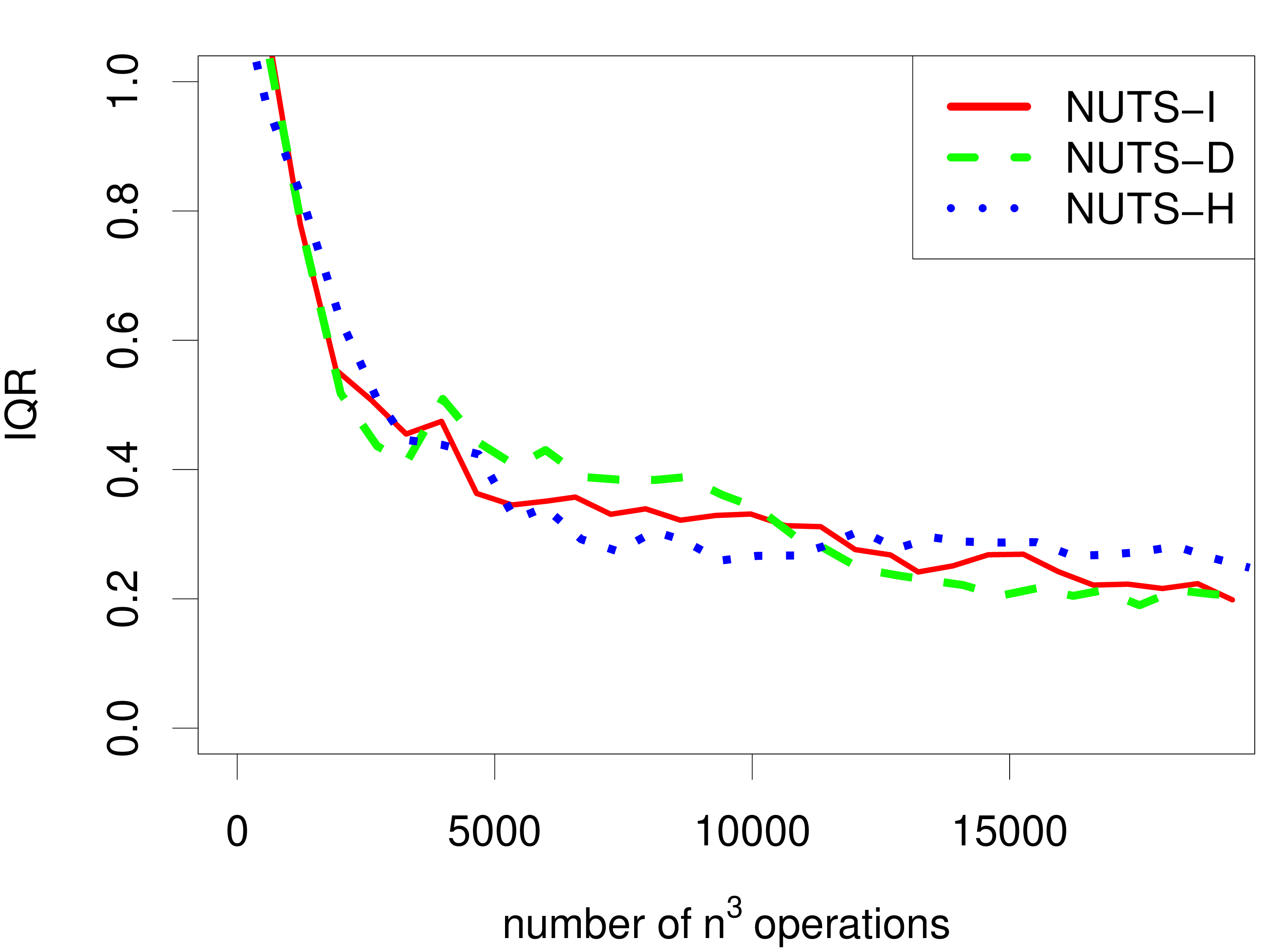}&
         \includegraphics[width=\figwidth,angle=0]{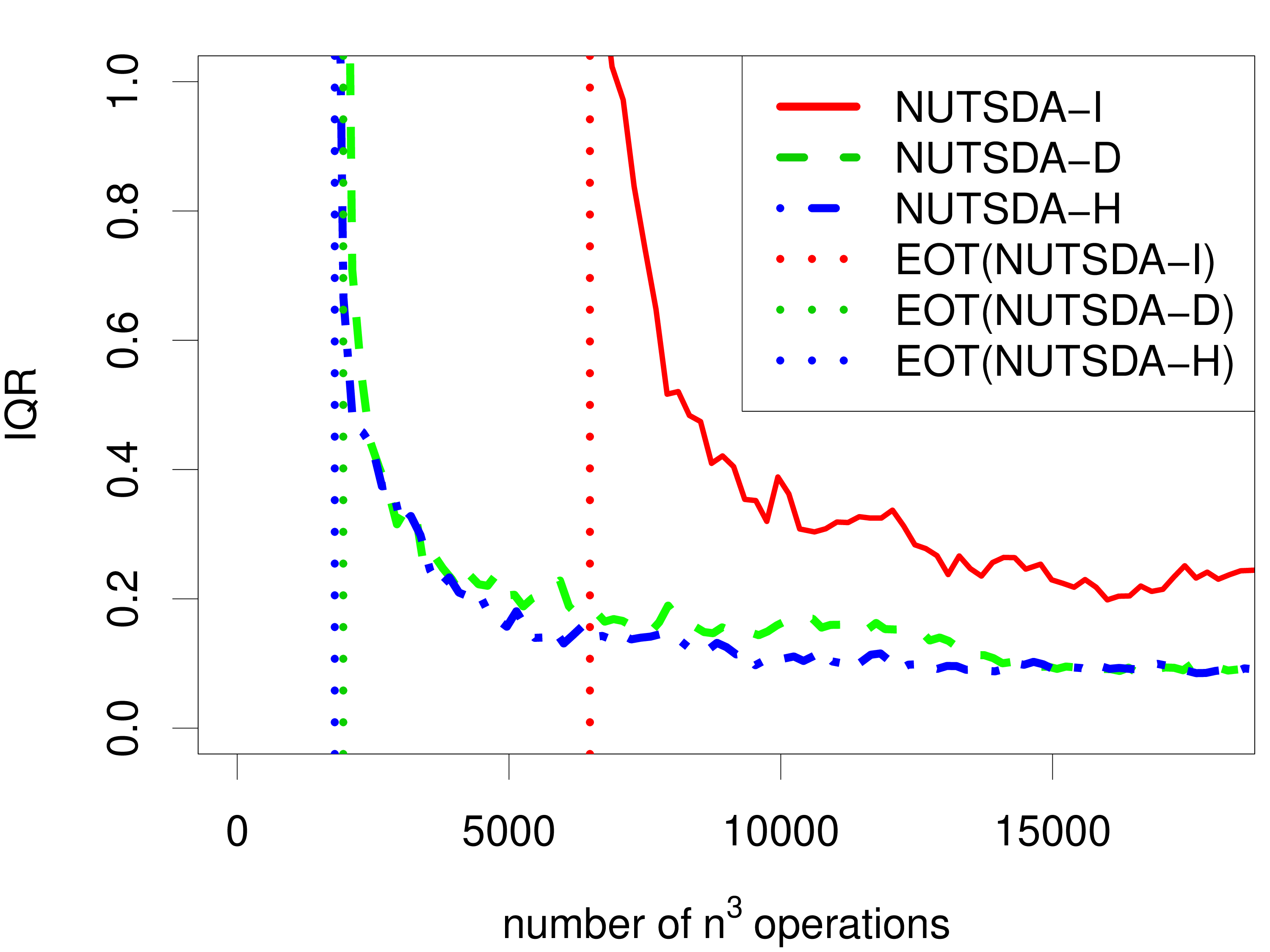}&
         \includegraphics[width=\figwidth,angle=0]{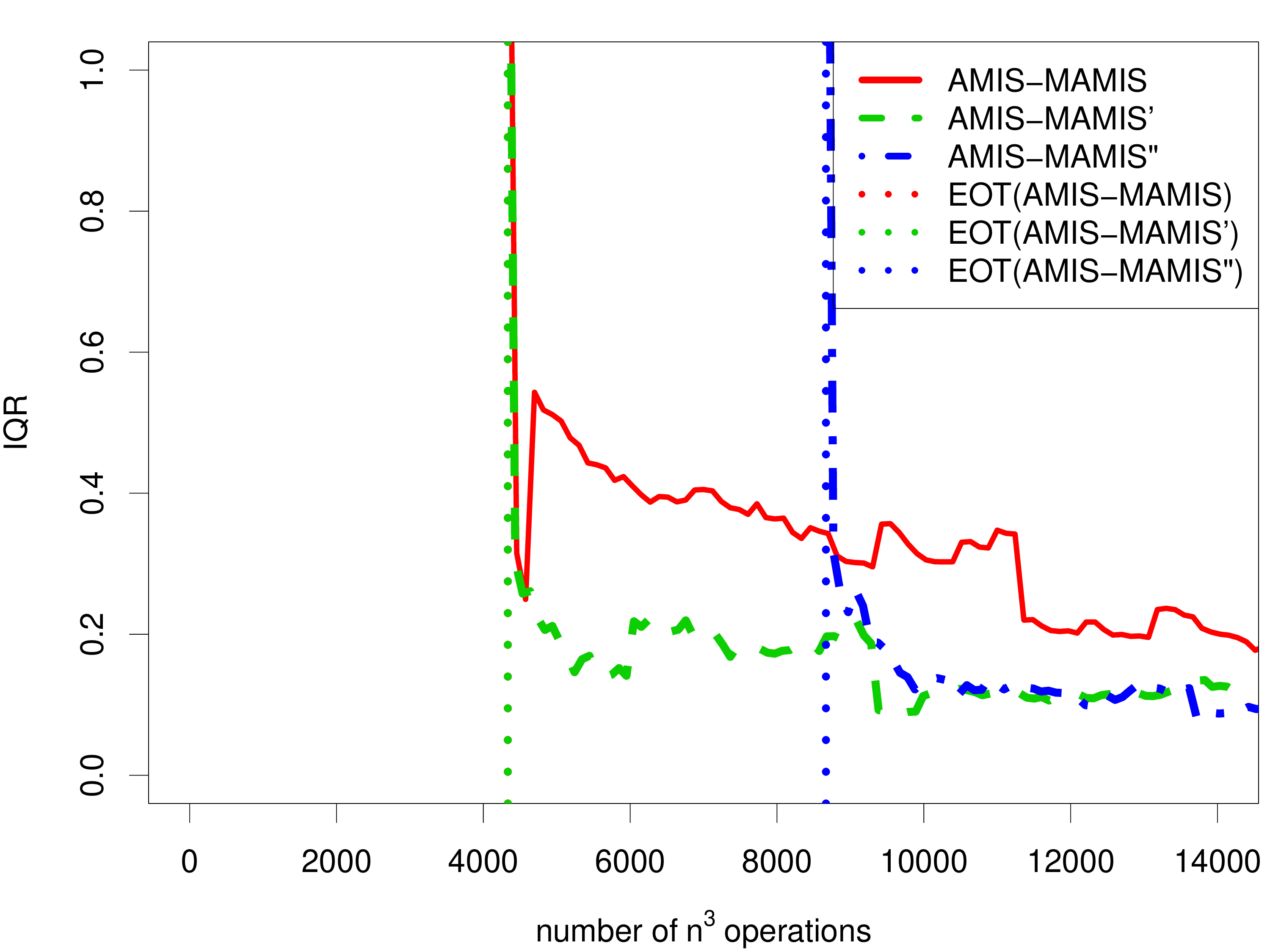}\\
          
        \end{tabular}
     \caption{Convergence of AMIS/MAMIS, MH, HMC, NUTS, NUTSDA for the Parkinsons dataset. EOT stands for "end of tuning". \label{fig:MH_HMC_NUT_NUTDA_PARK_ARD}}   
  \end{center}
  
\end{figure*}
\FloatBarrier

\clearpage
\section{Convergence of samplers for GP classification} \label{App:AppendixC}

\begin{figure*}[th]
  \begin{center}
  {\scriptsize  {\bf RBF }}\\
  \begin{tabular}{ccc}  
          \includegraphics[width=\figwidth,angle=0]{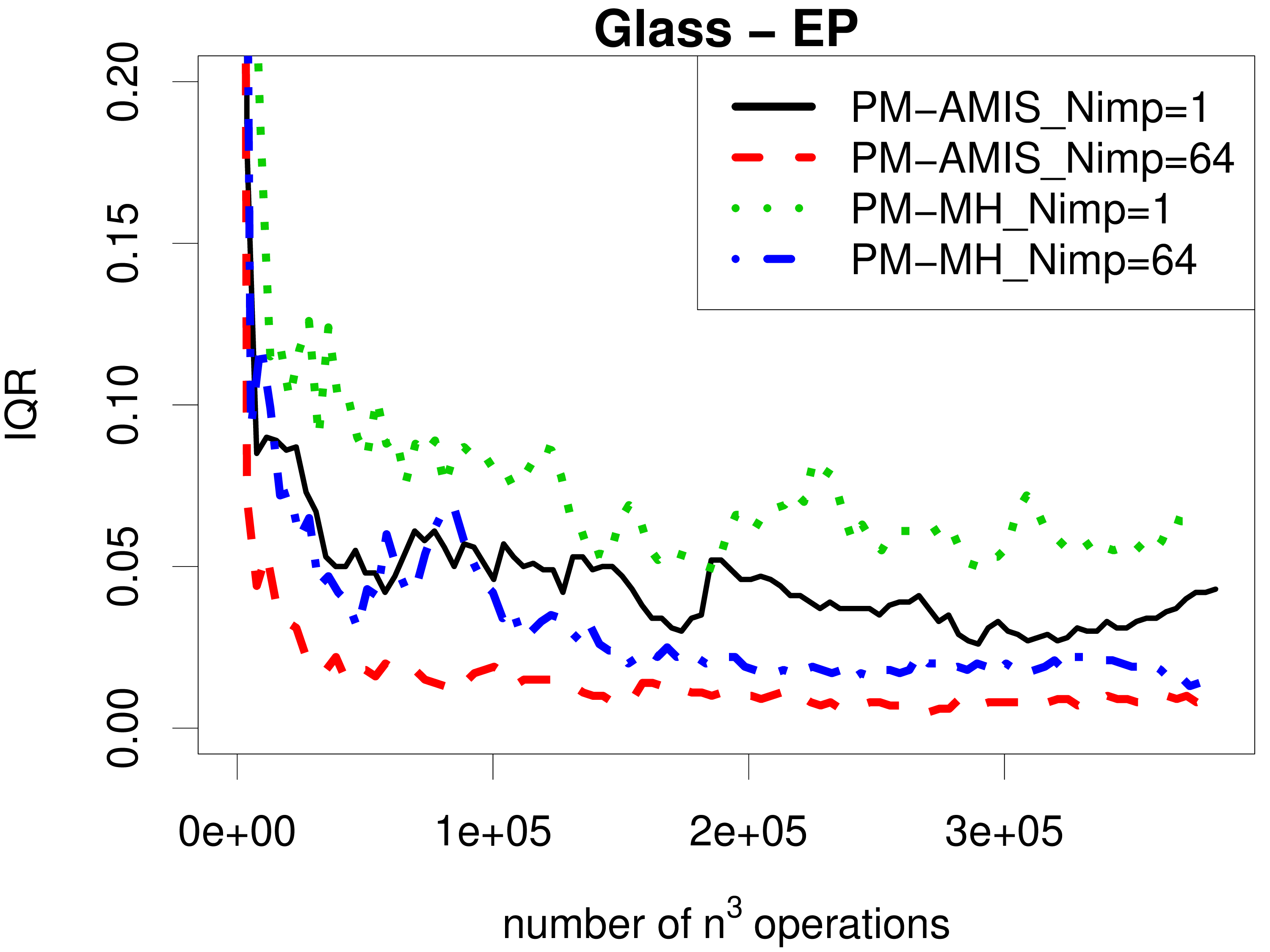}&
          \includegraphics[width=\figwidth,angle=0]{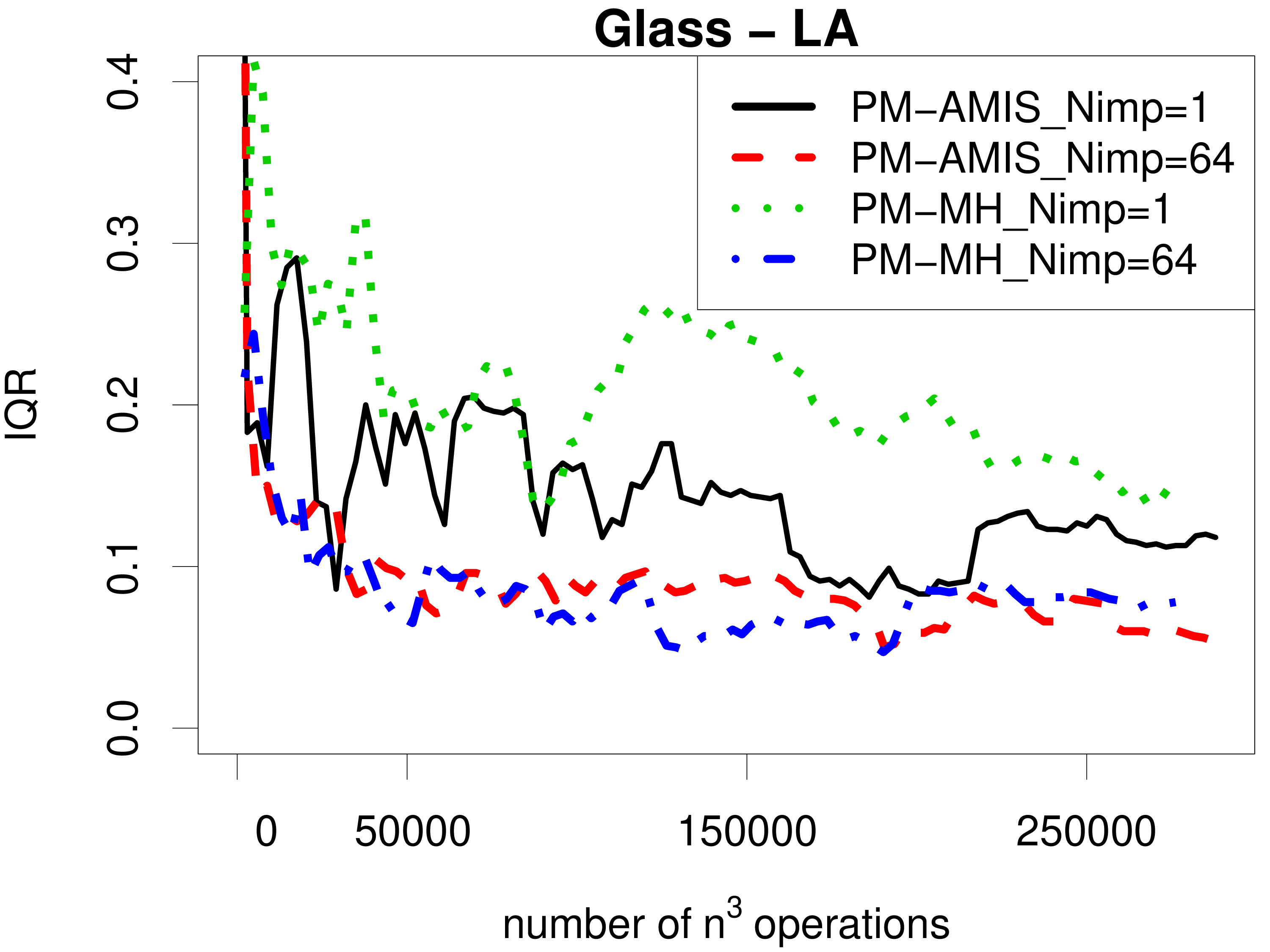}&
          \includegraphics[width=\figwidth,angle=0]{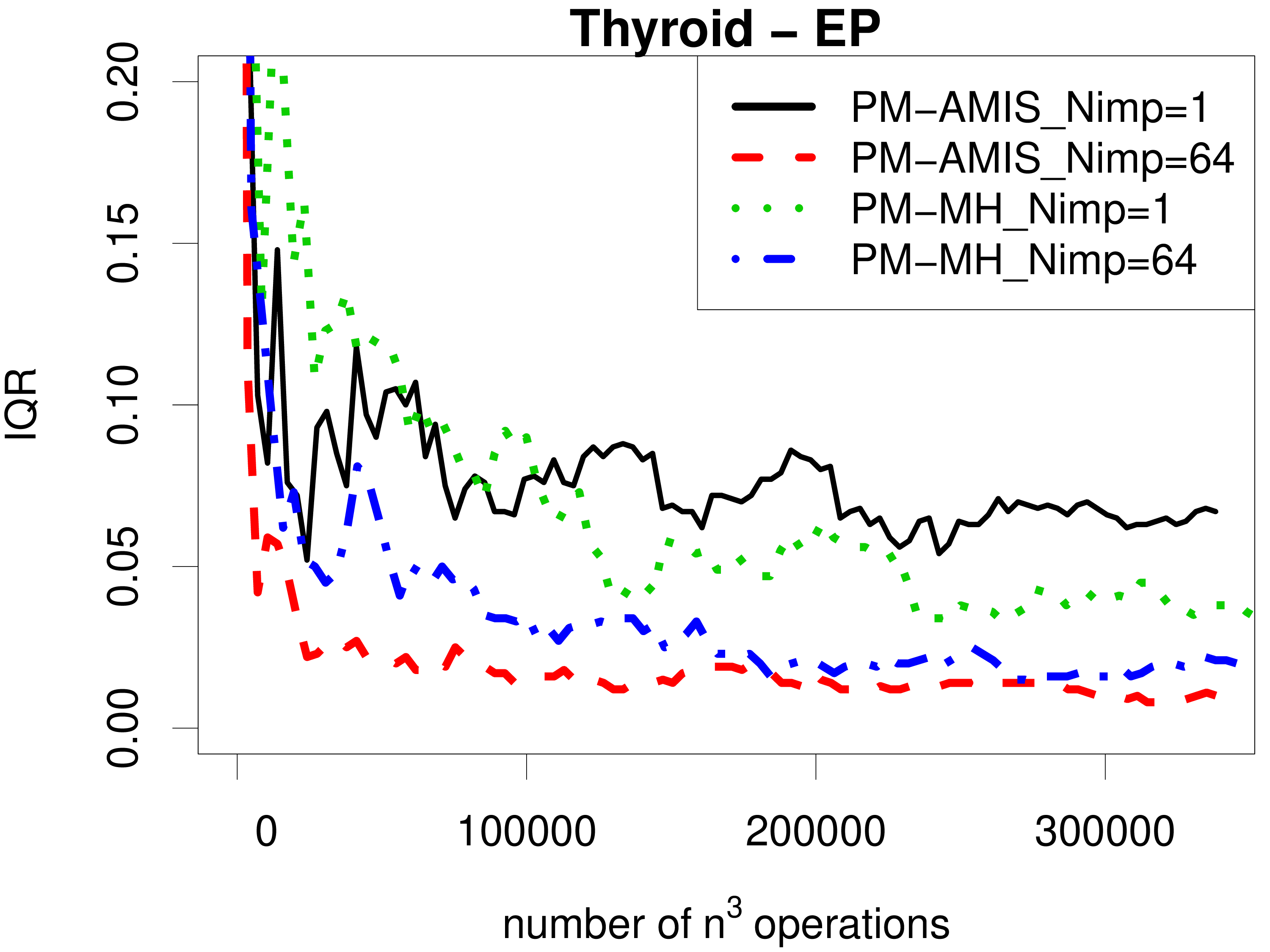}\\
          \includegraphics[width=\figwidth,angle=0]{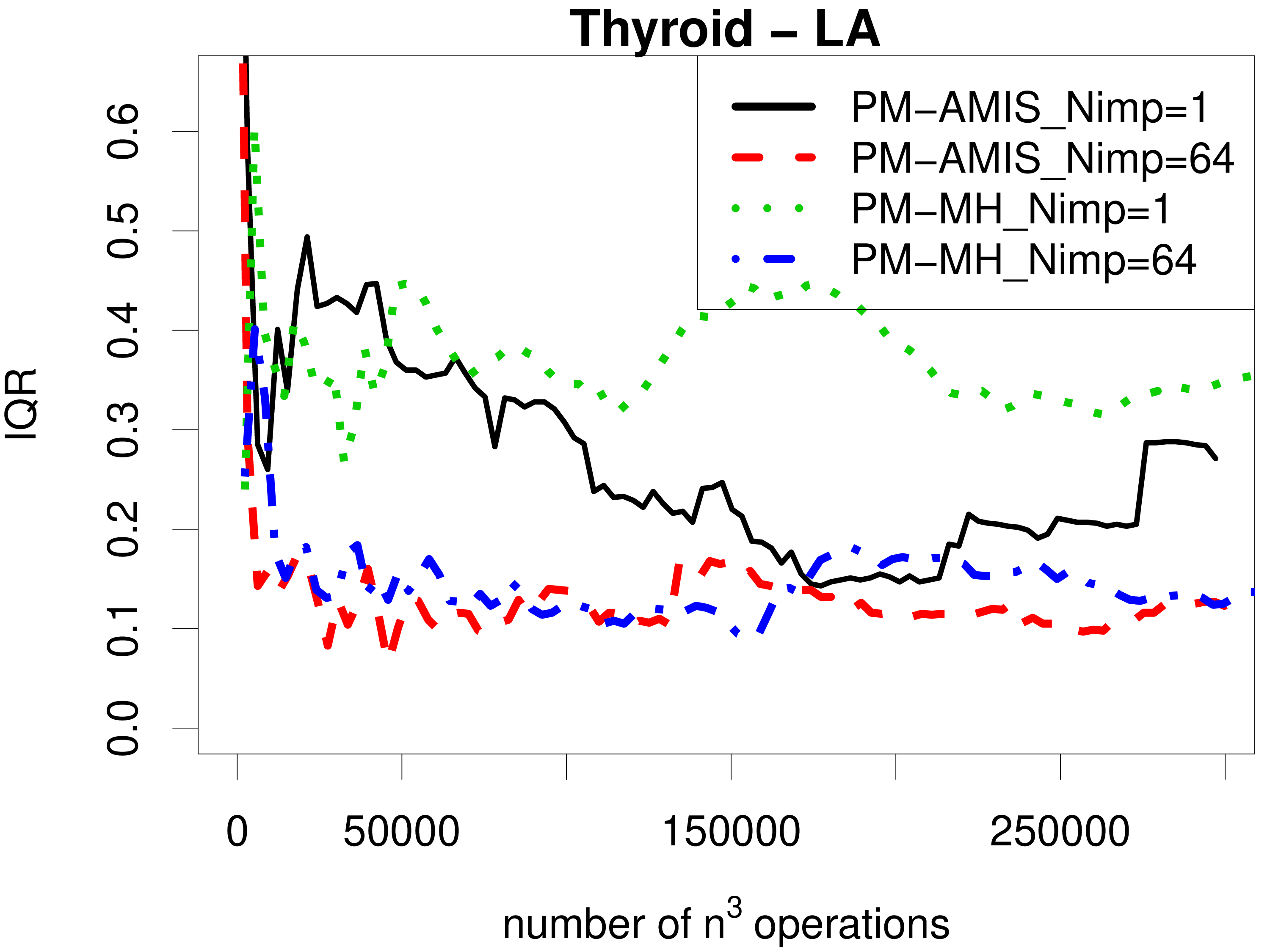}&
          \includegraphics[width=\figwidth,angle=0]{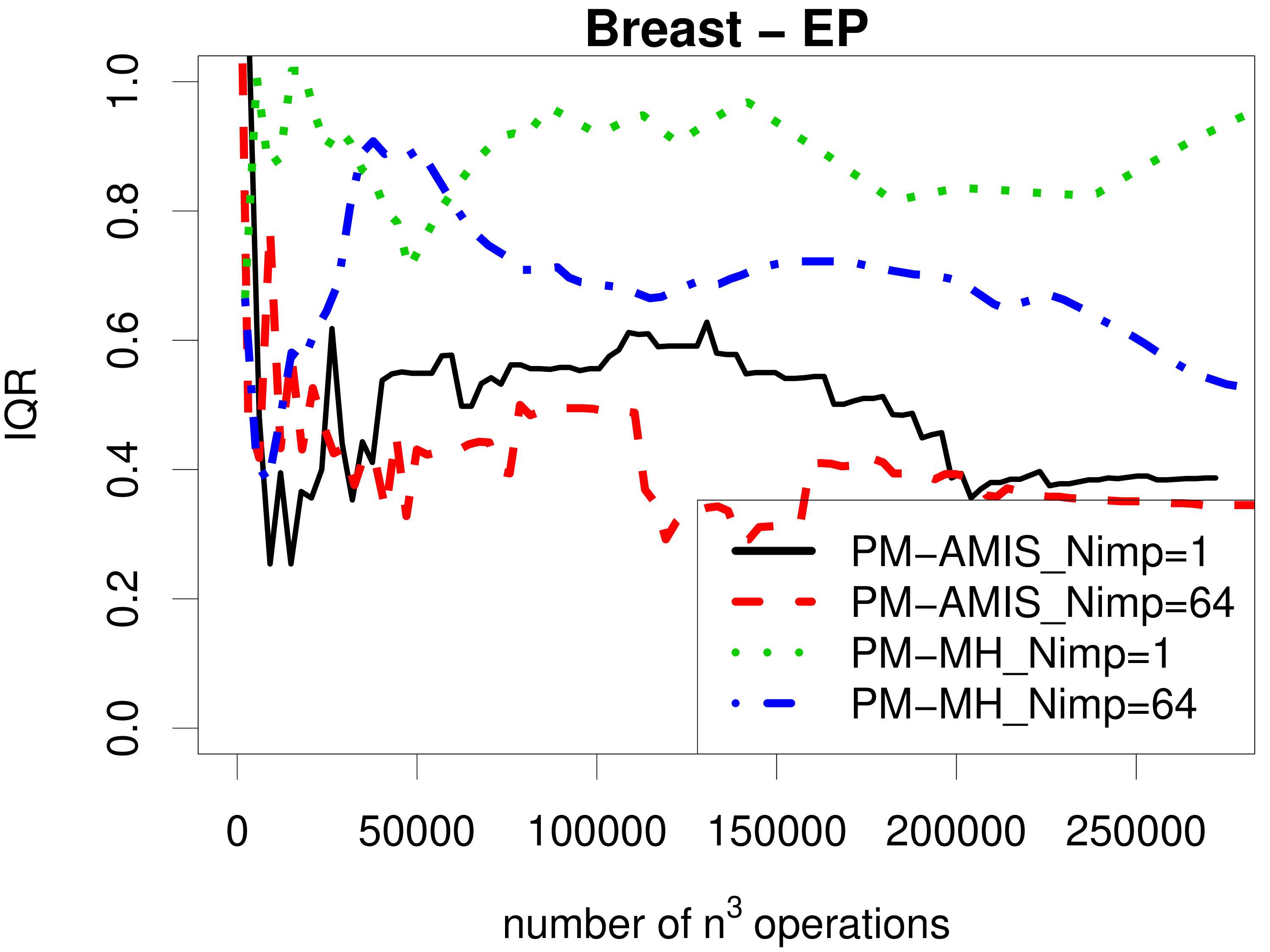}&
          \includegraphics[width=\figwidth,angle=0]{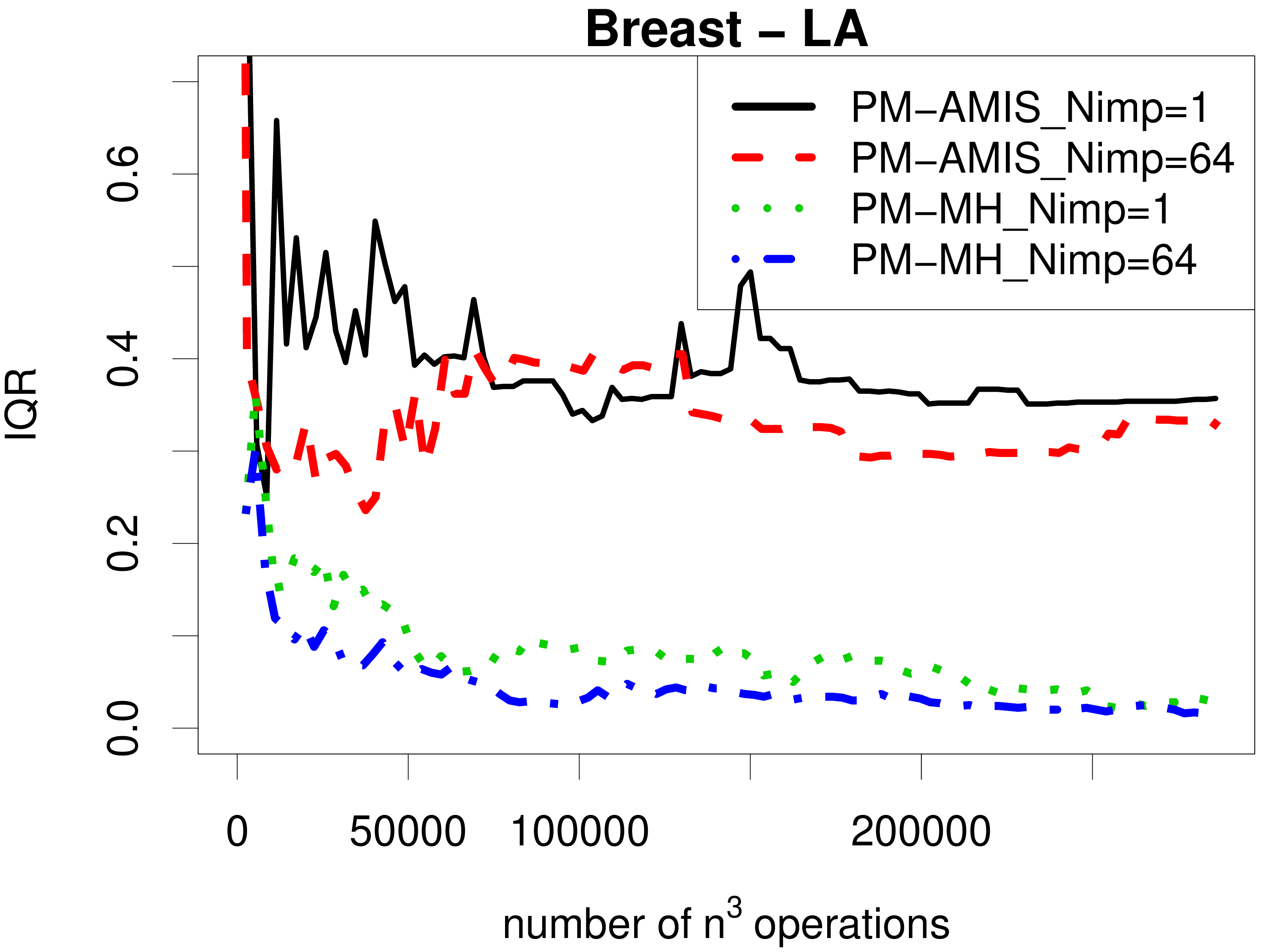}\\
         
   \end{tabular} 
   \caption{Convergence of PM-AMIS, PM-MH for the RBF case. LA indicates the Gaussian approximation to the posterior of latent variables $\mathbf{f}$ is obtained by LA approximation, whereas EP indicates the Gaussian approximation is obtained by EP approximation. Nimp denotes the number of importance samples of latent variables  to estimate the marginal likelihood $p(\mathbf{y} \mid \boldsymbol{\theta})$.  \label{fig:PM_RBF}}
  \end{center} 
\end{figure*}

\begin{figure*}[th]
  \begin{center}
  {\scriptsize  {\bf ARD}}\\
  \begin{tabular}{ccc}  
            \includegraphics[width=\figwidth,angle=0]{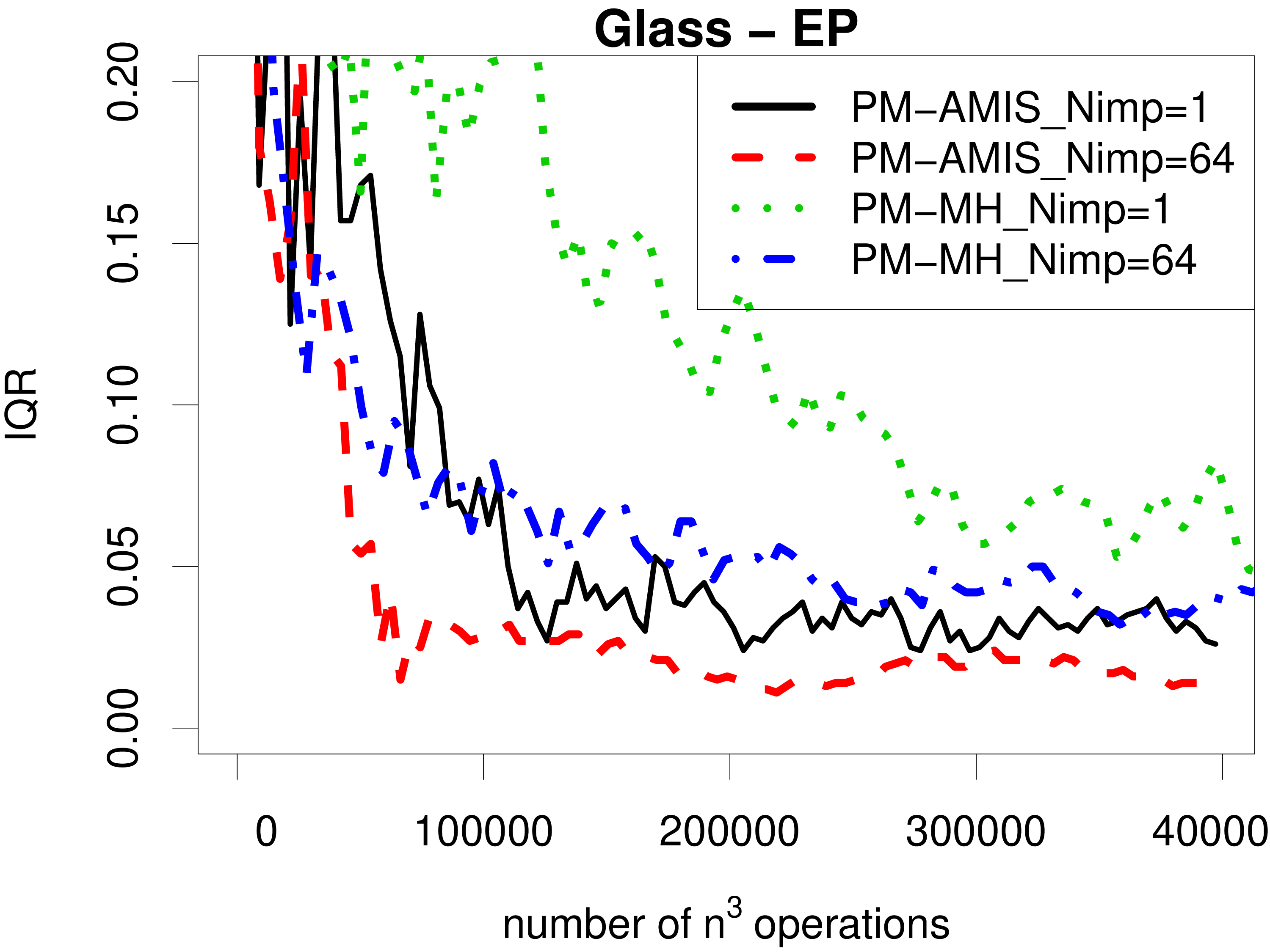}&
            \includegraphics[width=\figwidth,angle=0]{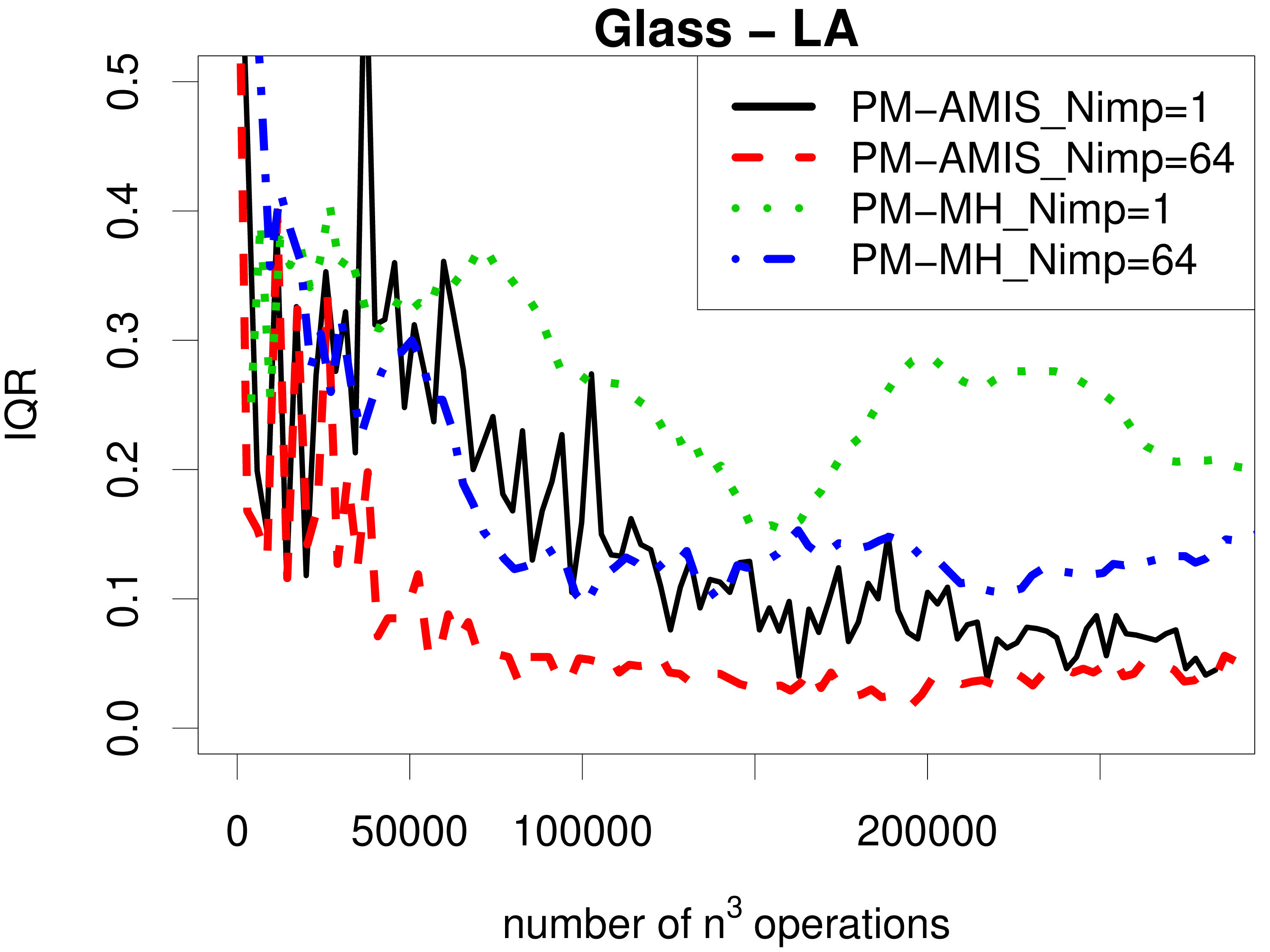}&
           \includegraphics[width=\figwidth,angle=0]{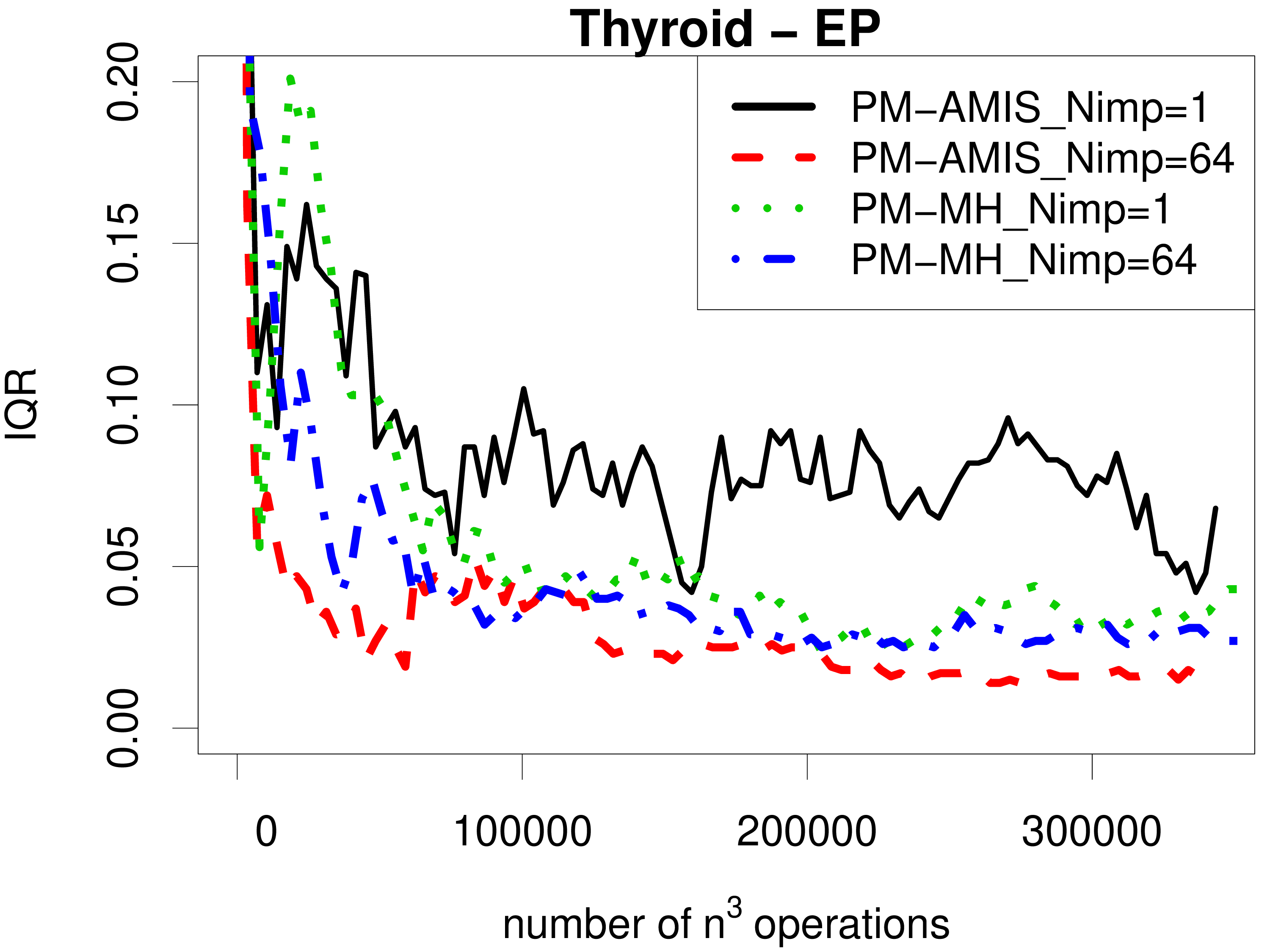}\\
           \includegraphics[width=\figwidth,angle=0]{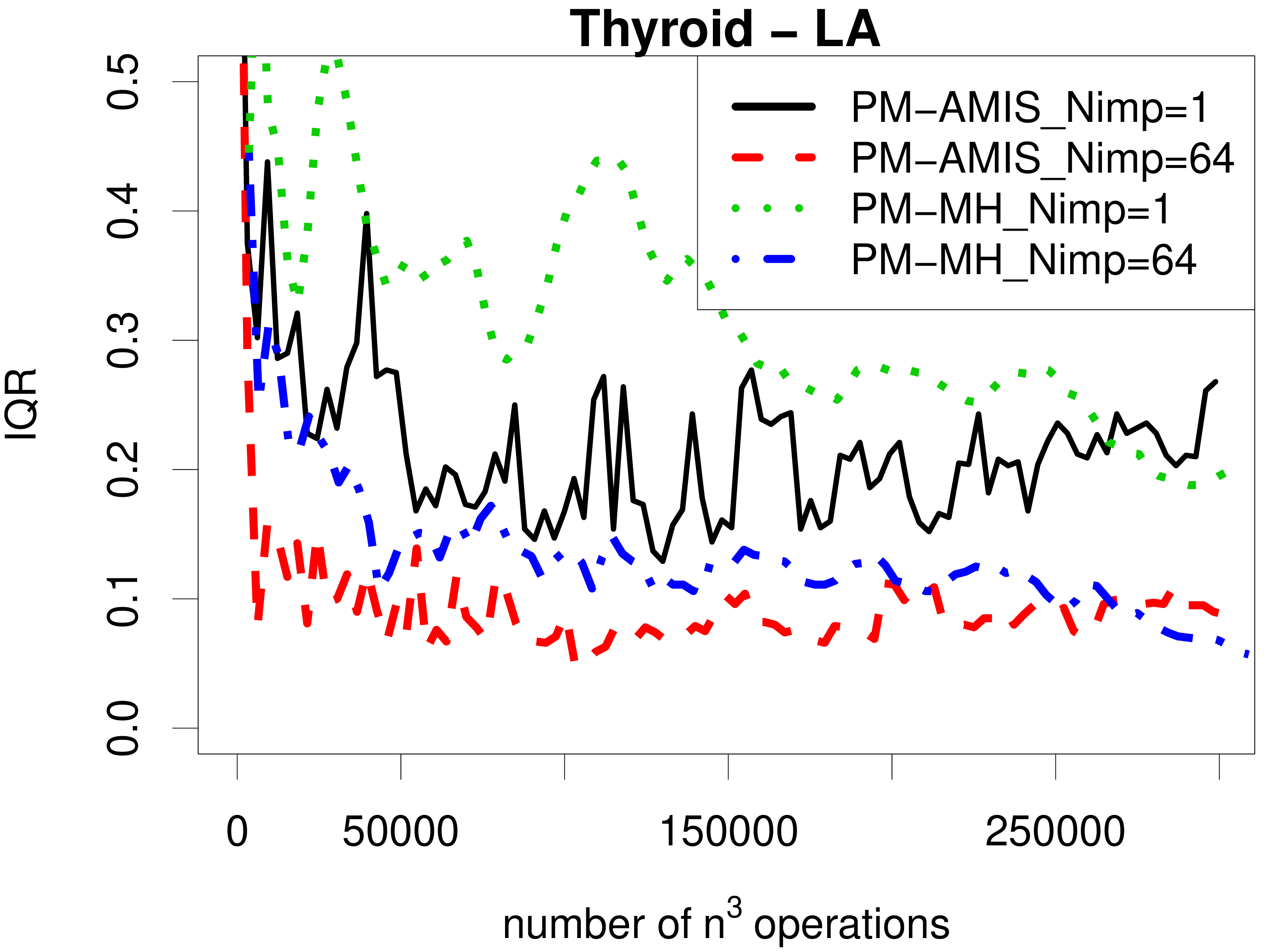}&
           \includegraphics[width=\figwidth,angle=0]{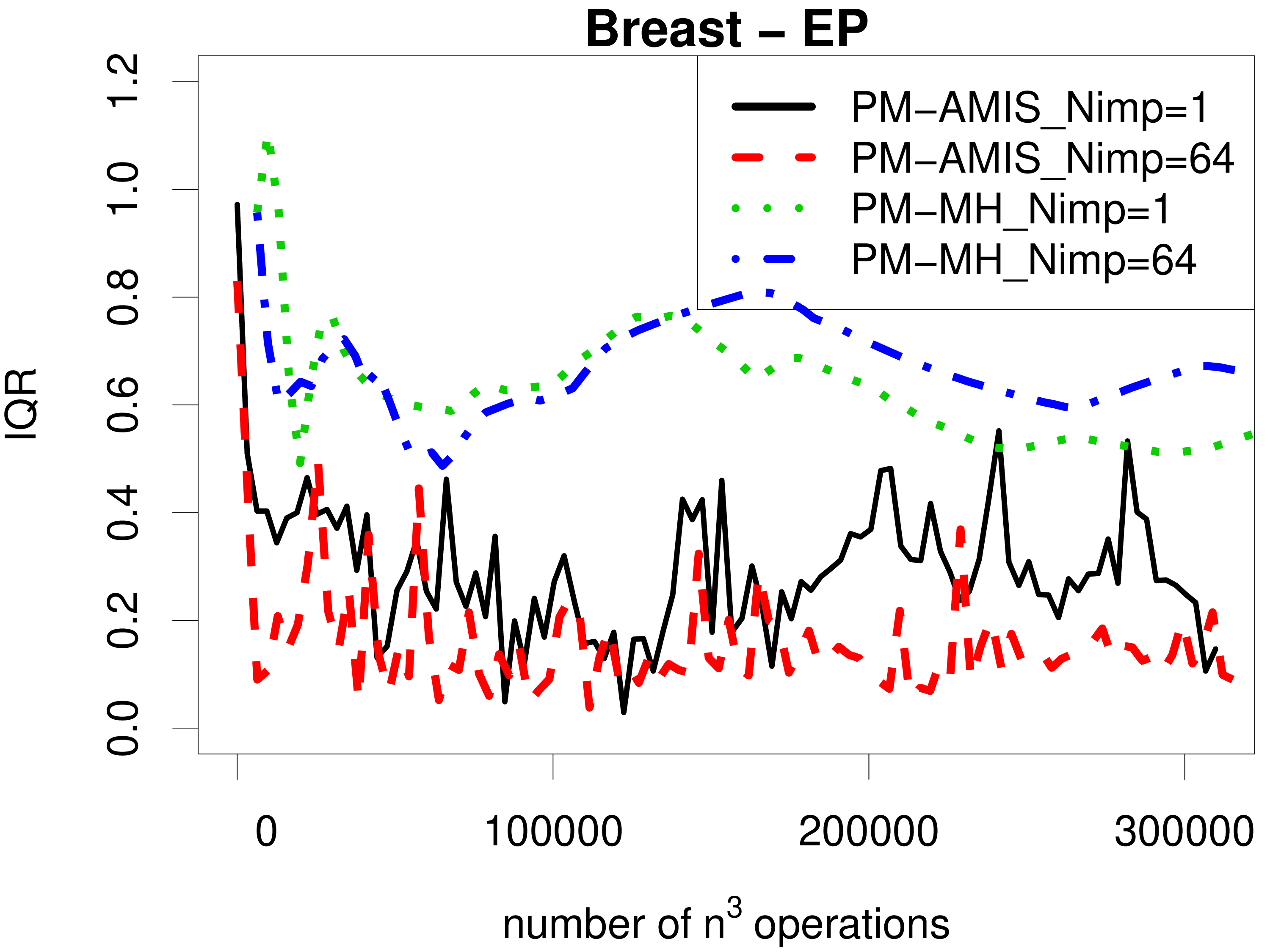}&
            \includegraphics[width=\figwidth,angle=0]{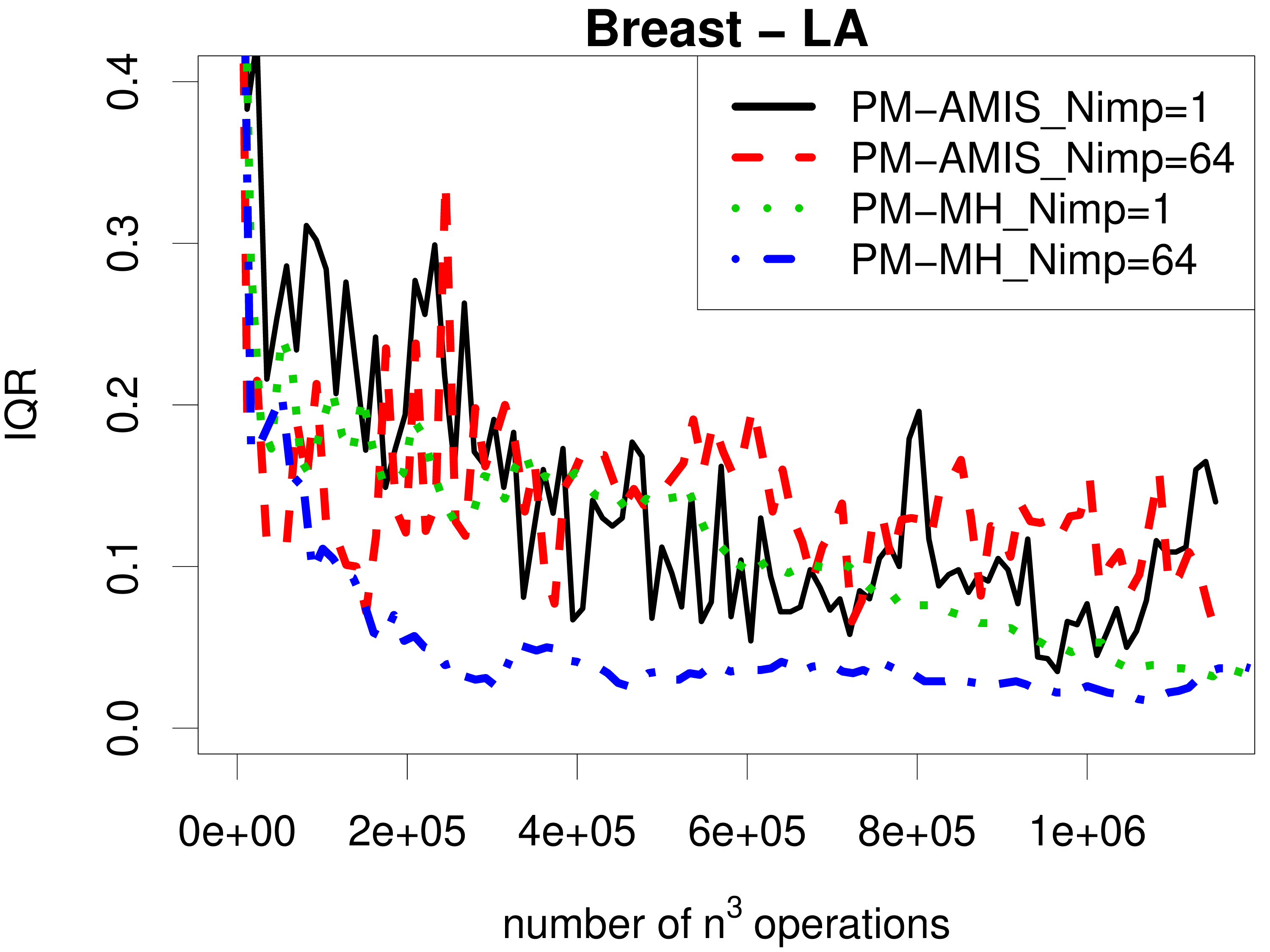}\\               
       
   \end{tabular} 
   \caption{Convergence of PM-AMIS, PM-MH for the ARD case. LA indicates the Gaussian approximation to the posterior of latent variables $\mathbf{f}$ is obtained by LA approximation, whereas EP indicates the Gaussian approximation is obtained by EP approximation. Nimp denotes the number of importance samples of latent variables to estimate the marginal likelihood $p(\mathbf{y} \mid \boldsymbol{\theta})$. \label{fig:PM_ARD}}
  \end{center} 
\end{figure*}

\end{document}